%% file: SMP-18-012_temp.tex
\pdfoutput=1

\documentclass[11pt,twoside,a4paper,cmspaper,final,collab]{cms-tdr}
\def\svnVersion{e2631f8-D}\def\svnDate{2020/12/04}\def\cmsCernNoTag{CERN-EP-2020-116}\def\cmsCernDate{\today}\def\cmsMessage{"Published in Physical Review D as \href{http://dx.doi.org/10.1103/PhysRevD.102.092012}{\doi{10.1103/PhysRevD.102.092012}.}"}

\begin{document}\cmsNoteHeader{SMP-18-012}

\hyphenation{had-ron-i-za-tion}
\hyphenation{cal-or-i-me-ter}
\hyphenation{de-vices}

\newlength\cmsFigWidth
\ifthenelse{\boolean{cms@external}}{\setlength\cmsFigWidth{0.49\textwidth}}{\setlength\cmsFigWidth{0.65\textwidth}}
\ifthenelse{\boolean{cms@external}}{\providecommand{\cmsLeft}{upper\xspace}}{\providecommand{\cmsLeft}{left\xspace}}
\ifthenelse{\boolean{cms@external}}{\providecommand{\cmsRight}{lower\xspace}}{\providecommand{\cmsRight}{right\xspace}}
\ifthenelse{\boolean{cms@external}}{\providecommand{\NA}{\ensuremath{\cdots}\xspace}}{\providecommand{\NA}{\ensuremath{\text{---}}\xspace}}

\cmsNoteHeader{SMP-18-012}

\title{Measurements of the \texorpdfstring{$\PW$}{W} boson rapidity, helicity, double-differential cross sections, and charge asymmetry in \texorpdfstring{$\Pp\Pp$}{pp} collisions at \texorpdfstring{$\sqrt{s} = 13\TeV$}{sqrt(s) = 13 TeV}}

\date{\today}

\newcommand{\lint}{35.9\fbinv\xspace}
\newcommand{\cost}{\ensuremath{\cos{\theta^{*}}}}
\newcommand{\yw}{\ensuremath{y_{\PW}}}
\newcommand{\ptw}{\ensuremath{\PT^{\PW}}\xspace}
\newcommand{\ptz}{\ensuremath{\PT^{\PZ}}\xspace}
\newcommand{\ptl}{\ensuremath{\PT^{\ell}}\xspace}
\newcommand{\ptmu}{\ensuremath{\PT^{\mu}}\xspace}
\newcommand{\ptel}{\ensuremath{\PT^{\Pe}}\xspace}
\newcommand{\leta}{\ensuremath{\eta^{\ell}}\xspace}
\newcommand{\etapt}{\ensuremath{(\ptl,\leta)}}
\newcommand{\etamu}{\ensuremath{\eta^{\mu}}\xspace}
\newcommand{\etael}{\ensuremath{\eta^{\Pe}}\xspace}
\newcommand{\absyw}{\ensuremath{\abs{y_{\PW}}}\xspace}
\newcommand{\abseta}{\ensuremath{\abs{\eta}}\xspace}
\newcommand{\absleta}{\ensuremath{\abs{\eta^{\ell}}}\xspace}
\newcommand{\wplus}{\ensuremath{\PWp}\xspace}
\newcommand{\wminus}{\ensuremath{\PWm}\xspace}
\newcommand{\wright}{\ensuremath{\PW_\mathrm{R}}\xspace}
\newcommand{\wleft}{\ensuremath{\PW_\mathrm{L}}\xspace}
\newcommand{\wlong}{\ensuremath{\PW_\mathrm{0}}\xspace}
\newcommand{\wplusr}{\ensuremath{\PW^+_\mathrm{R}}\xspace}
\newcommand{\wplusl}{\ensuremath{\PW^+_\mathrm{L}}\xspace}
\newcommand{\wminusr}{\ensuremath{\PW^-_\mathrm{R}}\xspace}
\newcommand{\wminusl}{\ensuremath{\PW^-_\mathrm{L}}\xspace}
\newcommand{\wplusmu}{\ensuremath{\PWp\to\Pgmp\nu}\xspace}
\newcommand{\wpluse}{\ensuremath{\PWp\to\Pep\nu}\xspace}
\newcommand{\wminusmu}{\ensuremath{\PWm\to\Pgmm\PAGn}\xspace}
\newcommand{\wminuse}{\ensuremath{\PWm\to\Pem\PAGn}\xspace}
\newcommand{\wplusell}{\ensuremath{\PWp\to\ell^+\nu}\xspace}
\newcommand{\wminusell}{\ensuremath{\PWm\to\ell^-\PAGn}\xspace}
\newcommand{\thetastar}{\ensuremath{\theta^{*}}\xspace}
\newcommand{\NNPDF} {\textsc{NNPDF3.0}\xspace}
\newcommand{\Zee}{\ensuremath{\cPZ\to\Pe\Pe} }
\newcommand{\mll}{\ensuremath{m_{2\ell}}}
\newcommand{\muR}{\ensuremath{\mu_\mathrm{R}}\xspace}
\newcommand{\muF}{\ensuremath{\mu_\mathrm{F}}\xspace}
\ifthenelse{\boolean{cms@external}}{\providecommand{\cmsTable}[1]{#1}}{\providecommand{\cmsTable}[1]{\resizebox{\textwidth}{!}{#1}}}

\abstract{
The differential cross section and charge asymmetry for inclusive \PW boson production at $\sqrt{s}=13\TeV$ is measured for the two transverse polarization states as a function of the \PW boson absolute rapidity. The measurement uses events in which a \PW boson decays to a neutrino and either a muon or an electron. The data sample of proton-proton collisions recorded with the CMS detector at the LHC in 2016 corresponds to an integrated luminosity of 35.9\fbinv. The differential cross section and its value normalized to the total inclusive \PW boson production cross section are measured over the rapidity range $\absyw<2.5$. In addition to the total fiducial cross section, the \PW boson double-differential cross section, $\rd^2\sigma/\rd\ptl\rd\absleta$, and the charge asymmetry are measured as functions of the charged lepton transverse momentum and pseudorapidity. The precision of these measurements is used to constrain the parton distribution functions of the proton using the next-to-leading order NNPDF3.0 set.  }

\hypersetup{
pdfauthor={CMS Collaboration},
pdftitle={Measurements of the W boson rapidity, helicity, double-differential cross sections, and charge asymmetry in pp collisions at sqrt(s)=13 TeV},
pdfsubject={CMS},
pdfkeywords={CMS, physics, W boson}}

\maketitle

\section{Introduction}
\label{sec:introduction}
The standard model (SM) of particle physics provides a description of
nature in terms of fundamental particles and their interactions
mediated by vector bosons. The electromagnetic and weak interactions
are described by a unified gauge theory based on the
$\mathrm{SU(2)_L{\times}U(1)_Y}$ symmetry group, where the photon, the
\PW boson, and the \PZ boson act as mediators of the unified
electroweak
interaction~\cite{Glashow:1961tr,Weinberg:1967tq,Salam:1964ry}.

Measurements of the kinematic properties of \PW bosons produced at
hadron colliders provide stringent tests of perturbative quantum
chromodynamics (QCD) calculations and probe the nature of the
electroweak interaction. In particular, the measurement of the
polarization of the \PW boson is fundamental in determining its
production mechanism.

At leading order (LO) in QCD, \PW bosons are produced at a hadron
collider with small transverse momentum (\pt) through the annihilation
of a quark and an antiquark: $\PQu\PAQd$ for the \wplus and
$\PAQu\PQd$ for the \wminus. At the CERN LHC, \PW bosons with
large rapidity (\absyw) are produced predominantly with momentum in
the same direction as the momentum of the quark that participates in
the hard scattering. This is because the parton distribution functions
(PDFs) of the proton favor the quark to carry a larger fraction ($x$)
of the proton momentum rather than the antiquark~\cite{Bjorken}.

Because of the V$-$A coupling of the \PW boson to fermions in the SM,
the spin of the \PW boson is aligned with that of the quark, \ie,
purely left-handed, and thus aligned opposite to the direction of the
momenta of both the \PW boson and the quark.  With smaller \absyw,
the \PW bosons produced at the LHC become a mixture of left-, and
right-handed polarization states at LO in QCD, and the rates of the
two polarizations become equal at $\absyw=0$.  With increasing \PW
boson \pt (\ptw), next-to-leading order (NLO) amplitudes contribute
in its production, and longitudinally polarized \PW bosons arise.
The relative fractions of the three polarization states depend on the
relative size of the amplitudes of the three main production
processes: $\PQu\PAQd\to\wplus \Pg$, $\PAQu\Pg \to\wplus \PQd$, and
$\Pg\PAQd \to\wplus\PAQu$, and are determined by the PDFs at high
values of $x$. Overall, left-handed \PW bosons are favored at the
LHC over right-handed and longitudinally polarized \PW bosons. The
relative fraction of positively (negatively) charged left-handed \PW
bosons is around 65 (60)\%, of right-handed \PW bosons around 28
(33)\%, and of longitudinally polarized \PW bosons around 7 (7)\% of
the total production cross section. The fraction of longitudinally
polarized \PW bosons increases monotonically with \ptw in the \ptw range relevant for this analysis.

At the LHC, \PW bosons are produced in large quantities, and it is
easy to trigger on their leptonic decays ($\PW\to\ell\nu$) with high
purity. Since the escaping neutrino means the momentum of the \PW
boson is not known, the direct measurement of the fully differential
cross section of the \PW boson is not possible. In particular, the
polarization and rapidity distributions of the \PW boson must be
inferred by using the PDFs. Uncertainties stemming from the imperfect
knowledge of these PDFs contribute a large fraction of the overall
uncertainties in recent measurements of the mass of the \PW
boson~\cite{Aaboud:2017svj} and in other high-precision measurements
at the LHC~\cite{Stirling:1989vx}.

Constraints on the PDFs and their uncertainties are possible through
many different measurements. Recently, the ATLAS and CMS
Collaborations published PDF constraints from double-differential
measurements of \PZ boson production and the accurate measurement of
$\sin^2\theta_{\PW}$~\cite{Aaboud:2017ffb,Sirunyan:2018swq,Sirunyan:2018owv}.
Studies of \PW bosons have been used by the ATLAS and CMS
Collaborations to set constraints on PDFs through the measurement of
charge asymmetries, in particular as a function of the charged lepton
pseudorapidity
\leta~\cite{CDFasymmetry,D0asymmetry,Aaboud:2016btc,Aad:2019rou,Chatrchyan:2012xt,Chatrchyan:2013mza,Khachatryan:2016pev,Aaij:2012vn,Aaij:2015zlq}.
Measurements of associated production of a \PW boson and a charm
quark by the ATLAS, CMS, and LHCb Collaborations at the
LHC~\cite{Sirunyan:2018hde,Aad:2014xca,Aaij:2015cha}, and by the CDF
and D0 Collaborations at the Fermilab
Tevatron~\cite{Aaltonen:2012wn,Abazov:2008qz}, also contribute to
constrain the strange quark distribution within the light quark sea in
the proton.

Previous measurements of the decay characteristics and polarization 
of $\PW$ bosons have been carried out by collaborations at the Tevatron
and the LHC~\cite{PhysRevD.70.032004,PhysRevD.63.072001,ATLAS:2012au,PhysRevLett.107.021802}.

Recently, a method has been proposed to directly measure the rapidity
spectrum differentially in three helicity states~\cite{Manca:2017xym}
for \PW bosons at the LHC. It exploits the fact that the three
helicity states of the leptonically decaying \PW boson behave
differently in the two-dimensional (2D) plane of observable lepton
transverse momentum \pt (\ptl) and \leta.

This paper describes an experimental implementation of this novel
method of measuring the \PW boson production differentially in its
helicity states, rapidity, and electric charge.  In addition, a
measurement of the charge asymmetry as a function of \absyw is
presented.  Furthermore, cross sections for \PW boson production are
provided as a function of the charged lepton kinematics in the 2D
plane of \ptl and \absleta, unfolded to particle level, along with the
fiducial cross section in the experimental phase space.

The paper is organized as follows. Section~\ref{sec:cms} gives a brief
description of the CMS detector, followed by
Sec.~\ref{sec:dataSimulation} detailing the data sample and the
simulated samples used for this
analysis. Section~\ref{sec:recoSelection} summarizes the physics
object and event selection. Section~\ref{sec:backgrounds} describes
the relevant background sources and the methods to estimate their
contributions. Section~\ref{sec:templatesFitting} explains the
procedure to define the simulated 2D templates for \ptl and \leta and
the fitting strategy to perform the statistical analysis. The
treatment of the systematic uncertainties is documented in
Sec.~\ref{sec:systematics}. The results are presented in
Sec.~\ref{sec:results} and a summary in
Sec.~\ref{sec:conclusions}.

\section{The CMS detector}
\label{sec:cms}
The central feature of the CMS apparatus is a superconducting solenoid
of 6\unit{m} internal diameter, providing a magnetic field of 3.8 T. A
silicon pixel and strip tracker, a lead tungstate crystal
electromagnetic calorimeter (ECAL), and a brass and scintillator
hadron calorimeter (HCAL), each composed of a barrel and two end cap
sections, reside within the solenoid volume. Muons are measured in
gas-ionization detectors embedded in the steel flux-return yoke
outside the solenoid. Extensive forward calorimetry complements the
coverage provided by the barrel and end cap section detectors. A more
detailed description of the CMS detector can be found in
Ref.~\cite{Chatrchyan:2008zzk}.

Events of interest are selected using a two-tiered trigger
system~\cite{Khachatryan:2016bia}. The first level (L1), composed of
custom hardware processors, uses information from the calorimeters and
muon detectors to select events at a rate of around 100\unit{kHz}
within a latency of 4\mus. The second level, known as the high-level
trigger (HLT), consists of a farm of processors running a version of
the full event reconstruction software optimized for fast processing,
and reduces the event rate to around 1\unit{kHz} before data
storage. In this paper the definition ``on-line'' refers to quantities
computed either in the L1 or in the HLT processing, while ``off-line''
refers to the ones evaluated later on the recorded events.

\section{Data and simulated samples}
\label{sec:dataSimulation}

The measurement is based on a data sample corresponding to an
integrated luminosity of 35.9\fbinv of proton-proton ($\Pp\Pp$) collisions at a
center-of-mass energy of 13\TeV recorded by the CMS experiment at the
LHC during 2016.

Candidate events are selected with single-lepton triggers with online
\pt thresholds of 24 (27)\GeV for muons (electrons) at the HLT. For
electrons, a higher threshold (up to about 40\GeV) for the L1 hardware
trigger was operational during the second half of the 2016 data-taking
period. These higher thresholds were present in the periods of highest
instantaneous luminosities at the beginning of the LHC fills.  Because
of the higher trigger thresholds for electrons, the data sample for
electrons is considerably smaller than that for muons and requires a
careful modeling of the trigger efficiencies as a function of electron
\pt. Identification and isolation criteria are applied for these
triggers to suppress backgrounds before full event reconstruction.

Several Monte Carlo (MC) event generators are used to simulate the
signal and background processes.  The signal sample of $\PW$+jets
events is simulated at NLO in perturbative QCD with the \MGvATNLO
event generator in version 2.2.2.~\cite{Alwall:2014hca}. Relevant
background processes are simulated with \MGvATNLO
($\cPZ\to\ell\ell$ and $\PW\to\tau\nu$ at NLO, and
diboson and top quark-antiquark pair (\ttbar) processes at LO), as
well as with \POWHEG2.0~\cite{Nason:2004rx,powheg2,Alioli:2010xd} at
NLO (single-top processes). All simulated events are interfaced with
the \PYTHIA~8.226~\cite{Sjostrand:2014zea} package and its
CUETP8M1~\cite{Khachatryan:2015pea} tune for parton showering,
hadronization, and underlying event simulation. The \NNPDF set of PDFs
at NLO in QCD is used for all simulated event
samples~\cite{Ball:2014uwa}. Additional $\Pp\Pp$ interactions in the
same or adjacent bunch crossings (pileup) are added to each simulated
event sample.  The events are weighted to match the pileup
distribution in simulation to that observed in data. The average
pileup in the data sample is 23.

Both simulated \PW and \PZ boson samples, generated at NLO
accuracy in perturbative QCD, are further reweighted by the ratio of
observed and predicted values in the \ptz spectrum, taken from a
measurement by the CMS Collaboration using the same
dataset~\cite{Sirunyan:2019bzr}.  While this procedure ensures
consistency for the \PZ background sample, reweighting \ptw by the
measured \ptz data versus the MC spectrum is not inherently
necessary. However, when adopting this weighting, the agreement
between the observed data and the MC prediction in \PZ events is
improved for the observable relevant to this analysis, namely \ptl.
In addition, the theoretical uncertainties for the boson \pt spectrum,
which will be described in Sec.~\ref{sec:systematics}, are large
enough to cover the difference between the raw and reweighted spectra.

The detector response is simulated using a detailed description of the
CMS detector implemented with the \GEANTfour
package~\cite{Agostinelli:2002hh}. Reconstruction algorithms are the
same for simulated events and data.

\section{Reconstruction and event selection}
\label{sec:recoSelection}

The analysis is performed by selecting $\PW\to\ell\nu$ candidate
events characterized by a single prompt, energetic, and isolated
lepton and missing transverse momentum (\ptmiss) due to the escaping
neutrino.  A particle-flow (PF) algorithm~\cite{Sirunyan:2017ulk} that
reconstructs all observable particles in the event is used. This
algorithm classifies particles into muons, electrons, photons, and charged
or neutral hadrons.  It optimally combines information from the
central tracking system, energy deposits in the ECAL and HCAL, and
tracks in the muon detectors to reconstruct these individual
particles. The algorithm also determines quality criteria, which are
used to select the particles used in the distributions of the
final-state observables.

Muon candidates are required to have a transverse momentum
$\PT^{\Pgm}>26\GeV$ and be within the geometrical acceptance of the
muon spectrometer, defined by $\abs{\eta^{\Pgm}}<2.4$. These values
are chosen so that the inefficiency due to the trigger is minimal,
once the full selection is applied.

Quality requirements on the reconstructed muons are applied to ensure
high purity of the selected events. These include requirements on the
matching of the tracker information to the information from the muon
system, as well as quality requirements on the combined track
itself. In addition, a requirement on the relative isolation of the
reconstructed muon is applied to suppress muons from background
processes, such as leptonic heavy-flavor decays. This isolation
variable is defined as the pileup-corrected ratio of the sum of the
\pt of all charged hadrons, neutral hadrons, and photons, divided by
the \pt of the muon itself~\cite{CMS-DP-2017-007}. It is calculated
for a cone around the muon of $\Delta R
= \sqrt{\smash[b]{(\Delta \phi)^2+(\Delta \eta)^2}}<0.4$, where $\phi$
is the azimuthal angle, and it is required to be smaller than 15\%.

Electron candidates are formed from energy clusters in the ECAL
(called superclusters) that are matched to tracks in the silicon
tracker.  Their \pt is required to exceed 30\GeV and they are selected
within the volume of the CMS tracking system up to $\abs{\eta^{\Pe}} <
2.5$. Electrons reconstructed in the transition region between the
barrel and the end cap sections, within $\abs{\eta^{\Pe}}>1.4442$ and
$\abs{\eta^{\Pe}} < 1.5660$, are rejected.

Electron identification is based on observables sensitive to
bremsstrahlung along the electron trajectory and geometrical and
momentum-energy matching between the electron trajectory and the
associated supercluster, as well as ECAL shower-shape observables and
variables that allow the rejection of the background arising from
random associations of a track and a supercluster in the ECAL.
Energetic photons produced in $\Pp\Pp$ collision may interact with the
detector material and convert into electron-positron pairs. The
electrons or positrons originating from such photon conversions are
suppressed by requiring that there is no more than one missing tracker
hit between the primary vertex and the first hit on the reconstructed
track matched to the electron; candidates are also rejected if they
form a pair with a nearby track that is consistent with a
conversion. Additional details of electron reconstruction and
identification can be found in
Refs.~\cite{Chatrchyan:2013dga,Khachatryan:2015hwa}.

A relative isolation variable similar to that for muons is constructed
for electrons, in a cone of $\Delta R < 0.3$ around their momenta~\cite{Khachatryan:2015hwa}.
This variable is required to be less than a value that varies from
around 20\% in the barrel part of the detector to 8\% in the end cap
part. The values used are driven by similar requirements in the HLT
reconstruction.

Off-line selection criteria are generally equal to or tighter than the
ones applied at the HLT. Despite this, differences in the definition
of the identification variables defined in the on-line system and
off-line selection create differences between data and simulation that
need dedicated corrections, as described in
Sec.~\ref{sec:efficiencies}.

The analysis is carried out separately for \wplus and \wminus
bosons and aims to measure the charge asymmetry in \PW boson
production, so any charge misidentification has to be reduced to a
minimum.  Thus, the off-line electron selection also employs a tight
requirement for the charge assignment, which reduces the charge
misidentification to 0.02 (0.20)\% in the barrel region (end cap
sections) in the \pt range of
interest~\cite{CMS_DPS_EGAMMA_2018_017}.

Events coming from $\PW\to\ell\nu$ decays are expected to contain one
charged lepton (muon or electron) and significant \ptmiss resulting
from the neutrino. The missing transverse momentum vector \ptvecmiss
is computed as the negative vector sum of the transverse momenta of
all the PF candidates in an event, and its magnitude is denoted
as \ptmiss~\cite{Sirunyan:2019kia}.  No direct requirement on
\ptmiss is applied, but a requirement is placed on the transverse
mass, defined as
$\mT=\sqrt{\smash[b]{2\pt\ptmiss(1-\cos\Delta\phi)}}$, where
$\Delta\phi$ is the angle in the transverse plane between the
directions of the lepton \pt and the \ptmiss.  Events are selected
with $\mT>40\GeV$.  This requirement rejects a large fraction of QCD
multijet backgrounds.

Events from background processes that are expected to produce multiple
leptons, mainly $\cPZ\to\ell\ell$, \ttbar, and diboson production are
suppressed by a veto on the presence of additional electrons or muons
in the event.  To maximize the rejection efficiency, these events are
rejected if additional leptons, selected with looser identification
and isolation criteria than the selected lepton, have $\pt>10\GeV$.

\subsection{Efficiency corrections}
\label{sec:efficiencies}

The measurement of differential cross sections relies crucially on the
estimation of the lepton selection efficiencies, both in the collision data and in the MC,
because these are among the dominant
contributions to the uncertainty. For the total absolute cross
sections, the uncertainties are dominated by the integrated luminosity
uncertainty. For normalized differential cross sections, the
correlation of the luminosity uncertainty between the inclusive and
differential measurements is such that it mostly cancels out in their
ratio. Thus, the dominant uncertainties are the ones related to
the lepton efficiency that are not fully correlated through the lepton
kinematics phase space.

The lepton efficiency is determined separately for three different
steps in the event selection: the trigger (L1+HLT), the off-line
reconstruction, and the off-line selection, which includes identification
and isolation criteria.
The lepton efficiency for each step is determined with respect to the
previous one.

A technique called \textit{tag-and-probe} is used, in which the
efficiency for each step is measured for MC simulation and collision
data using samples of $\cPZ \to \ell \ell$ events with very high
purity~\cite{Khachatryan:2010xn}. The sample is defined by selecting
events with exactly two leptons. One lepton candidate, denoted as
the \textit{tag}, satisfies tight identification and isolation
requirements. The other lepton candidate, denoted as the
\textit{probe}, is selected with the selection criteria that depend on
the efficiency of the above steps being measured. The number of probes
passing and failing the selection is determined from fits to the
invariant mass distribution with $\cPZ \to \ell \ell$ signal and
background components. The backgrounds in these fits stem largely from
QCD multijet events and are at the percent level. In certain regions
of phase space, especially in the sample of failing probes, these
backgrounds contribute significantly, requiring an accurate modeling
of the background components.  The nominal efficiency in collision
data is estimated by fitting the \PZ signal using a binned template
derived from simulation, convolved with a Gaussian function with 
floating scale and width to describe the effect of the detector
resolution. An exponential function is used for the background. The
nominal efficiency in MC simulation is derived from a simple ratio of
the number of passing probes over all probes.

For each step, the tag-and-probe method is applied to data and to
simulated samples, and the efficiency is computed as a function of
lepton \pt and $\eta$.  The ratio of efficiencies in data and
simulation is computed together with the associated statistical and
systematic uncertainties and is used to weight the simulated \PW
boson events.  The uncertainties in the efficiencies are propagated as
a systematic uncertainty in the cross section measurements. The
analysis strategy demands a very high granularity in the lepton
kinematics. Therefore, the efficiencies are computed in slices of
$\Delta\eta=0.1$ and steps of \pt ranging from 1.5 to 5.0\GeV.  A
smoothing is applied as a function of lepton \pt for each slice in
$\eta$, modeled by an error function.  Systematic uncertainties
associated with this method are propagated to the measurement and are
discussed in Sec.~\ref{sec:lepeff}. These include a correlated
component across \leta and an uncorrelated component related to
the statistical uncertainty in each of the slices in \leta.

\section{Background estimation}
\label{sec:backgrounds}

The selection requirements described in Sec.~\ref{sec:recoSelection}
result in a data sample of 114 (51)$\times 10^6$ \wplus and 88 (42)$\times 10^6$ \wminus 
candidate events in the muon (electron) final
state with small background. A summary of the inclusive
background-to-signal ratios is shown in
Table~\ref{tab:backgrounds}. The most significant residual background
is QCD multijet production, where the selected nonprompt leptons stem
from either semileptonic decays of heavy-flavor hadrons or are the
product of misidentified jets (usually from light quarks). The former
is the principal source of QCD background in the muon channel; the
latter dominates the background in the electron channel, along with
the production of electron-positron pairs from photon conversions.

The nonprompt-lepton background is estimated directly from data. A
control sample (the \textit{application} sample) is defined by one
lepton candidate that fails the standard lepton selection criteria,
but passes a looser selection. The efficiency,
$\epsilon_{\text{pass}}$, for such a loose lepton object to pass the
standard selection is determined using another independent sample (the
\textit{QCD-enriched sample}) dominated by events with nonprompt
leptons from QCD multijet processes. This QCD-enriched sample, which is
disjointed to the signal sample by means of the requirement
$\mT<40\GeV$, is defined by one loosely identified lepton and a jet
with $\pt>45\GeV$ recoiling against it.  The measured efficiency for
the leptons in this sample, parametrized as a function of \pt and
$\eta$ of the lepton, is used to weight the events in the
application sample by $\epsilon_{\text{pass}}/(1 - \epsilon_{\text{pass}}$) to
obtain the estimated contribution from the nonprompt-lepton
background in the signal region. The efficiency $\epsilon_{\text{pass}}$ is computed
with granularity of $\Delta\eta=0.1$, and in each $\eta$ bin it is
parametrized as a linear function of \pt.

A small fraction of the events passing the selection criteria are due
to other electroweak processes, and this contribution is estimated
from simulation. Drell--Yan (DY) events that produce a pair of muons
or electrons, and one of the two leptons falls outside the detector
acceptance, mimic the signature of \PW boson events rather closely. A
smaller effect from DY production stems from $\cPZ\to\tau\tau$ decays,
where one $\tau$ lepton decays leptonically and the other
hadronically. Additionally, events from $\PW\to\tau\nu$ decays are
treated as background in this analysis. The light leptons from the
$\tau$ decays typically exhibit lower \pt than that in signal
events and are strongly suppressed by the minimum \ptl
requirements.  Other backgrounds arise from \ttbar and single top
production, with one of the top quarks producing a \PW boson that
subsequently decays leptonically. There are small contributions to the
background from diboson ($\PW\PW$, $\PW\cPZ$, $\cPZ\cPZ$) production.
Finally, for the electron channel only, the background from
$\PW\to\Pe\nu$, where the lepton is reconstructed with the wrong
charge, is estimated. This background is completely negligible for the
muon final state.

\begin{table}
\centering
\topcaption{\label{tab:backgrounds}
  Estimated ratios of each background component to the total \PW boson signal in the $\PW\to\Pgm\nu$ and $\PW\to\Pe\nu$
  channels. The DY simulation includes $\ell=\Pe,\mu,\tau$.}
  \begin{scotch}{l c c}
{\multirow{2}{*}{Processes}} & \multicolumn{2}{c}{Bkg. to sig. ratio}  \\
                           & $\PW\to\mu\nu$ & $\PW\to\Pe\nu$ \\ 
\hline
$\cPZ\to\ell\ell$ (DY)        & 5.2\%  &   3.9\% \\
$\PW\to\tau\nu$               & 3.2\%  &   1.3\% \\
$\PW\PW$+$\PW\cPZ$+$\cPZ\cPZ$ & 0.1\%  &   0.1\%   \\
Top                           & 0.5\%  &   0.5\%   \\
Wrong charge                  & \NA    &  0.02\%   \\
QCD                           & 5.5\%  &   8.2\%  \\
  \end{scotch}
\end{table}

\section{Template construction and fitting procedure}
\label{sec:templatesFitting}

The measurement strategy is to fit 2D templates in the charged-lepton
kinematic observables of \ptl and \leta to the observed 2D
distribution in data. Whereas each of the background processes results
in a single template, the simulated \PW boson signal is divided into
its three helicity states, as well as into slices of the \PW boson
rapidity \absyw. The procedure of constructing these helicity- and
rapidity-binned signal templates is described below.

\subsection{Construction of helicity and rapidity signal templates}
\label{sub:templateReweighting}

The inclusive \PW boson production cross section at a hadron
collider, with its subsequent leptonic decay, neglecting the small
terms which are exclusively NLO in QCD, is given
by~\cite{Bern:2011ie}:
\begin{equation}\begin{aligned}
  \label{eq:coeff}
  \frac{\rd N}{\rd\cos\theta^{*}\rd\phi^{*}} &\propto (1+\cos^2\theta^{*}) + \frac{1}{2}A_0 (1-3\cos^2\theta^{*}) \\
  & + A_1 \sin 2\theta^{*}\cos\phi^{*} + \frac{1}{2}A_2 \sin^2\theta^{*}\cos2\phi^{*} \\
  & + A_3 \sin\theta^{*}\cos\phi^{*} + A_4\cos\theta^{*},
\end{aligned}
\end{equation}
where \thetastar and $\phi^{*}$ are the polar and azimuthal decay
angles of the lepton in the Collins--Soper frame of
reference~\cite{Collins:1977iv}, where the lepton refers to the
charged lepton in the case of \PWm and the neutrino in the case of
\PWp.  The angular coefficients $A_0$ to $A_4$ in
Eq.~(\ref{eq:coeff}) depend on the \PW boson charge, \ptw, and
\yw, and receive contributions from QCD at leading and higher
orders. When integrating Eq.~(\ref{eq:coeff}) over $\phi^{*}$, the
cross section is written as:
\begin{equation}\begin{aligned}
  \label{eq:coeff2}
  \frac{\rd N}{\rd\cos\theta^{*}} &\propto (1+\cos^2\theta^{*}) \\
  & + \frac{1}{2}A_0 (1-3\cos^2\theta^{*}) + A_4\cos\theta^{*}.
\end{aligned}\end{equation}
This expression can equivalently be written as a function of the helicity
amplitudes~\cite{Ellis:1991qj}:
\begin{equation}\begin{aligned}
\label{eq:helfrac}
    \frac{1}{N}\frac{\rd N}{\rd\cos{\theta^{*}}\rd \ptw \rd \yw} &= \frac{3}{8}(1\mp\cos{\theta^{*}})^2  f_\mathrm{L}^{\left(\ptw,\yw\right)} \\
                                                                 &+  \frac{3}{8}(1 \pm \cos{\theta^{*}})^2  f_\mathrm{R}^{\left(\ptw,\yw\right)}\\
                                                                 &+  \frac{3}{4} \sin^2{\theta^{*}}  f_\mathrm{0}^{\left(\ptw,\yw\right)},
\end{aligned}
\end{equation}
where the coefficients $f_i$ are the helicity fractions, and the upper
(lower) sign corresponds to \PWp (\PWm) boson,
respectively. Thus, the fractions of left-handed, right-handed, and
longitudinal \PW bosons ($f_\mathrm{L}$, $f_\mathrm{R}$ and
$f_\mathrm{0}$, respectively) are related to the coefficients $A_i$ of
Eq.~(\ref{eq:coeff2}), with $A_0\propto f_\mathrm{0}$ and
$A_4 \propto \mp (f_\mathrm{L} - f_\mathrm{R})$ depending on the \PW
boson charge, where by definition $f_i > 0$ and $f_\mathrm{L} +
f_\mathrm{R} + f_\mathrm{0} = 1$.  The generated leptons are
considered before any final-state radiation (``pre-FSR leptons'') and
are called pre-FSR leptons.

Since there is no helicity information in the simulated MC signal
sample, the reweighting procedure is implemented based on the
production kinematics of the \PW boson and the kinematics of the
leptonic decay of the \PW boson.

\begin{figure}[ht!]
\centering
\includegraphics[width=0.47\linewidth]{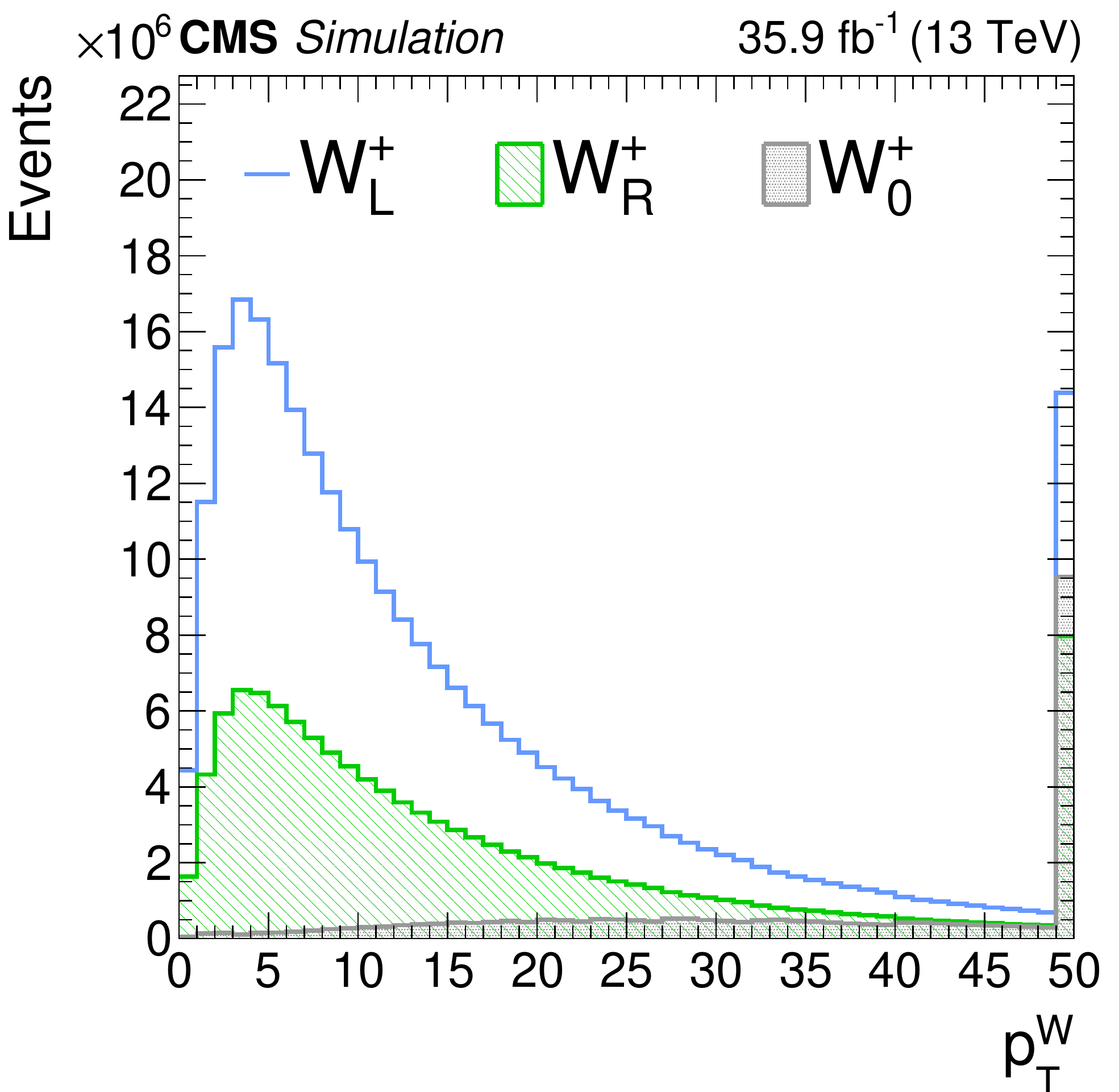}
\includegraphics[width=0.47\linewidth]{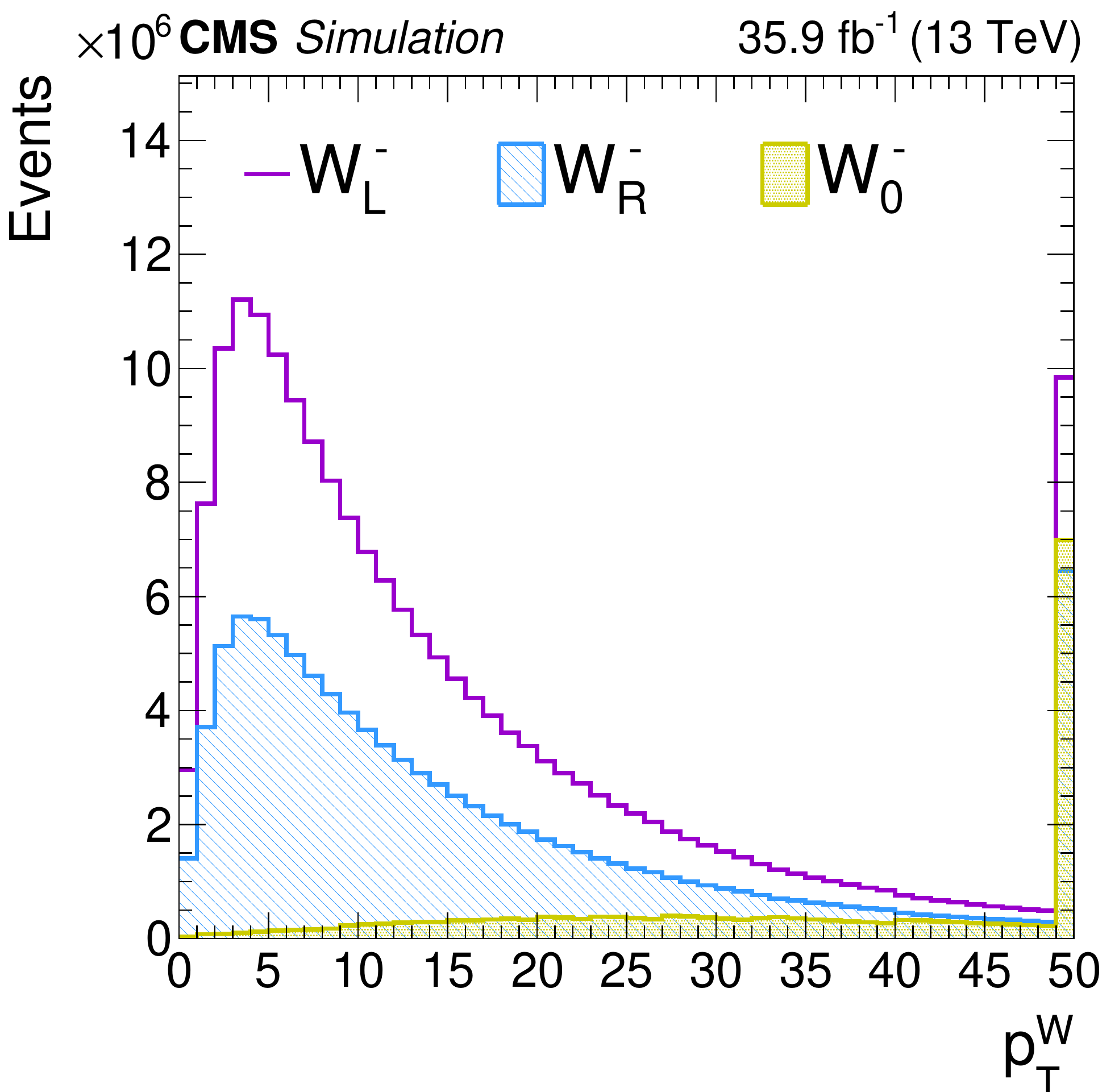} \\
\includegraphics[width=0.47\linewidth]{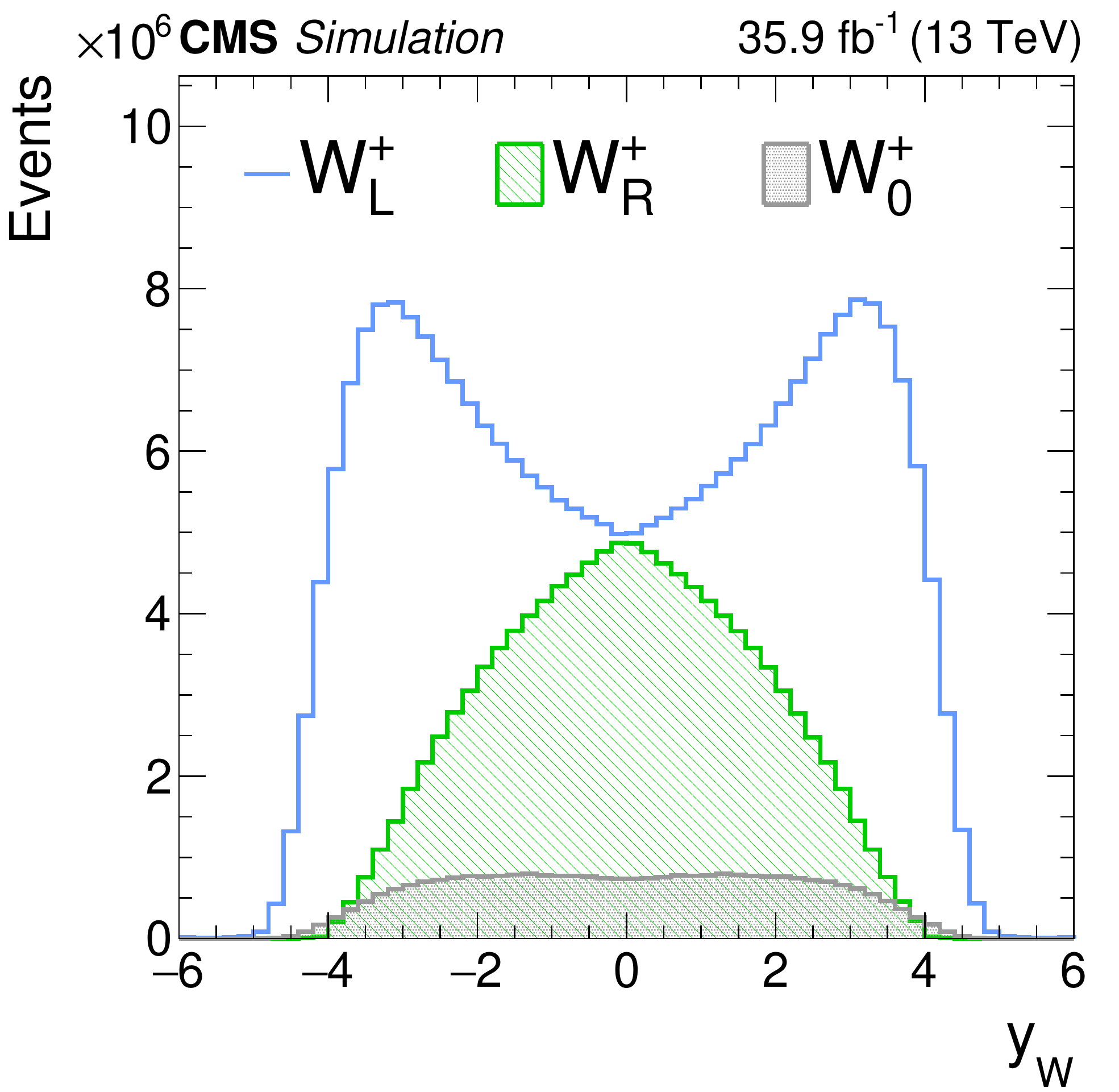}
\includegraphics[width=0.47\linewidth]{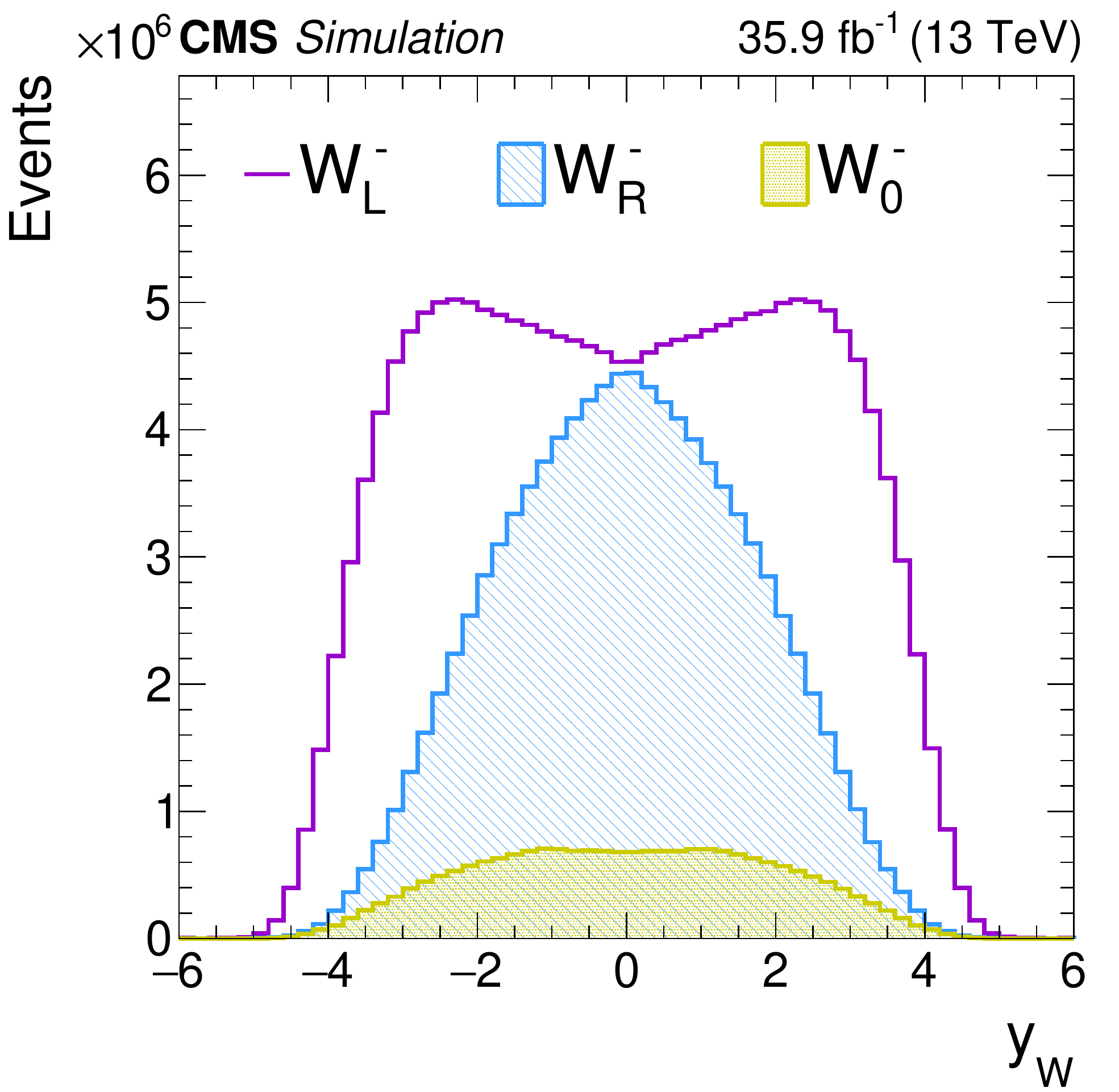} \\
\includegraphics[width=0.47\linewidth]{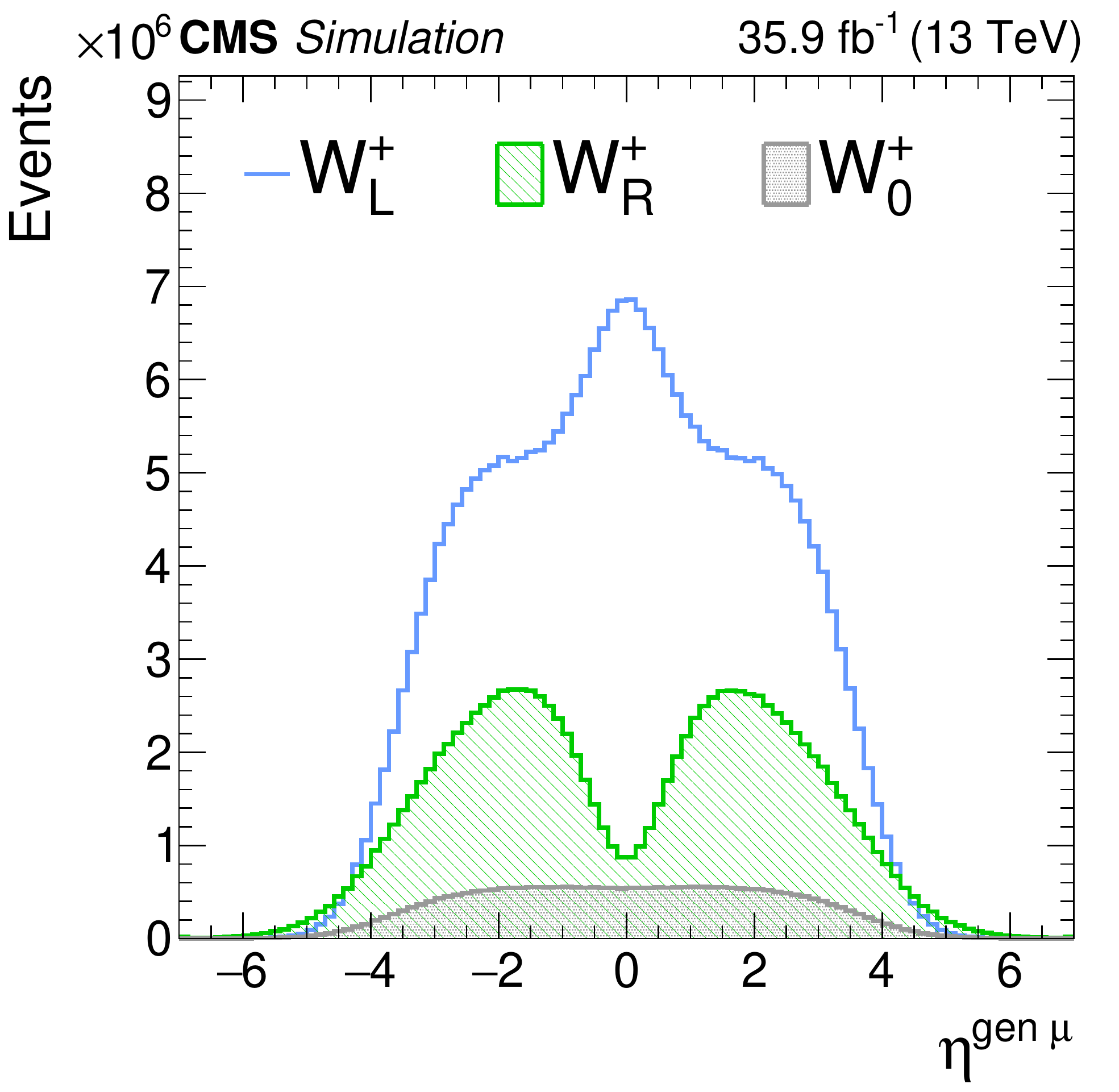} 
\includegraphics[width=0.47\linewidth]{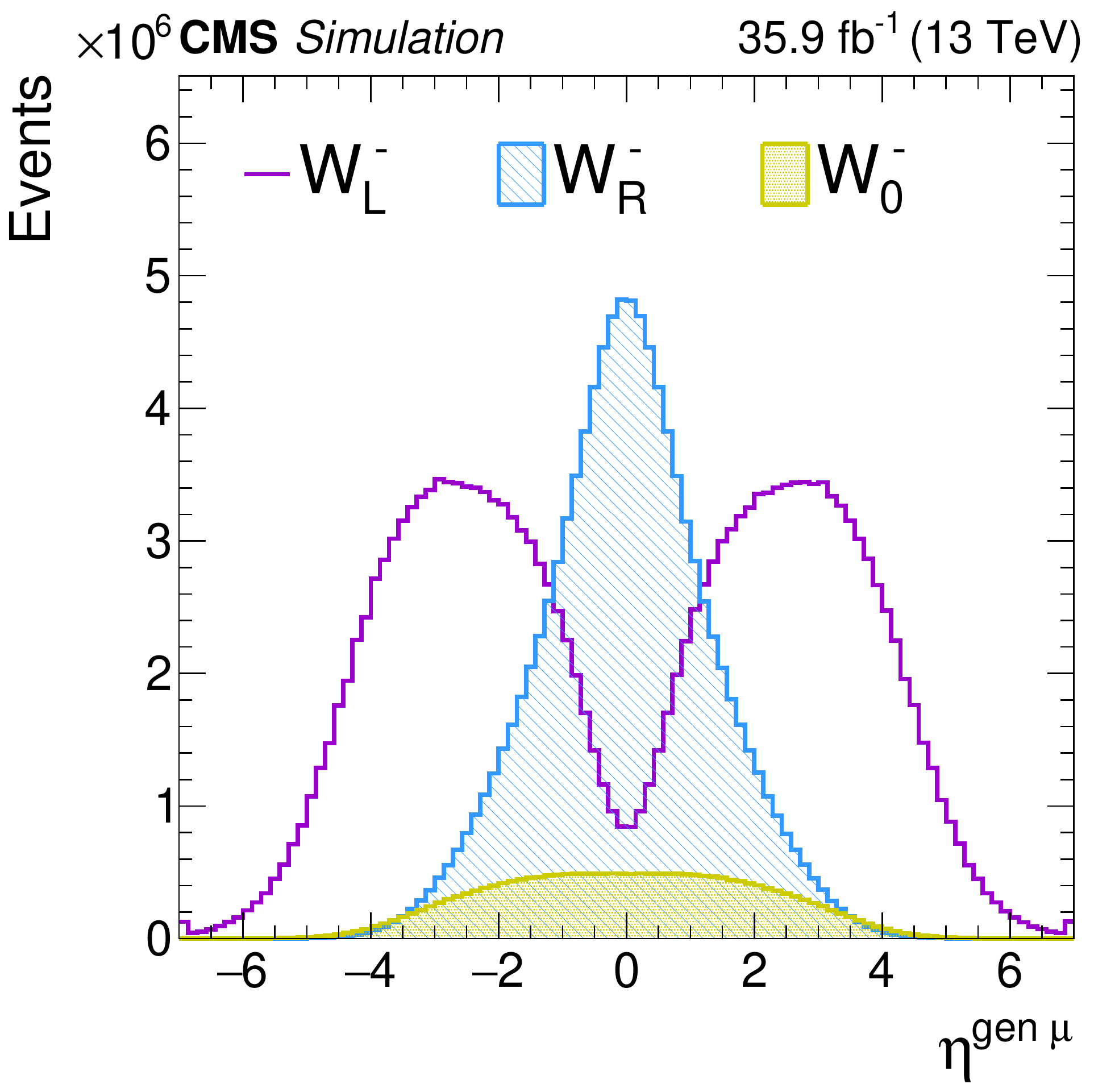} \\
\caption{Generator-level distributions of the \PW boson \ptw (top), \absyw (center),
  and the resulting $\eta$ distribution of the charged lepton (bottom)
  after reweighting each of the helicity components for positively
  (left) and negatively (right) charged \PW bosons.}
\label{fig:reweighting}
\end{figure}

The coefficients $f_i$ depend strongly on the production kinematics of
the \PW boson, namely \ptw, \absyw, and its charge. Therefore, a
reweighting procedure is devised in which the
\cost~distribution is fitted in bins of \ptw and \absyw, separately
for each charge, to extract the predicted $f_i$. These spectra of the 
decay angle are constructed in the full phase space of the \PW boson
production. Each simulated event
is reweighted three separate times to obtain pure samples of
left-handed, right-handed, and longitudinally polarized \PW bosons.
The results of this procedure are illustrated in
Fig.~\ref{fig:reweighting}, where the simulated signal is split into
the three helicity states by reweighting by the extracted helicity
fractions $f_i$. Distributions of \ptw and \absyw are shown for both
charges of \PW bosons, along with the resulting distribution of the
charged lepton $\eta$.

The distributions of \ptw and \absyw are substantially different for
the three helicity components.  Whereas the left-handed \PW bosons
(\wleft) and the right-handed \PW bosons (\wright) behave in the
same way as a function of \pt, their behavior in \absyw is
significantly different. Their production cross sections are equal at
$\absyw=0$, but that of the \wleft component increases up to a
maximum at \absyw between 3.0 and 3.5, whereas the \wright component
decreases monotonically with
\absyw.  The longitudinally polarized \PW bosons (\wlong)
have an overall much lower production cross section, which is
relatively flat in \absyw and increases as a function of \pt, as
expected in the Collins--Soper reference frame. The different
distributions in \absyw of the \wright and \wleft components, paired
with the preferential decay direction of the charged lepton for these
two helicity states, results in distinctly different \leta
distributions. For positively charged \PW bosons at a given \absyw,
the \wleft component causes the charged lepton to have values of
\leta closer to zero.  In contrast, the positively charged \wright
component tends to have larger values of \absleta. The opposite is
true for negatively charged \PW bosons, \ie, the charged lepton
\absleta will tend to be large for left-handed \PWm bosons,
whereas right-handed \PWm bosons lead to leptons observed mostly at
small \absleta.

\subsection{Fitting strategy for the rapidity-helicity measurement}
\label{sub:fitStrategy}

\begin{figure}[ht!]
\centering
\includegraphics[width=0.97\linewidth]{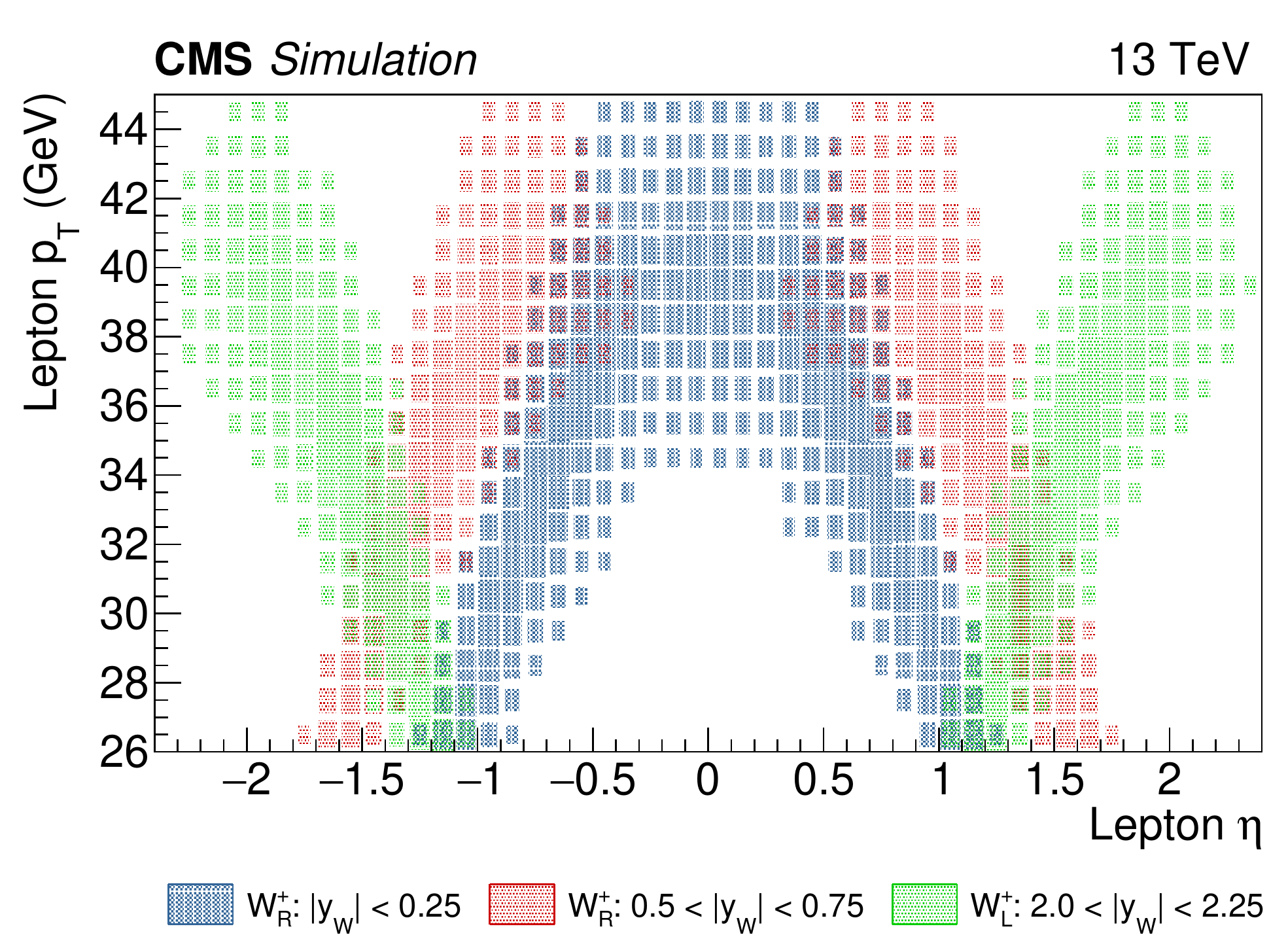}
\caption{Distributions of 2D templates of \ptl versus \leta for
  simulated positively charged \PW bosons events in different
  helicity or rapidity bins. Templates for the muon channel are
  shown. Blue: \wplusr with $\absyw<0.25$, red: \wplusr with
  $0.50<\absyw<0.75$, and green: \wplusl with
  $2.00<\absyw<2.25$. \label{fig:templateExample}}
\end{figure}

The characteristic behavior of the lepton kinematics for different
polarizations of the \PW boson can be exploited to measure the cross
section for \PW boson production differentially in \absyw and
separately for the three helicity components. This is done by
splitting each of the three helicity states into bins of \absyw and
constructing the charged lepton \ptl versus \leta
templates for each of the helicity and charge components from the MC
as described above.  Example 2D templates are shown in
Fig.~\ref{fig:templateExample}, where three different templates are
shown for \wplus bosons. The blue template is obtained from events
with a \wright produced from 0.00 to 0.25 in \absyw, the red
template from events with a \wright produced between 0.50 and 0.75
in
\absyw, and the green template from events with a \wleft produced between 2.00 and
2.25 in \absyw. The behavior described above is clearly seen. 
Another important aspect of the underlying physics may also be
understood from Fig.~\ref{fig:templateExample}: while the \PW bosons
are produced in orthogonal regions of phase space, the resulting
templates for the observable leptons overlap considerably for the
different helicity and rapidity bins. This overlap is most striking
for adjacent bins in \absyw in a given helicity state.  In
Fig.~\ref{fig:templateExample}, the two distributions for the right-handed \PW boson and
the distribution for the left-handed \PW boson show sizeable overlap,
albeit with contrasting shapes as a function of the observable lepton
kinematics. A consequence of the large overlaps in general, and in
neighboring bins in rapidity in particular, are large
(anti-)correlations in the fitted differential cross sections in
helicity and rapidity.

The 2D templates in the observable lepton kinematics extend from the
minimum \ptl requirement of 26 (30)\GeV for muons (electrons) to a
maximum value of 45\GeV in bins with width of 1\GeV. In the
observable \leta, the width of the bins is 0.1, extending from $-2.4$
($-2.5$) to 2.4 (2.5) for muons (electrons).

To extract the differential cross sections in \PW boson rapidity for
the three helicity states, the full sample of simulated \PW boson
events is divided using the method described earlier into the three
helicity components and 10 bins of \absyw of width 0.25 up to
$\absyw=2.5$. These separate signal processes are left freely floating
in a maximum likelihood (ML) fit to the observed 2D distribution
for \ptl versus \leta. All events above the threshold $\absyw=2.5$ are
fixed to the prediction from simulation and are treated as background
because of the rapid loss in acceptance for certain charge and
helicity combinations at higher rapidity. Additionally, the
longitudinally polarized states are fixed to the MC prediction. This
results in 40 freely floating cross sections in the fit, corresponding
to the 10 bins in \PW boson rapidity for each charge, and for the
left- and right-handed polarizations.

\subsection{Fitting strategy for the double-differential \texorpdfstring{$\PW$}{W} boson cross section}
\label{sub:fitDiffXsec}

The double-differential \PW boson production cross sections, as functions of
\ptl and \absleta, are measured with an analogous technique. The double-differential 
cross section for each charge of the \PW boson is denoted by
\begin{equation}
  \label{eq:xsec2d}
  \sigma^\pm = \frac{\rd\sigma}{\rd\absleta \rd\ptl}(\Pp\Pp\to\PW^\pm + X \to \ell^\pm\nu + X), 
\end{equation}
and can be measured in very fine bins of \leta and \ptl.  Current
theoretical calculations predict these cross sections with next-to-NLO
(NNLO) accuracy in perturbative QCD, and such a measurement is a more
rigorous test of these calculations than the previous studies
performed by the CDF and D0 Collaborations at the Fermilab Tevatron
$\Pp\Pap$ collider~\cite{CDFasymmetry,D0asymmetry}, or by the ATLAS,
CMS, and LHCb Collaborations at the LHC~\cite{Aaboud:2016btc,Aad:2019rou,Chatrchyan:2012xt,Chatrchyan:2013mza,Khachatryan:2016pev,Aaij:2012vn,Aaij:2015zlq},
which all measured the cross section as a function of reconstructed
\leta only. The CDF Collaboration has also inferred the charge
asymmetry as a function of \absyw in Ref.~\cite{CDFasymmetry}.  When
integrating either over the \absleta or in the \ptl
dimension, the usual one-dimensional differential cross section measurement can be
recovered.

This measurement is performed by fitting the same 2D distributions of
\ptl versus \leta, with different freely floating signal processes. As opposed 
to the rapidity-helicity measurement, where each signal template corresponds
to one bin in the underlying \absyw and
helicity state of the generated $\PW$ boson, each signal process in the
double-differential measurement corresponds to
a bin in the underlying generated lepton \pt and lepton \abseta. The
generated leptons in this measurement are subject to a so-called
``dressing'' procedure, where electroweak radiation is added back to
the charged-lepton momentum within a cone of $\Delta R < 0.1$.  The
unfolding corrects for bin-by-bin differences in generated versus
reconstructed \ptl and \leta. The resulting number of underlying
signal processes increases from the 40 processes in the
helicity/rapidity fit to a total of 324, corresponding to 18 bins in
the \ptl times 18 bins in \absleta. The generated \ptl ranges
from 26 to 56\GeV. The bin widths in \ptl are 2\GeV from 26 to 30\GeV, 1.5\GeV
from 30 to 48\GeV, and 2\GeV above. The bin
width in \absleta is 0.1 up to $\absleta=1.3$, followed by 4 bins of
width 0.2, and a final bin ranging from $\absleta=2.1$ to 2.4.  Events
in which the generated leptons are outside of the reconstructed
acceptances are treated as a background component in this fit.  The
treatment of the backgrounds and the systematic uncertainties remains
the same as for the rapidity/helicity fit.

\subsection{Likelihood construction and fitting}

A ML fit is performed to extract the parameters of interest.  The
construction and calculation of the likelihood, as well as the
minimization are implemented using the \textsc{TensorFlow} software
package originally developed for machine learning
applications~\cite{TensorFlow}.  The benefit of such an implementation
is that the gradients required for minimization are computed
automatically by backpropagation, which is both faster and more
numerically accurate and stable than finite difference approaches used
in existing tools.  The calculation of the likelihood, and the
additional linear algebra associated with the minimization algorithm,
can also be parallelized on vector processing units and/or multiple
threads, as well as using graphics processing units, for a further
improvement in the speed of the fit.  The implementation is also
optimized to keep memory usage acceptable, given the large number of
measurement bins and parameters, with a sparse tensor representation
used where appropriate.

The negative log-likelihood function can be written as follows:
\begin{widetext}
  \begin{equation}
    \label{eqn:nll}
      L = -\ln({\cal L}(\mathrm{data} | \vec{\mu}, \vec{\theta}))
      = \sum_\mathrm{i}\left(-n^\text{obs}_{i}\ln n^\text{exp}_{i}(\vec{\mu}, \vec{\theta})
      + n^\text{exp}_{i}(\vec{\mu}, \vec{\theta}) \right)  + \frac{1}{2}\sum_{k}\left(\theta_{k} - \theta^0_{k}\right)^2,
  \end{equation}
\end{widetext}
with
\begin{equation}   
n^\text{exp}_{i}(\vec{\mu}, \vec{\theta}) = \sum_{p} \mu_{p} n^\text{exp}_{i,p}\prod_{k}\kappa_{i,p,k}^{\theta_{k}},
\end{equation}
where: $n^\text{obs}_{i}$ is the observed number of events in
each bin, assumed to be independently Poisson-distributed;
$n^{\text{exp}}_{i,p}$ is the expected yield per bin per process;
$\mu_{p}$ is the freely floating signal strength multiplier per signal
process fixed to unity for background processes; $\theta_{k}$ are the
nuisance parameters associated with each systematic uncertainty; and
$\kappa_{i,p,k}$ is the size of the systematic effect per bin, per
process, and per nuisance parameter. The systematic uncertainties are
implemented with a unit Gaussian constraint on the nuisance parameter
$\theta_{k}$ such that the factor $\kappa_{i,p,k}^{\theta_{k}}$
multiplying the yield corresponds to a log-normal distribution with
the mean equal to 0 and the width equal to $\ln\kappa_{i,p,k}$. All nuisance
parameters are fully profiled in the fit. This
parametrization corresponds to the one used by the LHC Higgs
Combination Working Group~\cite{HiggsComb}.

The signal strength modifiers and nuisance parameters are extracted
directly from the ML fit, with the corresponding covariance matrix
computed from the Hessian of the likelihood at the minimum, which can
also be calculated to high numerical accuracy using backpropagation.
The unfolded cross sections are extracted simultaneously in the ML fit
by including the dependence of the predicted cross section on the
nuisance parameters associated with the theoretical uncertainties.
The cross sections and corresponding covariance matrix are extracted
based on the postfit values of the signal strength modifiers and
nuisance parameters and their covariance.

While the cross section vectors $\vec\sigma$ are left freely floating
when fitting for the rapidity/helicity or the double-differential
cross sections, it is also possible to fix these parameters to their
expected values. Performing the fit in such a way allows for the
direct measurement of the constraints set by the data on every
nuisance parameter. This is especially interesting for the case of the
PDF uncertainties, as the large and quite pure selected sample of
\PW bosons can place strong constraints on the PDF uncertainties by
using the charged lepton kinematics.

\subsection{Measurement of the charge asymmetry and unpolarized cross sections}
\label{sub:asymmetry}

The fit to the data is performed simultaneously for the two charge
categories and to the three helicity states. Therefore, the
minimization can yield combinations of the measured cross sections
with the proper propagation of the uncertainties through the fit
covariance matrix, either differentially in rapidity or
double-differentially in \ptl and \absleta.

One of the additional quantities considered is the polarized
\PW boson charge asymmetry, defined as follows:
\begin{widetext}
  \begin{equation}
    \mathcal{A}^\text{pol}(\absyw) = \frac{\rd\sigma^\text{pol}/\rd\absyw(\wplus\to\ell^+\nu) - \rd\sigma^\text{pol}/\rd\absyw(\wminus\to\ell^-\PAGn)}
                                   {\rd\sigma^\text{pol}/\rd\absyw(\wplus\to\ell^+\nu) + \rd\sigma^\text{pol}/\rd\absyw(\wminus\to\ell^-\PAGn)},
                                   \label{eq:asymmetry}  
  \end{equation}
\end{widetext}
where pol represents the \PW polarization state.  The charge
asymmetry, as a function of \absyw as extracted from the ML fit,
differentially in the three polarizations, provides a more direct
constraint on the PDF than the previous measurements at the LHC, which
are performed differentially in the reconstructed lepton
pseudorapidity~\cite{Khachatryan:2016pev,Aaboud:2016btc}.  In the CDF
Collaboration measurement~\cite{CDFasymmetry}, the \PW boson charge
asymmetry was extracted as a function of \absyw, but not separately
in the \PW boson helicity state.

The charge asymmetry of \PW bosons, which is also determined from the
double-differential cross section measurement, is written as follows:
\begin{equation}
  \label{eq:asymmetry2d}
  \mathcal{A}(\absleta,\ptl) = \frac{\rd^2\sigma^+/\rd\absleta \rd\ptl - \rd^2\sigma^-/\rd\absleta \rd\ptl}{\rd^2\sigma^+/\rd\absleta \rd\ptl + \rd^2\sigma^-/\rd\absleta \rd\ptl}.
\end{equation}
When the distribution is integrated over \ptl, the results may be
compared directly with previous measurements of $\mathcal{A}(\absleta)$
at hadron colliders. Similarly, when integrating over \absleta,
$\mathcal{A}(\ptl)$ is obtained.  These one-dimensional (1D)
distributions as functions of \ptl and \leta are obtained by
integrating over the other variable after performing the fully
differential 2D fit. Associated uncertainties are included properly
from the full 2D covariance matrix of the fit.

\section{Systematic uncertainties}
\label{sec:systematics}

This section describes the treatment of systematic uncertainties from
experimental sources, as well as from modeling and theoretical
uncertainties. In general, systematic uncertainties are divided into
two types: those affecting only the normalization of the templates and
those affecting their shape.

Normalization uncertainties are treated as log-normal nuisance
parameters acting on a given source of background or signal. They
change the overall normalization of the process by the given value,
while retaining the relative contributions of the process in each of
the \ptl and \leta bins.

Shape uncertainties do the exact opposite. While the integral of a
background or signal component is kept constant at the central value,
the relative shape of the 2D template is allowed to float. This
necessitates both an up and down variation of each shape nuisance
parameter.  These uncertainties are incorporated by means of vertical
interpolation of the event count in each bin of the template.

Uncertainties can also be a combination of the two, \ie, change the
normalization, as well as the shape of the 2D templates simultaneously.

\subsection{Experimental uncertainties}
\label{sec:expunc}

\subsubsection{QCD multijet background}
The QCD multijet background is estimated from data sidebands in the
lepton identification and isolation variables, as described in
Sec.~\ref{sec:backgrounds}.

The uncertainty in the method itself is estimated from closure tests
in a background-dominated region, obtained by inverting the \mT
requirement, \ie, $\mT<40$ (30)\GeV for the $\mu$ (\Pe) channel.  The
level of agreement in this background-dominated region is an estimate
of the uncertainty in the normalization of this process. The agreement
in the 2D \etapt plane is rather good for both muons and electrons,
and varies with lepton $\eta$ and \pt. In the case of electrons,
where this background is larger than in the muon case, the central
value of the QCD background is also rescaled by the values derived in
this closure test.

The nonclosure amounts to about 5\% in the muon final state for all
the \absleta bins, and 0.5 to 5.0\% in the electron final state, with
larger uncertainties at higher \absleta. The smaller uncertainty for
electrons is related to the increased size of the misidentified-lepton
dominated control sample used for closure.  Each of these
normalization uncertainties is treated as uncorrelated with the
others.

A systematic uncertainty in the normalization of the QCD multijet
background is also estimated by a closure test in the
background-dominated region in bins of \ptl 3 (5)\GeV wide for the muon (electron) final state. 
The uncertainties range from 30 to 15\% (10 to 20\%), depending on the
\ptl region for the muon (electron) final state. Although the
uncertainty is related to differences in the composition of
misidentified leptons in the control and signal regions, which are
common across the whole \ptl range, the fraction of real leptons from
jets and random combinations of tracks and ECAL deposits within jets
might change across the phase space. Thus, conservatively, these
normalization uncertainties are also considered uncorrelated among
each other.

The closure test is also evaluated for the two charges separately,
weighting the events with the charge-independent
$\epsilon_{\text{pass}}$ misidentification efficiency. The two
estimates are consistent within the uncertainties, with a similar
dependency on \ptl and \leta. A further check was carried out by
computing a charge-dependent $\epsilon_{\text{pass}}^\pm$. Based on
these checks, an additional charge-dependent uncertainty of 2\% is
introduced in the muon case, in the same coarse bins of \absleta, to
include possible charge asymmetries in the production of true muons
from decays in flight of heavy quarks.  No additional uncertainty for
electrons is added, since the dominating source of misidentified
electrons is random geometric association of energy deposits in the
ECAL with tracks within jets, which is charge-symmetric.

The uncertainty in the extraction of the QCD multijet efficiency
$\epsilon_{\text{pass}}$ is evaluated as follows.
This lepton misidentification rate,
$\epsilon_{\text{pass}}$, is extracted through a linear fit to
\ptl, which has an uncertainty associated with it. While
a variation of the offset parameter of this fit is absorbed by the
normalization uncertainty, the linear parameter of the fit is varied,
which therefore varies the QCD multijet background as a function of
\ptl.  This uncertainty is applied in the same uncorrelated bins
of \absleta as the normalization uncertainty.

In total, 46 (55) nuisance parameters that affect the QCD multijet
background estimation are considered for each charge of the muon
(electron) final state. The larger number of parameters for the
electrons is due to a more granular binning and the larger acceptance
in \leta.

\subsubsection{Lepton momentum scale}
The lepton momentum scales are calibrated and corrected using events
from \PZ boson decays. Closure tests are performed by fitting the
invariant mass spectrum in data and simulation with a Breit--Wigner
line shape, convolved with a Crystal Ball function. The data-to-MC
difference in the fitted mass of the \PZ boson is taken as the
nonclosure. Small values of nonclosure may arise because the lepton
selection, fitting model, and invariant mass range are different in the
derivation of the lepton momentum scale calibrations, as compared to
the analysis.
This nonclosure is of the order of 10$^{-4}$ in the muon
case.  For such a precision, a detailed nuisance model was implemented
to cover residual effects~\cite{Bodek:2012id} that can remain after
the calibration procedure is applied.

Systematic uncertainties in the derivation of the muon momentum scale
corrections are included. These uncertainties are related to: the
modeling of \ptz, electroweak effects on the \PZ boson line shape,
and the effect of the acceptance on the dimuon invariant mass.  Hence,
they are finely grained in muon $\eta$ and \pt. Furthermore, the
uncertainty in the limited data and simulated \PZ sample is
estimated from 100 statistical replicas of the two data sets. 
Every such replica is constructed from a subset of the total event ensemble
through a case resampling using a replacement method~\cite{efron1979}.
Each of them is also finely binned in muon
$\eta$ and \pt. The 99 independent statistical uncertainties are
diagonalized with the procedure of Ref.~\cite{Carrazza:2015aoa}, and
their independent contributions are included as shape nuisance
effects.

For electron candidates, the observed residual differences in the
energy scales for the data and the simulated $\cPZ$ sample are of the
order of 10$^{-3}$. A procedure similar to that used for the muon
momentum scale is adopted.  Two systematic effects are included in
fine bins of \etael and \ptel. The first is the difference in the
\PZ boson mass value obtained by fitting the mass peak for
$\PZ\to\Pep\Pem$ events in two different ways. The first fit uses a
MC template convolved with a Gaussian resolution function and the
second with a functional form consisting of a Breit--Wigner line shape
for a \PZ boson, convolved with a Crystal Ball function, with
floating mean and width
parameters~\cite{Oreglia:1980cs,Gaiser:1982yw}. The effect is the main
contribution to the systematic uncertainty, and ranges from 0.1 to
0.2\% for $\pt^{\Pe}<45\GeV$ and 0.2--0.3\% at higher values.  The
second smaller systematic effect comes from the modeling of \ptz.  In
the muon case, the limited size of the samples used to derive the
energy scale corrections is accounted for by the means of 100 replicas
of the data and MC samples, diagonalized to get 99 independent
nuisance parameters.

For both lepton flavors, the precision in the estimate of the momentum
scale decreases when increasing \absleta. The \PW boson sample with
a lepton in the more forward regions of the detector still has
sufficient statistical power to allow the fit to constrain the
momentum scale nuisance parameters. If the systematic effect related
to the momentum scale is fully correlated across the full \leta
acceptance, then its constraint in the profiling procedure, driven by
the large effect on the templates at high \absleta, may result in an
unphysical constraint in the central region.  This is avoided by
decorrelating the nuisance parameters related to the various momentum
scale systematics in wide bins of \leta, for both muons and
electrons. In contrast, the parameters relating to the statistical
part of this uncertainty are kept fully correlated across
\leta.

Since the systematic uncertainty in the momentum scale of the leptons
allows the \pt of a lepton to be changed and, therefore, for bin-to-bin
migration, it is applied as a shape uncertainty.

\subsubsection{Lepton efficiency scale factors}
\label{sec:lepeff}
Data-to-simulation efficiency scale factors are derived through the
tag-and-probe method, also using $\cPZ\to\ell\ell$ events.  Two types of
systematic uncertainties are considered for the tag-and-probe method. 

The first uncertainty comes
from the scale factors themselves and depends on the functional forms used to
describe the background and signal components when fitting the
efficiencies in each bin of \leta as a function of \ptl of the probe
lepton. In order to estimate it, alternative fits are performed by using different models for
the dilepton invariant mass line shape for either the \PZ boson
events or for the combinatorial background events, resulting in
different efficiencies. The alternative signal shape is a
Breit--Wigner function with the nominal $\cPZ$ boson mass and width,
convolved with an asymmetric resolution function (Crystal Ball
function) with floating parameters. The alternative background
description is done with a function modeling the invariant mass of
random combinations of two leptons satisfying the minimum \pt
criteria. Overall, this alternative signal and background systematic uncertainty 
is assumed to be correlated among all
bins in \leta, and the size of it ranges from a few per mill
at low \absleta, to around 1--2\% in the very forward region. 

The second type of systematic uncertainty in the lepton efficiency 
scale factors arises from the statistical
uncertainties in the event count in each \leta bin in which the efficiencies are measured.
These statistical uncertainties are derived by varying the parameters of the error
function that is used to interpolate between the measured efficiency
values as a function of \ptl, described in
Sec.~\ref{sec:efficiencies}, by their uncertainties. These statistical uncertainties
are uncorrelated between each bin in \leta.
In total, this procedure of estimating the statistical uncertainty introduces three 
nuisance parameters for each bin in \leta,
resulting in a total of 144 (150) nuisance parameters per charge in
the muon (electron) final state. The larger number of parameters for
the electrons is due to the larger acceptance in \leta.  These
systematic uncertainties are considered uncorrelated for the two
charges since they are measured independently, and the statistical
uncertainty of the data and MC sample in each bin is large. 

One additional uncertainty in the trigger efficiency is included for
events with electrons in the end cap sections of the detector. This
uncertainty is due to a radiation-induced shift in the ECAL timing in
the 2016 data-taking period, which led to an early event readout
(referred to as prefiring) in the L1 trigger and a resulting
reduction in the efficiency for events with significant energy
deposits in the ECAL end cap sections.  The correction is estimated
using a set of the $\PZ\to\Pep\Pem$ events collected in collisions
where, because of L1 trigger rules, the event is saved regardless of
the L1 trigger decision for the in-time bunch crossing (BX).  This
sample is composed of events where the L1 decision is positive for the
third BX before the in-time BX: this records only about 0.1\% of the
total $\PZ\to\Pep\Pem$ events and is thus statistically limited.  The
uncertainty ranges from 0.5\% for $\abseta \approx 1.5$ to 10\% at
$\abseta \approx 2.5$ for electrons from \PW boson decays.

\subsubsection{Extra lepton veto}
To reduce multilepton backgrounds, especially $\cPZ\to\ell\ell$,
a veto on additional leptons is implemented. The efficiency of this veto
depends on the differences in the lepton selection efficiencies between the data and MC simulation.
Since more background survives
the selection at higher \absleta, where the uncertainties in the
lepton efficiencies are larger, a normalization uncertainty is applied, equal to
2 (3)\% for the muon (electron) channel.  In the electron channel, an
additional uncertainty is included to account for the L1 trigger
prefiring effect, described previously in Sec.~\ref{sec:lepeff},
in $\PZ\to\Pep\Pem$ events in which one electron is in one of the ECAL
end cap sections. This uncertainty ranges from 2\% at low electron
\pt to 10\% in the highest \absleta and \ptl bins.

\subsubsection{Charge misidentification}
The probability of mistakenly assigning the incorrect charge to a muon
in the \ptl range considered is
negligible~($10^{-5}$)~\cite{Chatrchyan:2011jz}, thus no uncertainty
is introduced for this effect. For the electrons, the
statistical uncertainty in the estimate of wrong charge assignment in
$\PZ\to\Pep\Pem$ events reconstructed with same-sign or opposite-sign
events is used. It is dominated by the limited sample of same-sign
events in the 2016 dataset.  The uncertainty assigned to this small
background component, in the electron channel only, is
30\%~\cite{CMS_DPS_EGAMMA_2018_017}.

\subsubsection{Integrated luminosity}
Because of the imperfect knowledge of the integrated luminosity, a fully correlated 
normalization uncertainty is assigned to all processes estimated from a MC simulation.
Its value is set to 2.5\%~\cite{CMS:2017sdi}.

\subsection{Modeling and theoretical uncertainties}
\label{sec:theounc}

\subsubsection{\texorpdfstring{\ptw}{pT(W)} modeling and missing higher orders in QCD}

Imperfect knowledge of the \ptw spectrum results in an uncertainty
that affects the \ptl spectrum. It is most important in the region of
low \ptw, where fixed-order perturbative calculations lead to
divergent cross sections as \ptw approaches zero, which can be fixed
by using resummation.  The nominal templates are evaluated from
the \MGvATNLO simulated sample with the \ptw spectrum reweighted by
the measured data versus MC corrections in the \ptz distribution
obtained in data, as described in Sec.~\ref{sec:dataSimulation}. 

The theoretical uncertainties resulting from missing higher orders in
the QCD calculations, associated with the \ptw modeling, are implemented
in such a way as to reduce the sensitivity to the
theoretical prediction, at the cost of increasing the statistical
uncertainty of the results. They are implemented in the following way.

Renormalization and factorization scales, \muR and \muF,
respectively, are changed to half and twice their original value. This
change is propagated to the resulting weight for each simulated event
in three variations: the uncorrelated ones in which either
\muR or \muF is varied, and the correlated one
in which both are varied simultaneously but in the same direction,
\ie, both up or down by a factor of two. 
This uncertainty is applied
to all signal processes, as well as to the simulated $\cPZ\to\ell\ell$
background.  For the signal processes, these variations lead to a
normalization shift that is largely independent of \leta. The impact
on the shape of the \ptl distribution is within 0.5\% up to
$\ptl<35\GeV$; however, for $\ptl>35\GeV$ a significant modification
of the predicted \ptl distribution is seen.  These uncertainties
change both the normalization and the shape of the overall 2D
templates. In the case of the signal, they are split into several
components.
The uncertainties in \muR and \muF are divided into ten bins of \ptw:
[0.0, 2.9, 4.7, 6.7, 9.0, 11.8, 15.3, 20.1, 27.2, 40.2, and
13\,000]\GeV. These nuisance parameters are uncorrelated for each
charge. In the case of the polarized cross section measurement, an
uncorrelated uncertainty is used for each helicity state to account
for the different production mechanisms of the longitudinally, left,
and right polarized \PW bosons. The \muR and
\muF uncertainties in the $\PW\to\tau\nu$ process are
binned in the same \ptw bins, albeit integrated in polarization, and
so are uncorrelated with the signal processes.

\subsubsection{Parton distribution functions}
Event weights in a MC simulation derived from 100 variations of the
\NNPDF PDF set, referred to as replica sets, are used to evaluate the
PDF uncertainty in the predictions.  These 100 replicas are
transformed to a Hessian representation to facilitate the treatment of
PDF uncertainties in the analysis via the procedure described in
Ref.~\cite{Carrazza:2015aoa} with 60 eigenvectors and a starting
scale of 1\GeV.  Because the PDFs determine the kinematics and the
differential polarization of the \PW boson, variations of the PDFs
alter the relative contribution of the \PW boson helicity states
in \ptw and \absyw. Thus, the alternative weighting of the signal
templates described in Sec.~\ref{sub:templateReweighting} is
repeated independently for each of the 60 Hessian variations.  Each
signal process is reweighted once for each of the 60 independent
variations as the \textit{up} variation, corresponding to one positive
standard deviation. The corresponding \textit{down} variation is
obtained by mirroring the up variation with respect to the
nominal template. Since the underlying PDF uncertainties also affect
the DY and $\PW\to\tau\nu$ backgrounds, the same procedure is applied
to the simulated events for these backgrounds, and the uncertainties
are treated as fully correlated between the signal and these two
background processes.  This procedure changes the overall
normalization of the templates as well as their shapes. The magnitudes
of the Hessian variations are 1\% or lower for the normalization, but
show significantly different behavior in the \ptl versus \leta plane,
from which a constraint on these PDF uncertainties is expected.

\subsubsection{Choice of \texorpdfstring{\alpS}{alpha S} value}
The 100 PDF replicas of the \NNPDF set are accompanied by two
variations of the strong coupling. The central value of
\alpS at the mass of the \PZ boson of 0.1180 is varied
from 0.1195 to 0.1165. Both normalization and shape are affected by
this variation.

\subsubsection{Simulated background cross sections}
The backgrounds derived from simulation, namely DY, diboson, and
$\PW\to \tau \nu$ production, and all top quark backgrounds
are subject to an overall normalization-only uncertainty. The main
contributions to the theoretical uncertainty in the \PZ and \PW
boson production cross section arise from the PDF uncertainties,
\alpS, and \muR and \muF.  These
are included as shape nuisance parameters affecting the templates of
such processes, and they are fully correlated with the same parameters
affecting the signal. For the $\PW\to\tau\nu$ process, a further 4\%
normalization uncertainty is assigned, to address the residual
uncertainty because of the much lower \pt of the decay lepton.

For the top quark and diboson backgrounds, the kinematic distributions
are well modeled by the higher-order MC generators.  The uncertainties
assigned to the normalization are 6 and 16\%, respectively, motivated
by the large theoretical cross section uncertainty for each of the
contributing processes.  Because these processes make a small
contribution to the selected sample of events, the effect of these
relatively large uncertainties is small.

\subsubsection{Choice of the \texorpdfstring{$m_{\PW}$}{mW} value}
Events are reweighted to two alternative values of $m_{\PW}$ with
values $\pm$50\MeV, with respect to the default $m_{\PW}$ value in
the generator of 80.419\GeV, using a Breit--Wigner assumption for the
invariant mass distribution at the generator level. Since the central
value of $m_{\PW}$ does not significantly influence the \PW boson
cross sections, the impact of this uncertainty is very small.

\subsubsection{Modeling of QED radiation}
The simulation of the signal processes models the lepton FSR through
the quantum electrodynamic (QED) showering in \PYTHIA within the
\MGvATNLO MC generator.  An uncertainty in this modeling is assessed
by considering an alternative showering program,
\PHOTOS~3.56~\cite{Golonka:2005pn}. A large sample of $\PW\to\ell\nu$
($\ell=\Pep,\Pem,\mu^+,\mu^-$ separately) events is produced at the
generator level only at NLO in QCD, and is interfaced to either
\PYTHIA or \PHOTOS. The variable sensitive to FSR, which accounts for
the different radiation rate and, in case of radiation, for the harder
FSR photon spectrum produced by \PHOTOS with respect to \PYTHIA, is
the ratio
$r_\mathrm{FSR}=\pt^\text{dress}/\pt^\text{bare}$ between the
dressed lepton \pt and the bare lepton \pt (after radiation).
Alternative templates are built by reweighting the nominal \MGvATNLO
events by the ratio between \PHOTOS and \PYTHIA, as a function of
$r_\mathrm{FSR}$.

The effect of QED FSR is largely different for the two lepton flavors
because of the differences in the lepton masses and the estimate of
the lepton momentum. For the muons, only the track is used, and there
is no explicit recovery of the FSR. For these reasons, the nuisance
parameters related to this effect are kept uncorrelated between the
two lepton flavors.  For the electrons, the effect is derived from a
combination of the measurements using the track and the ECAL
supercluster. The latter dominates the estimate for the energy range
exploited in this analysis, and its reconstruction algorithm,
optimized to gather the bremsstrahlung photons, also efficiently
collects the FSR photons.

\subsubsection{Statistical uncertainty in the \texorpdfstring{$\PW$}{W} simulation}
An uncertainty is assigned to reflect the limited size of the MC
sample used to build the signal templates. The sample size, when
considering the negative weights of the NLO corrections, corresponds
to approximatively one fifth of the data sample. This is included in
the likelihood with the Barlow--Beeston Lite
approach~\cite{Barlow:1993dm} and represents one of the dominant
contributions to the systematic uncertainty.

A summary of the systematic uncertainties is shown in
Table~\ref{tab:systematics}.  They amount to 1176 nuisance parameters
for the helicity fit.

\begin{table*}[t]
\centering
\topcaption{Systematic uncertainties for each source and process. Quoted numbers correspond to the size of log-normal nuisance parameters 
applied in the fit, while a ``yes'' in a given cell corresponds to the
given systematic uncertainty being applied as a shape variation over
the full 2D template space.\label{tab:systematics}}
\cmsTable{
\begin{scotch}{l c c c c c c c }

Source/process         & Signal & DY    & $\PW\to\tau\nu$             & QCD     & Top      & Dibosons & Charge flips \\

\hline
\multicolumn{8}{c}{Normalization uncertainty for $\PW\to\ell\nu$ ($\ell=\mu,\Pe$)}                        \\
\hline
Integrated luminosity       & 2.5\%  & 2.5\%     & 2.5\%        &\NA        & 2.5\%    & 2.5\%    & \NA              \\
DY cross section   &\NA     & 3.8\%  &\NA          &\NA        &\NA       &\NA       & \NA              \\
\ttbar, single-t cross section &\NA     &\NA     &\NA          &\NA        & 6\%      &\NA       & \NA              \\
Diboson cross section  &\NA     &\NA      &\NA         &\NA        &\NA       & 16\%     & \NA              \\

\hline
\multicolumn{8}{c}{Normalization uncertainty for $\PW\to\mu\nu$}                        \\
\hline
QCD normalization vs. \leta    &\NA     &\NA &\NA               & 5\%     &\NA       &\NA       & \NA              \\
QCD charge asymmetry vs. \leta    &\NA     &\NA &\NA               & 2\%     &\NA       &\NA       & \NA              \\
QCD normalization vs. \ptl     &\NA     &\NA   &\NA            & 15--30\% &\NA       &\NA       & \NA        \\
Lepton veto           &\NA     & 2\%  &\NA&\NA        &\NA       &\NA       & \NA              \\

\hline
\multicolumn{8}{c}{Normalization uncertainty for $\PW\to\Pe\nu$}                        \\
\hline
QCD normalization vs. \leta    &\NA     &\NA     &\NA         & 1--6\%      &\NA       &\NA       & \NA              \\
QCD normalization vs. \ptl     &\NA     &\NA     &\NA         & 10--30\%      &\NA       &\NA       & \NA        \\
Charge-flip normalization   &\NA     &\NA           &\NA    &\NA        &\NA       &\NA       &  30\%            \\
Lepton veto           &\NA     & 3\% &\NA &\NA      &\NA       &\NA       & \NA              \\
\hline
\multicolumn{8}{c}{Shape uncertainty for $\PW\to\ell\nu$ ($\ell=\mu,\Pe$)}                                \\
\hline
Lepton efficiency \syst     & yes    & yes & yes              &\NA        &\NA       &\NA       & \NA              \\
Lepton efficiency \stat     & yes    & yes & yes              &\NA        &\NA       &\NA       & \NA              \\
L1 trigger pre-firing    & yes    & yes & yes              &\NA        &\NA       &\NA       & \NA              \\
60 PDF variations              & yes    & yes     & yes          &\NA        &\NA       &\NA       & \NA              \\
\alpS            & yes    & yes     & yes          &\NA        &\NA       &\NA       & \NA              \\
\muF (binned in \ptw)  & yes    &\NA  & yes            &\NA        &\NA       &\NA       & \NA              \\
\muR (binned in \ptw)  & yes    &\NA  & yes            &\NA        &\NA       &\NA       & \NA              \\
$\mu_\mathrm{{F+R}}$ (binned in \ptw)  & yes    &\NA & yes             &\NA        &\NA       &\NA       & \NA              \\
\PW boson mass  & yes    &\NA  &\NA           &\NA        &\NA       &\NA       & \NA              \\

\muF                &\NA   & yes   &\NA           &\NA        &\NA       &\NA       & \NA              \\
\muR                &\NA   & yes   &\NA           &\NA        &\NA       &\NA       & \NA              \\
$\mu_{\mathrm{F+R}}$            &\NA   & yes    &\NA          &\NA        &\NA       &\NA       & \NA              \\
$\mu$ momentum scale  \syst   & yes    & yes     & yes        &\NA        &\NA       &\NA       & \NA              \\
$\mu$ momentum scale  \stat   & yes    & yes     & yes        &\NA        &\NA       &\NA       & \NA              \\
 \Pe  momentum scale  \syst   & yes    & yes     & yes        &\NA        &\NA       &\NA       & \NA              \\
 \Pe  momentum scale  \stat   & yes    & yes     & yes        &\NA        &\NA       &\NA       & \NA              \\
Lepton misidentification vs. \ptl      &\NA     &\NA  &\NA             & yes       &\NA       &\NA       & \NA              \\
QED radiation               & yes    &\NA &\NA              &\NA        &\NA       &\NA       & \NA              \\   
Simulated sample size       & yes    & yes & yes &\NA& yes & yes & yes  \\
\end{scotch}
}
\end{table*}

\subsection{Impact of uncertainties in the measured quantities}
\label{sec:impacts}

\begin{figure}[ht!]
\centering
\includegraphics[width=0.97\linewidth]{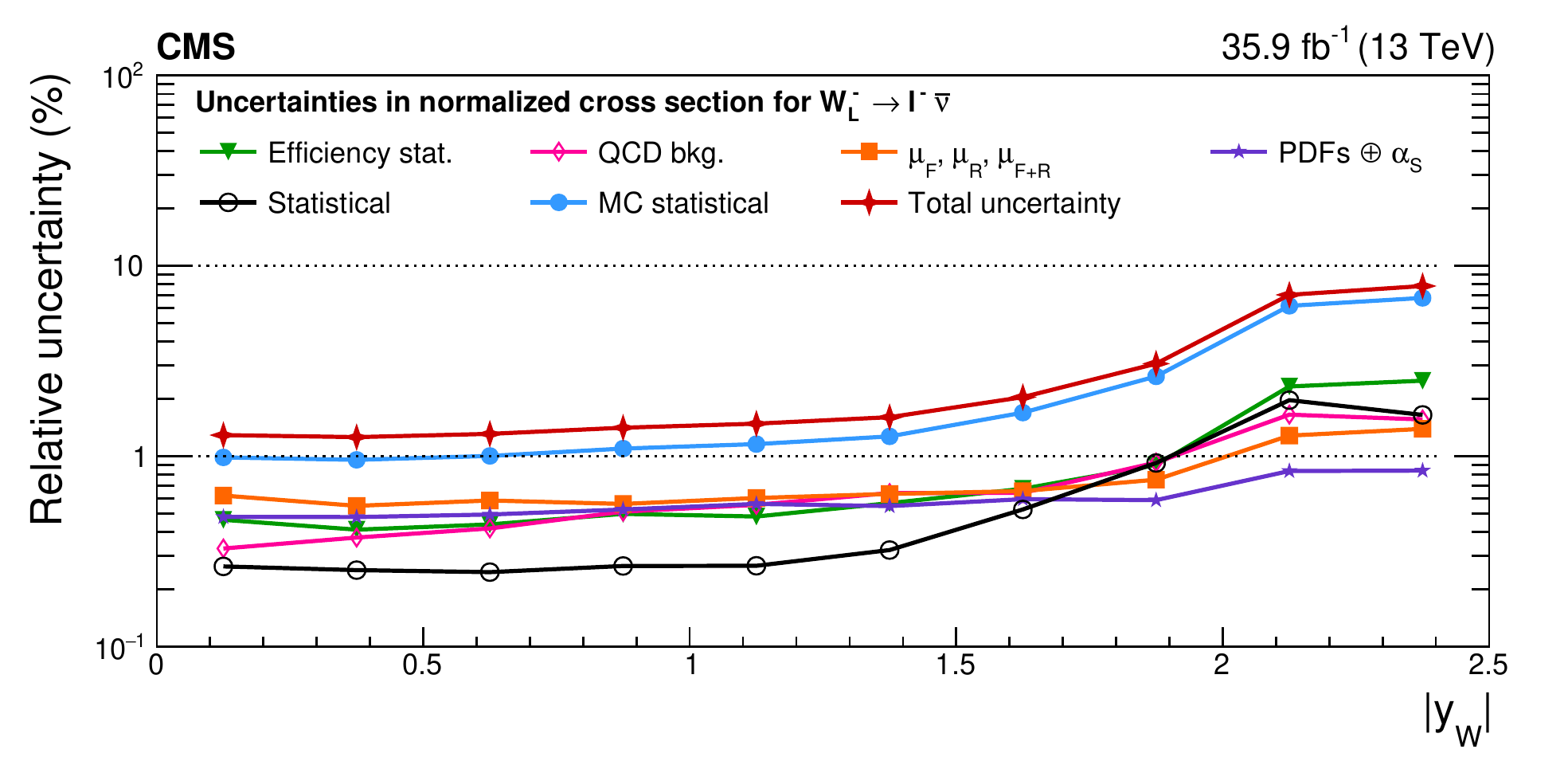} \\
\includegraphics[width=0.97\linewidth]{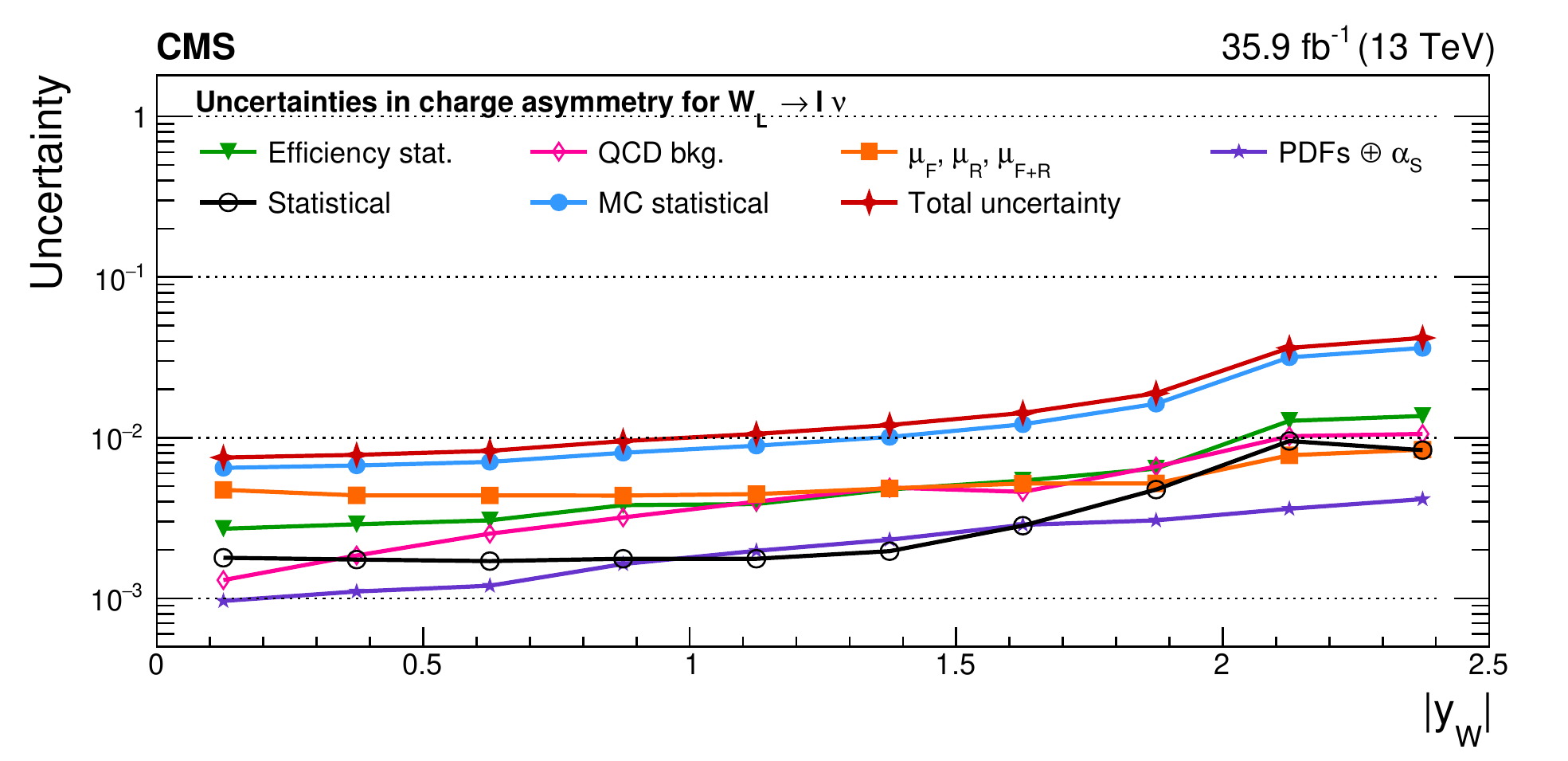}
\caption{Upper: relative impact of groups of uncertainties (as defined
  in the text) on the normalized signal cross sections as functions of
  the \PW boson rapidity for the \wminusl case. Lower: absolute
  impact of uncertainties on the charge asymmetry of the \wleft
  boson. All impacts are shown for the combination of the muon and
  electron channels in the helicity fit. The groups of
  uncertainties subleading to the ones shown are suppressed
  for simplicity.\label{fig:impacts}}
\end{figure}

The effects of the systematic uncertainties on the measured quantities
(signal strength modifiers for one process, $\mu_{p}$ in
Eq.~(\ref{eqn:nll}), absolute cross sections $\sigma_{p}$, or
normalized cross sections $\sigma_{p}/\sigma_\text{tot}$) are
presented as the impact of an uncertainty in the parameter of
interest.  The impact on a given measured parameter $\mu_p$ from a
single nuisance parameter, $\theta_k$ in Eq.~(\ref{eqn:nll}), is
defined as $C_{pk}/\sigma(\theta_k)$, where $C_{pk}$ is the covariance
for the nuisance parameter and the parameter of interest, and
$\sigma(\theta_k)$ is the postfit uncertainty on the nuisance
parameter.  In the limit of Gaussian uncertainties, this is equivalent
to the shift that is induced as the nuisance parameter $\theta_k$ is
fixed and brought to its $+1\sigma$ or $-1\sigma$ postfit values,
with all other parameters profiled as normal. The procedure is
generalized to groups of uncertainties, gathered such that each group
includes conceptually related and/or strongly correlated sources.
Groups are defined for:
\begin{itemize}
\item \textit{luminosity ---\/} uncertainty in integrated luminosity,
\item \textit{efficiency stat. ---\/} uncorrelated part (in \leta) of
  the lepton efficiency systematics,
\item \textit{efficiency syst. ---\/} correlated part (in \leta) of
  the lepton efficiency systematics (coming from the tag-and-probe
  method), L1 prefiring uncertainty for the signal electron or the
  second electron from $\PZ\to\Pep\Pem$ events,
\item \textit{QCD bkg. ---\/} includes both the normalization and shape
  uncertainties related to the misidentified lepton background from
  QCD multijet events,
\item \textit{lepton scale ---\/} uncertainty in the lepton momentum scale,
\item \textit{other experimental ---\/} systematic uncertainties estimated from simulation and the extra-lepton veto,
\item \textit{other bkg ---\/} normalization uncertainties for all
  backgrounds, except for the nonprompt background,
\item \textit{PDFs $\oplus~\alpS$ ---\/} 60 Hessian variations of the
  NNPDF3.0 PDF set and \alpS,
\item \textit{\muF, \muR, $\mu_\mathrm{F+R}$ ---\/} separate \muR and
  \muF variations, plus the correlated variation of both
  \muR and \muF,
\item \textit{FSR ---\/} modeling of final state radiation,
\item \textit{MC sample size ---\/} statistical uncertainty per bin
  of the template for all the samples,
\item \textit{statistical ---\/} the statistical uncertainty in the data sample.
\end{itemize}
The impact of each group is the effect of the combined variation of
all the parameters included in it. It is evaluated as
$\sqrt{\smash[b]{v^\mathrm{T} C^{-1} v}}$, where $v$ ($v^\mathrm{T}$) is (the
transpose of) the matrix of the correlations between the measured
parameter and the nuisance parameters within the group, and $C$ is
the subset of the covariance matrix corresponding to the nuisance
parameters in the group.  This is equivalent to computing the combined
impact of the eigenvectors for the postfit nuisances within a
group. These groups cover all the nuisance parameters included in the
likelihood and are mutually exclusive.  Figure~\ref{fig:impacts}
summarizes the relative impact of groups of systematic uncertainties
for two illustrative measurements: the normalized cross sections and
the charge asymmetry for \wleft, both for the combination of the muon and
electron final states.  The total uncertainty is not expected to be
exactly equal to the sum in quadrature of the impacts due to remaining
correlations between groups.  The impact of uncertainties that are
strongly correlated among all the rapidity bins mostly cancel when
considering either the cross section normalized to the total cross
section or in the charge asymmetry. In these plots, the groups of
subleading uncertainties, with respect to the ones shown, are suppressed
for simplicity.

\begin{figure}[ht!]
\centering
\includegraphics[width=0.97\linewidth]{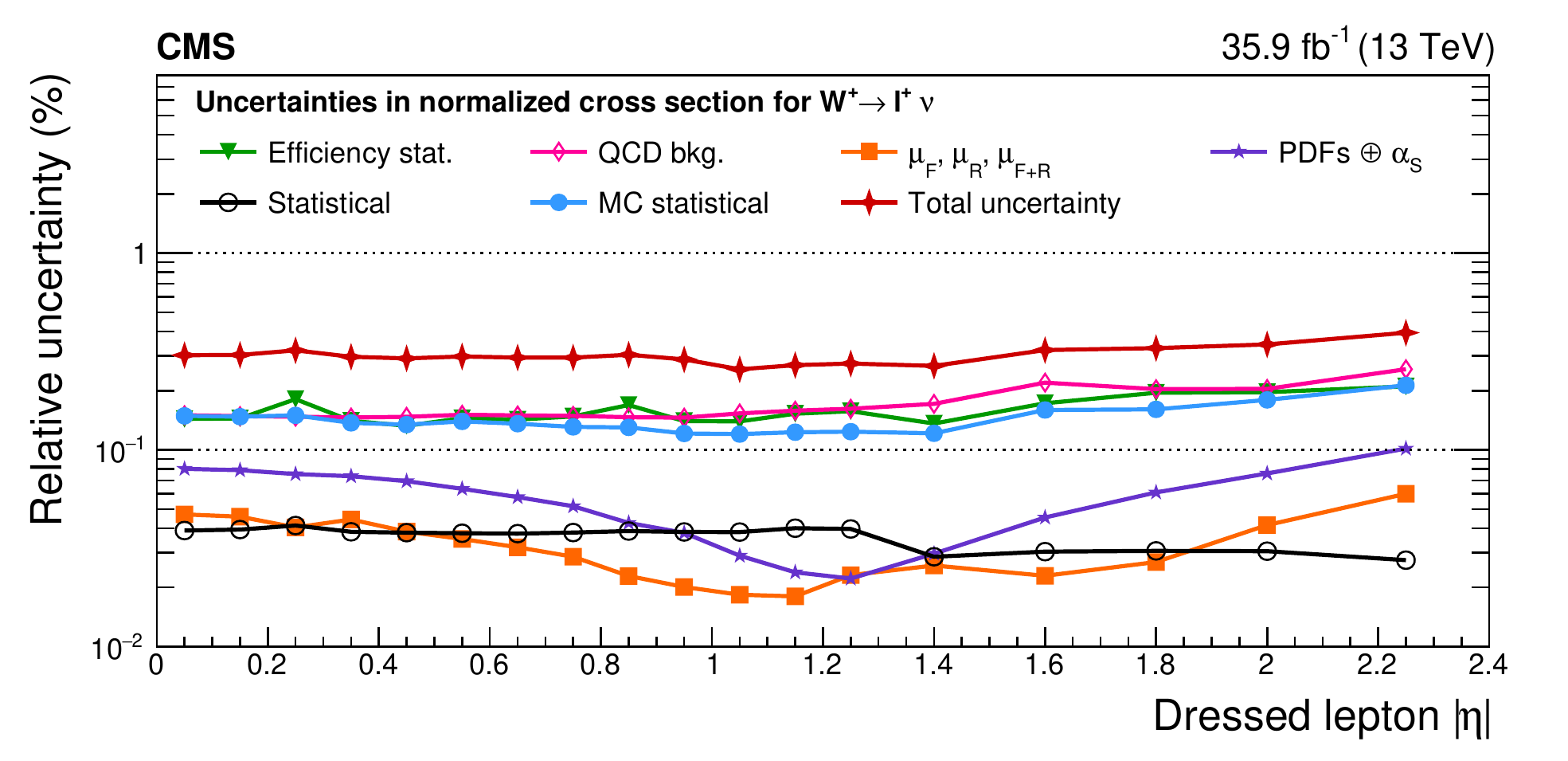} \\
\includegraphics[width=0.97\linewidth]{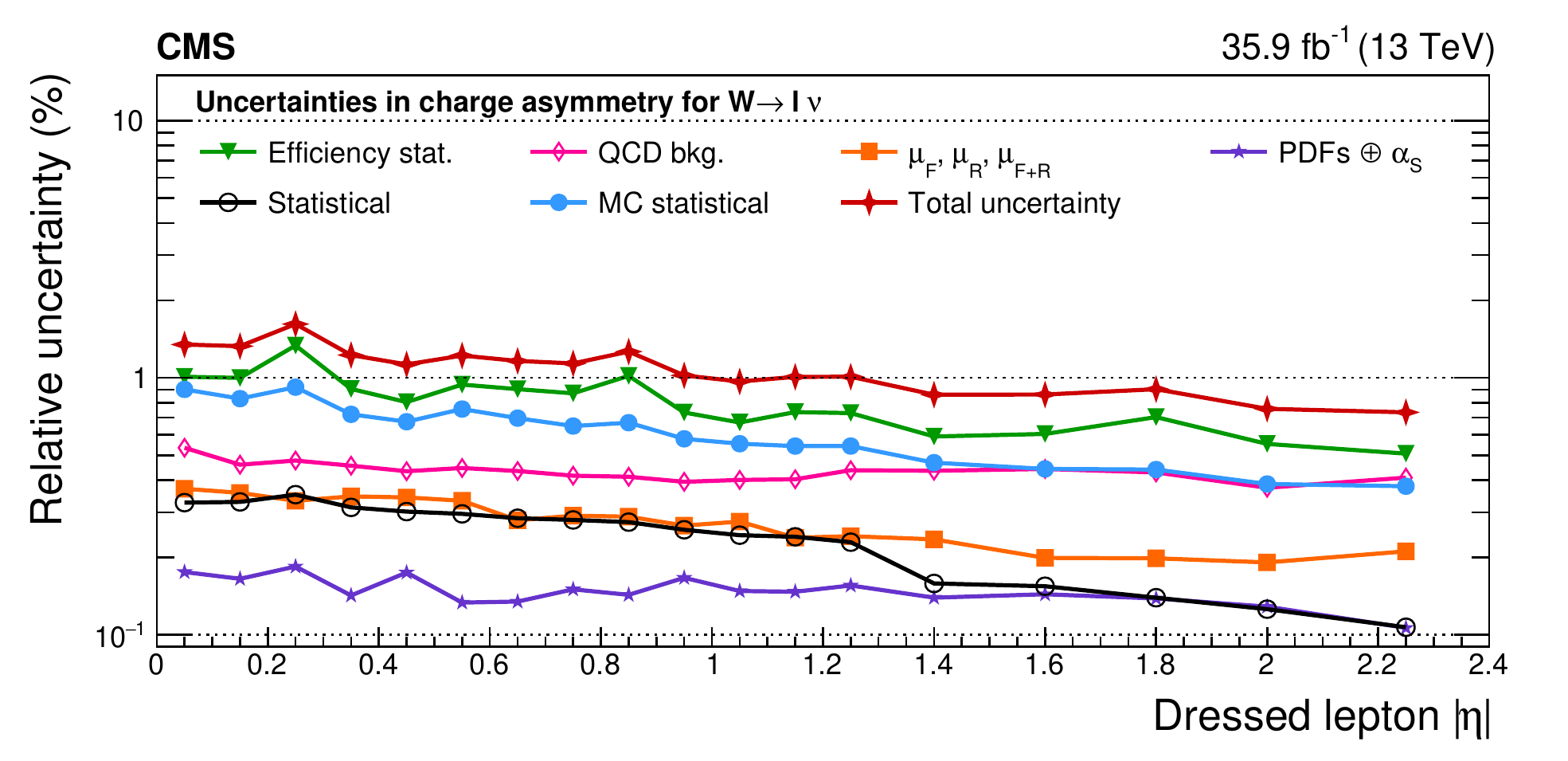}
\caption{Upper: relative impact of groups of systematic uncertainties
  (as defined in the text) on the normalized cross sections for the
  \wplus case as functions of \absleta. Lower: relative impact of
  uncertainties on charge asymmetry. All impacts are shown for the
  combination of the muon and electron channels in the
  double-differential cross section fit. The groups of uncertainties
  subleading to the ones shown are suppressed for
  simplicity. \label{fig:impacts2D}}
\end{figure}

In a similar manner, the effect of the statistical and systematic
uncertainties is shown for the normalized double-differential cross
section and for its charge asymmetry.  For simplicity, the
distribution is integrated over \ptl, and it is shown as a
function of \absleta in Fig.~\ref{fig:impacts2D}.

The two most dominant sources of uncertainties are the uncertainty in
the integrated luminosity and the uncertainty due to the limited size
of the MC sample compared with the size of the recorded data set. The
latter dominates for all normalized quantities, while the former is
the largest contribution to the total uncertainty in most regions of
the phase space for absolute quantities.

\section{Results and interpretations}
\label{sec:results}

\begin{figure*}[t]
\centering
\includegraphics[width=0.4\linewidth]{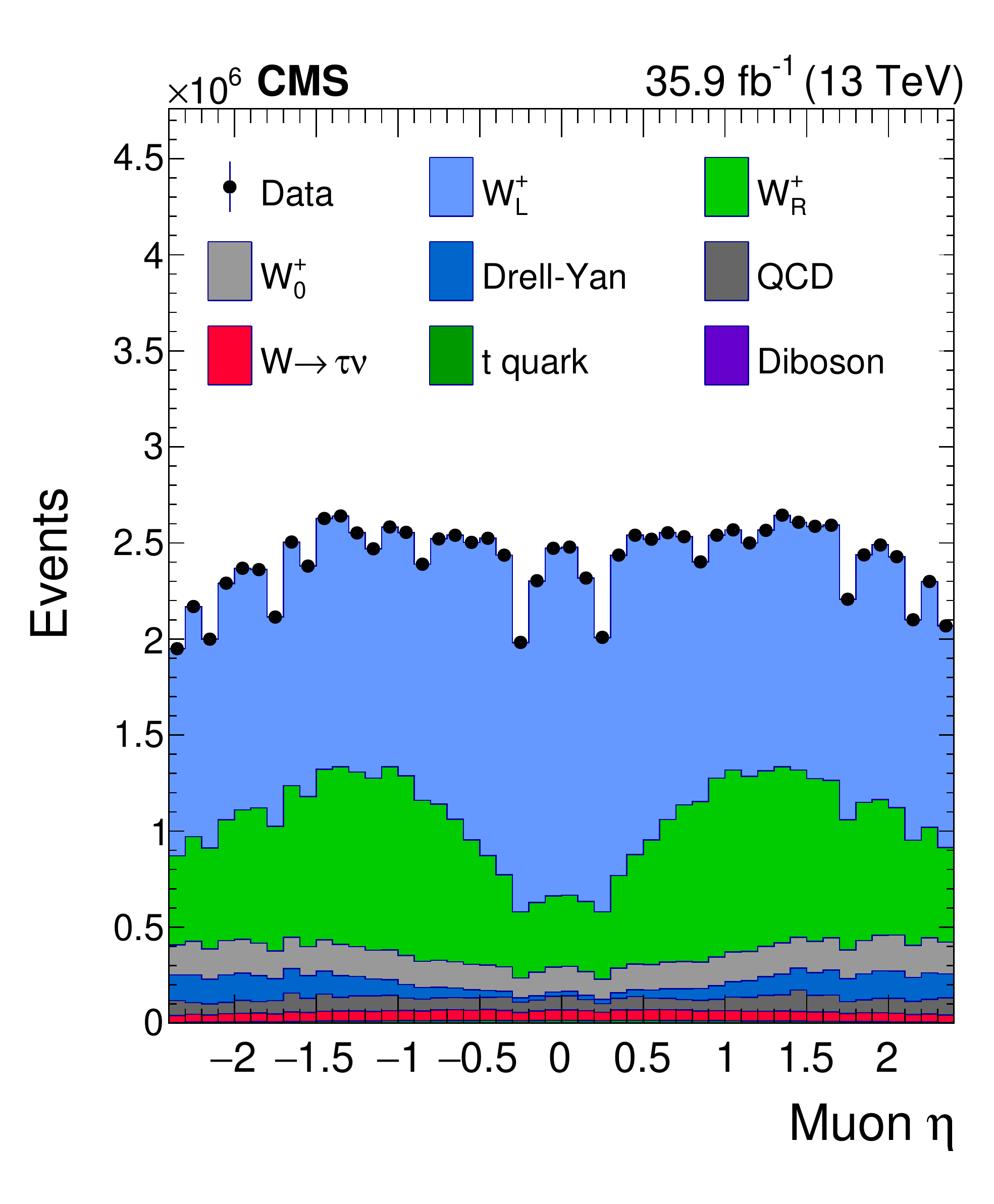}
\includegraphics[width=0.4\linewidth]{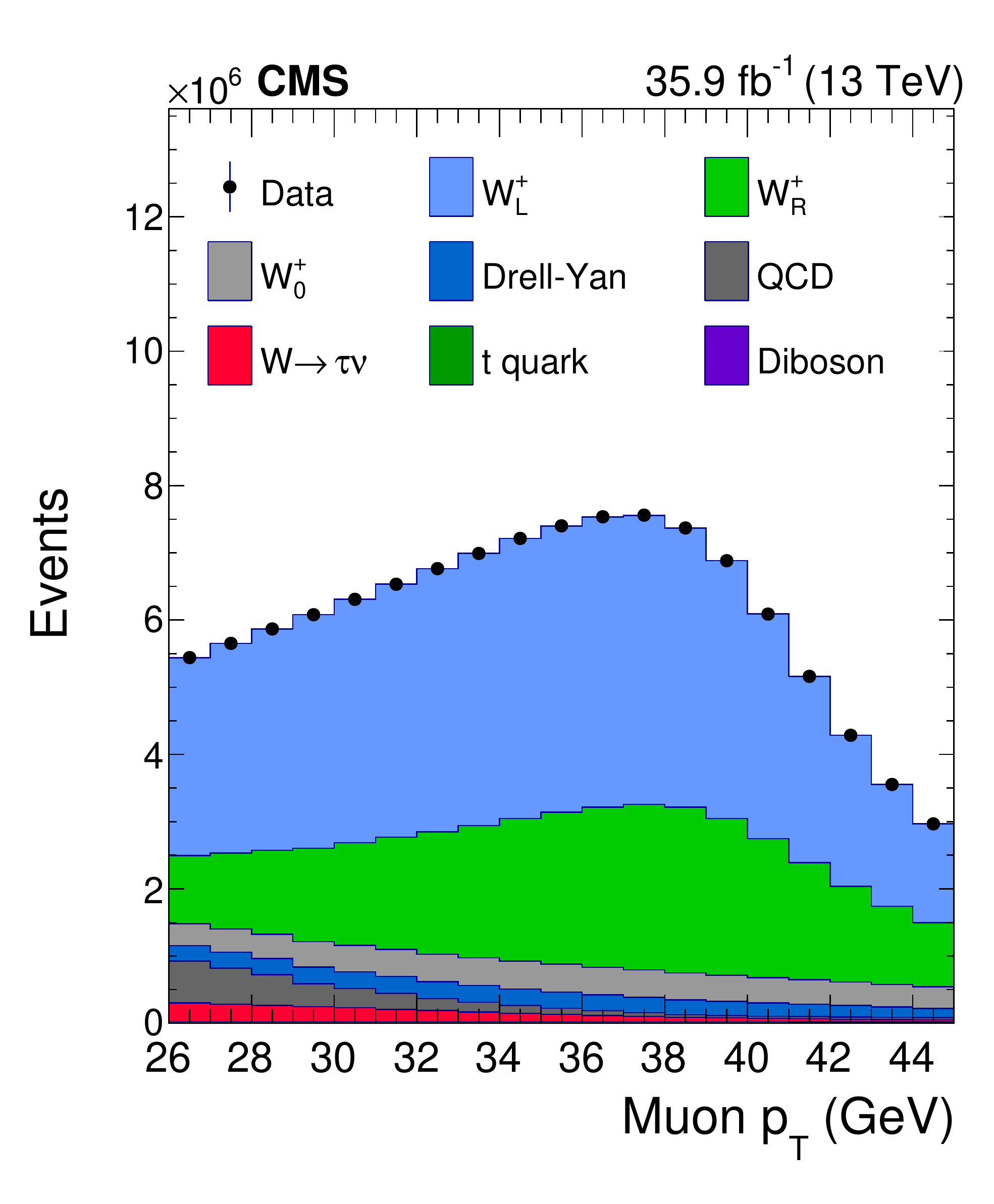} \\
\includegraphics[width=1.0\linewidth]{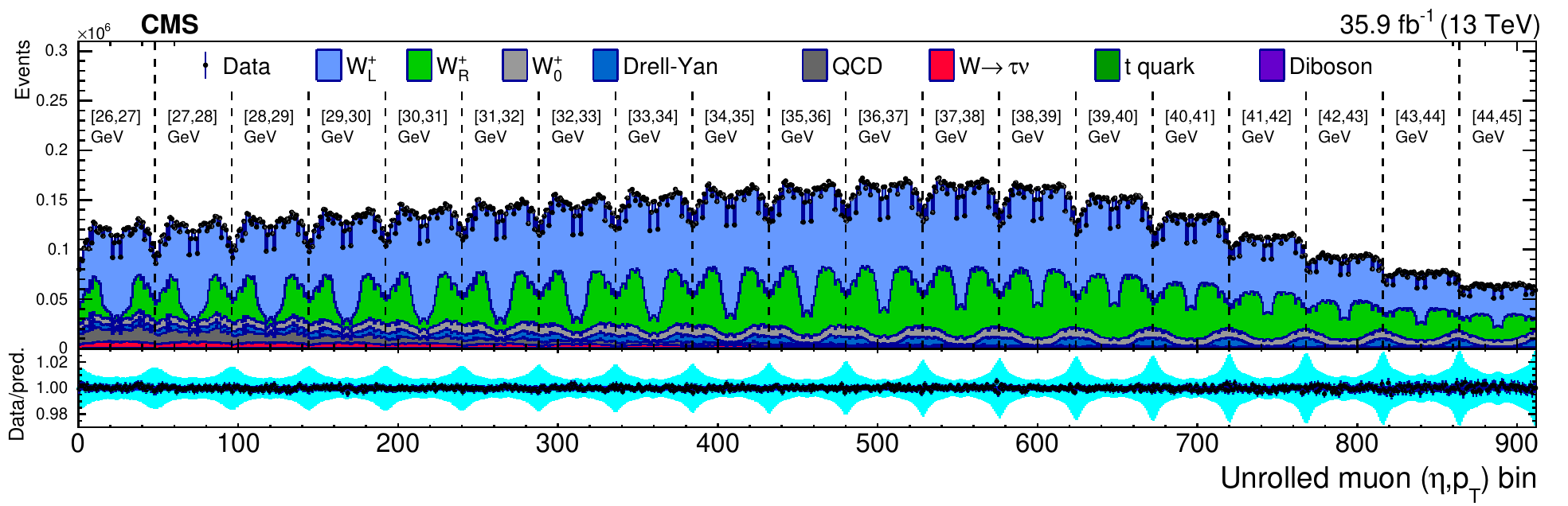}
\caption{Distributions of \etamu (upper left), \ptmu (upper right), and
  $\text{bin}_\text{unrolled}$ (lower) for $\wplus\to\mu^+\nu$ events for
  observed data superimposed on signal plus background events. The
  signal and background processes are normalized to the result of the
  template fit. The cyan band over the data-to-prediction ratio
  represents the uncertainty in the total yield in each bin after the
  profiling process. \label{fig:postfit_muplus}}
\end{figure*}

\begin{figure*}[t]
\centering
\includegraphics[width=0.4\linewidth]{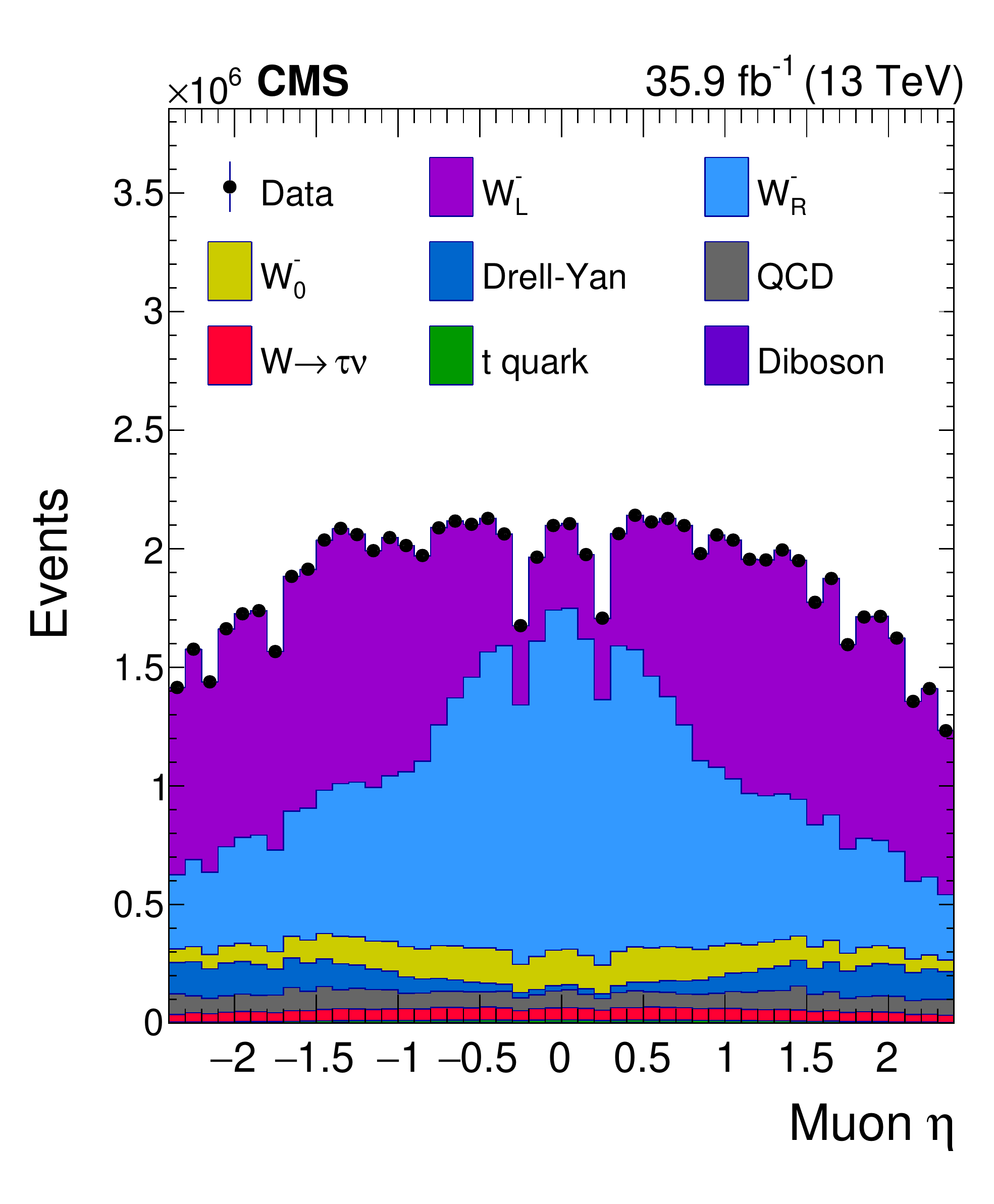}
\includegraphics[width=0.4\linewidth]{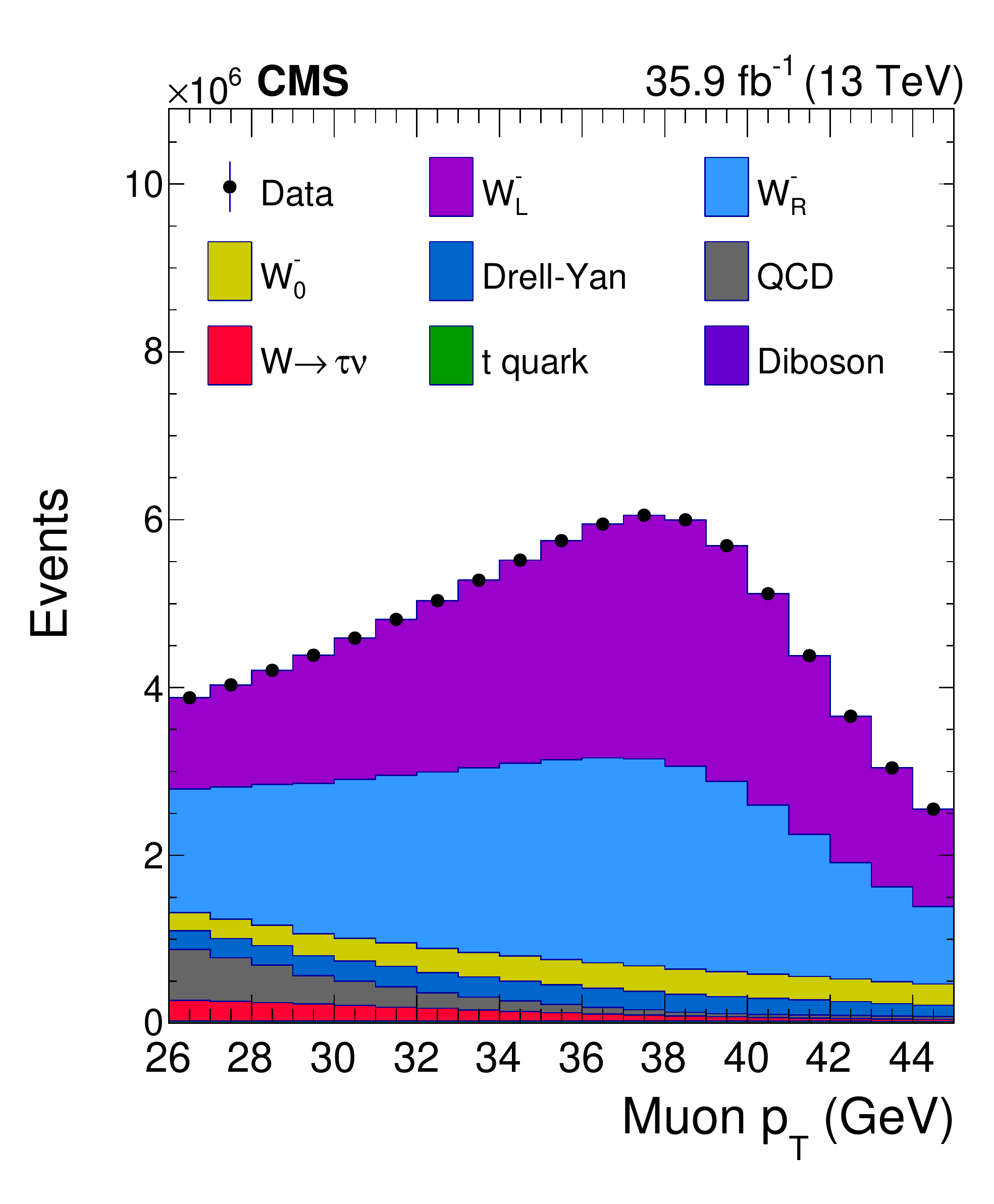} \\
\includegraphics[width=1.0\linewidth]{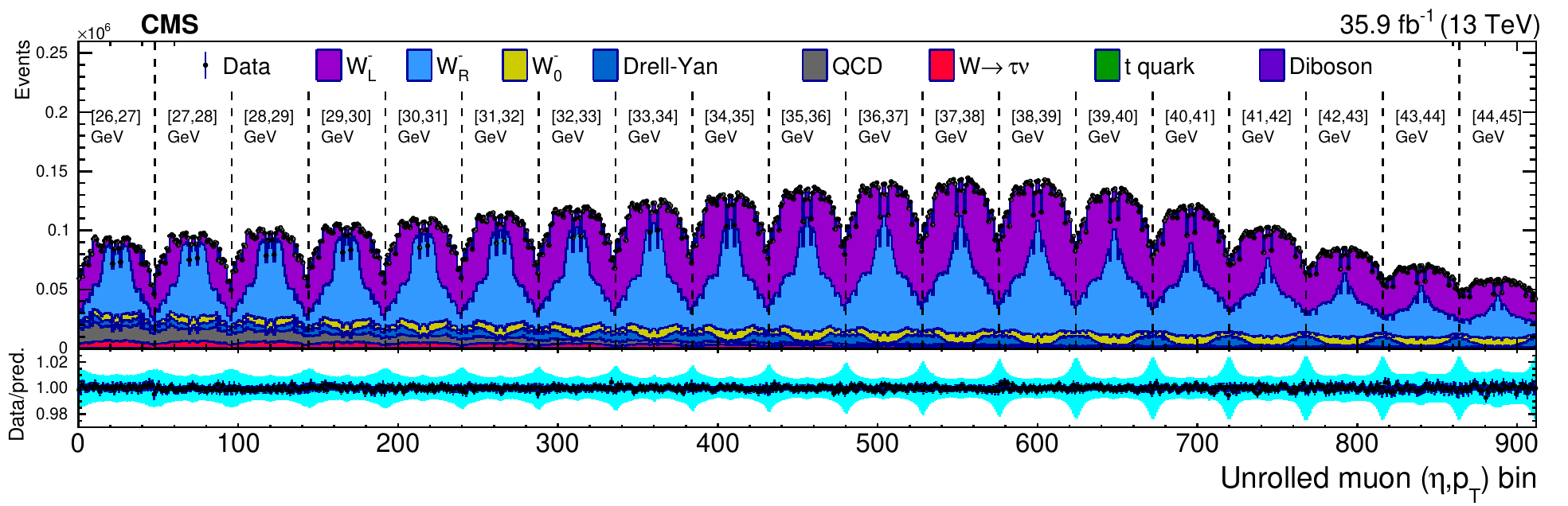}
\caption{Distributions of \etamu (upper left), \ptmu (upper right), and
  $\text{bin}_\text{unrolled}$ (lower) for $\wminus\to\mu^-\PAGn$ events
  for observed data superimposed on signal plus background events. The
  signal and background processes are normalized to the result of the
  template fit. The cyan band over the data-to-prediction ratio
  represents the uncertainty in the total yield in each bin after the
  profiling process. \label{fig:postfit_muminus}}
\end{figure*}

\begin{figure*}[t]
\centering
\includegraphics[width=0.4\linewidth]{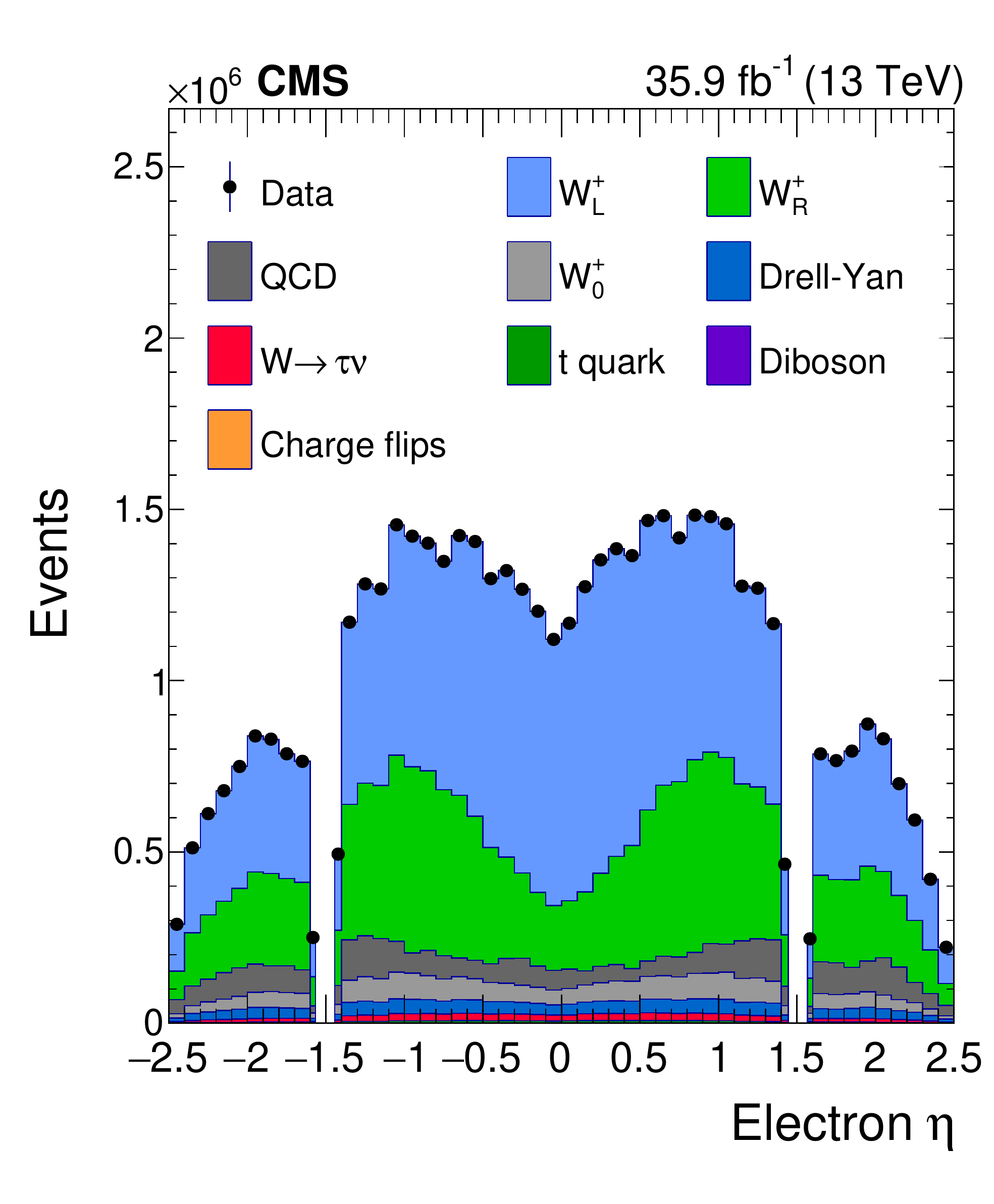}
\includegraphics[width=0.4\linewidth]{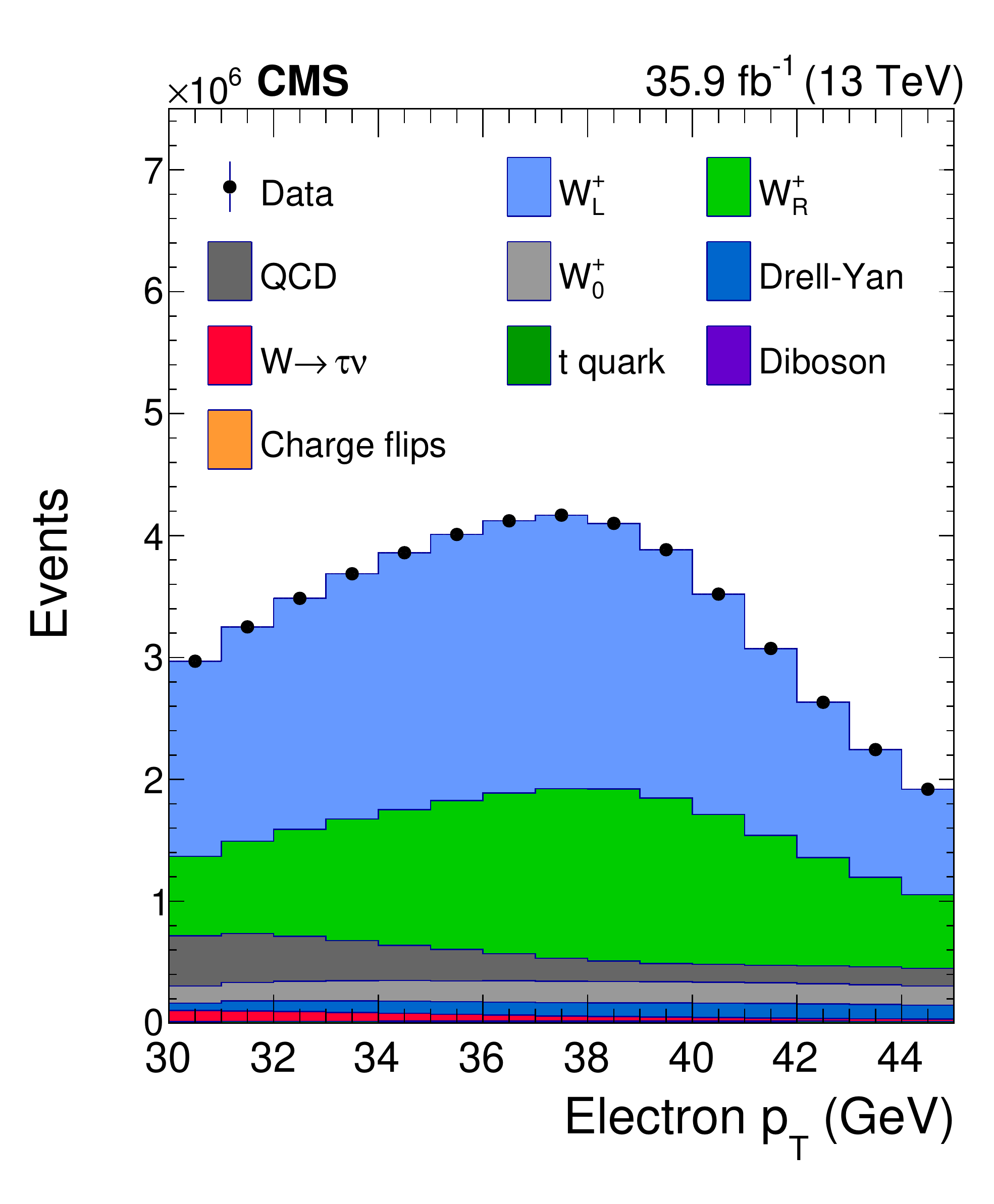} \\
\includegraphics[width=1.0\linewidth]{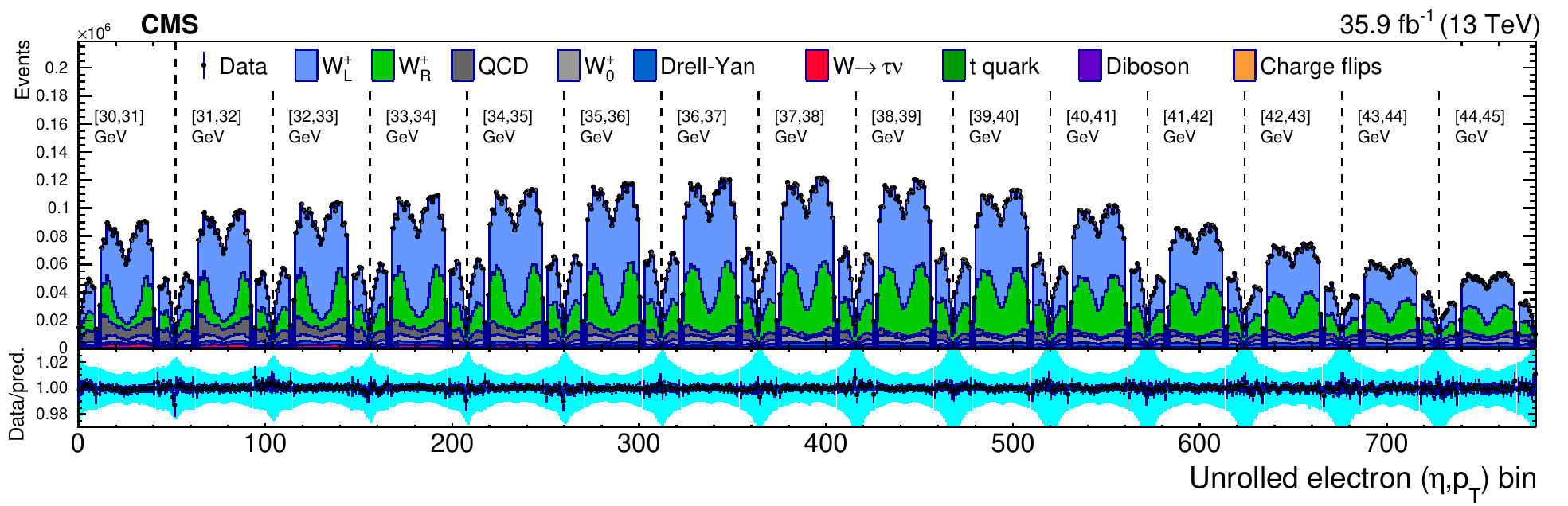}
\caption{Distributions of \etael (upper left), \ptel (upper right), and
  $\text{bin}_\text{unrolled}$ (lower) for $\wplus\to\Pe^+\nu$ events for
  observed data superimposed on signal plus background events. The
  signal and background processes are normalized to the result of the
  template fit. The cyan band over the data-to-prediction ratio
  represents the uncertainty in the total yield in each bin after the
  profiling process.
  \label{fig:postfit_elplus}}
\end{figure*}

\begin{figure*}[t]
\centering
\includegraphics[width=0.4\linewidth]{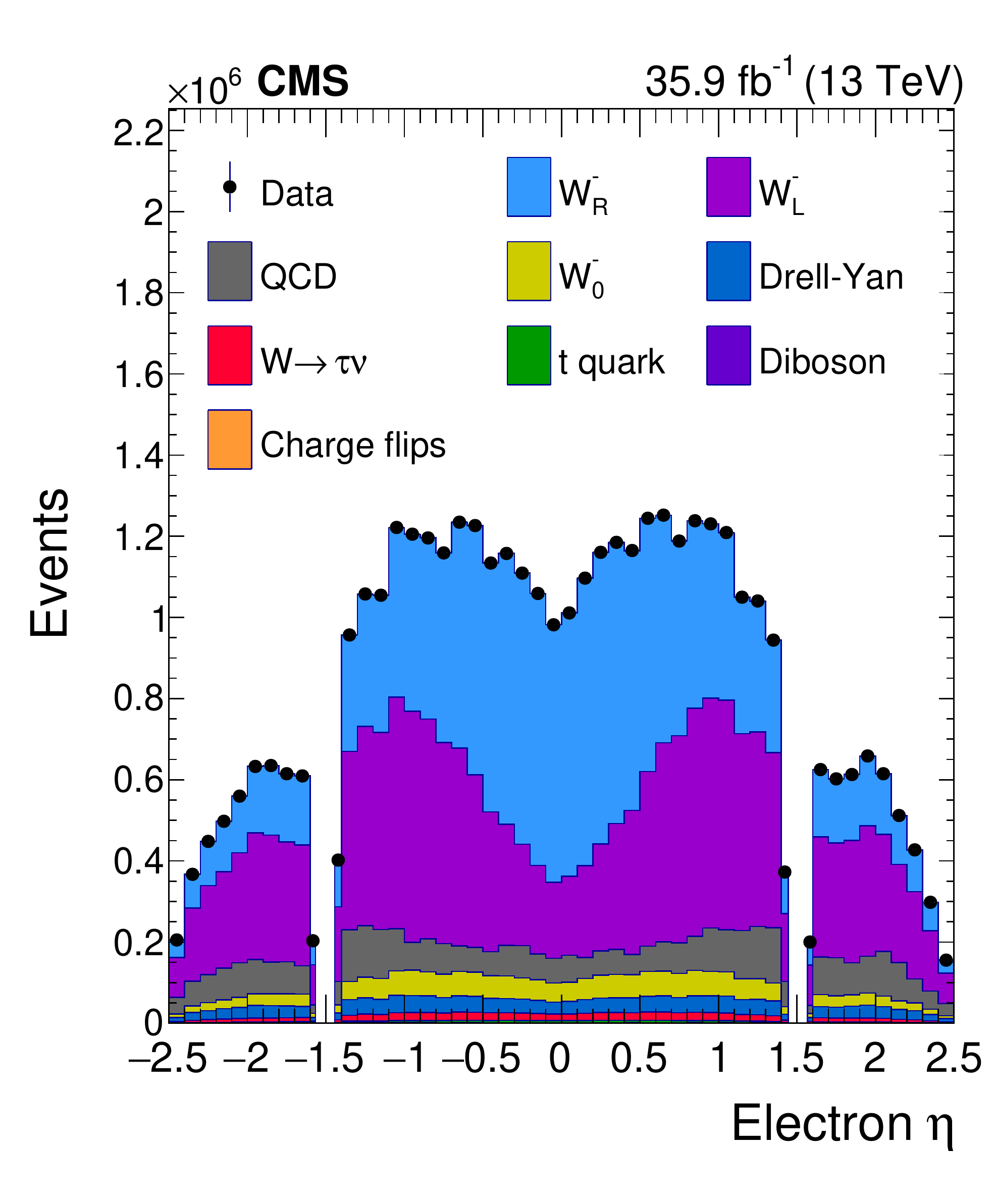}
\includegraphics[width=0.4\linewidth]{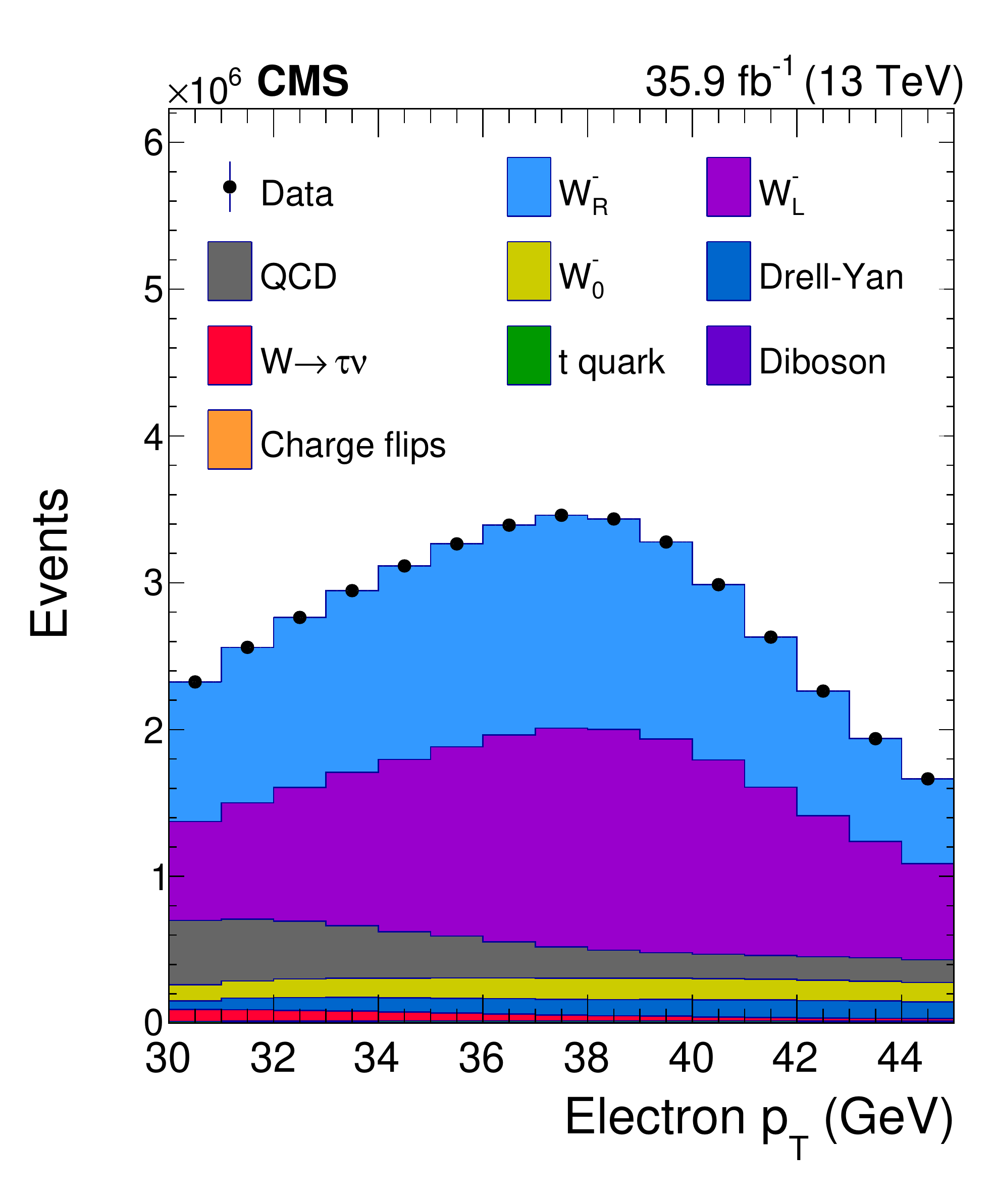} \\
\includegraphics[width=1.0\linewidth]{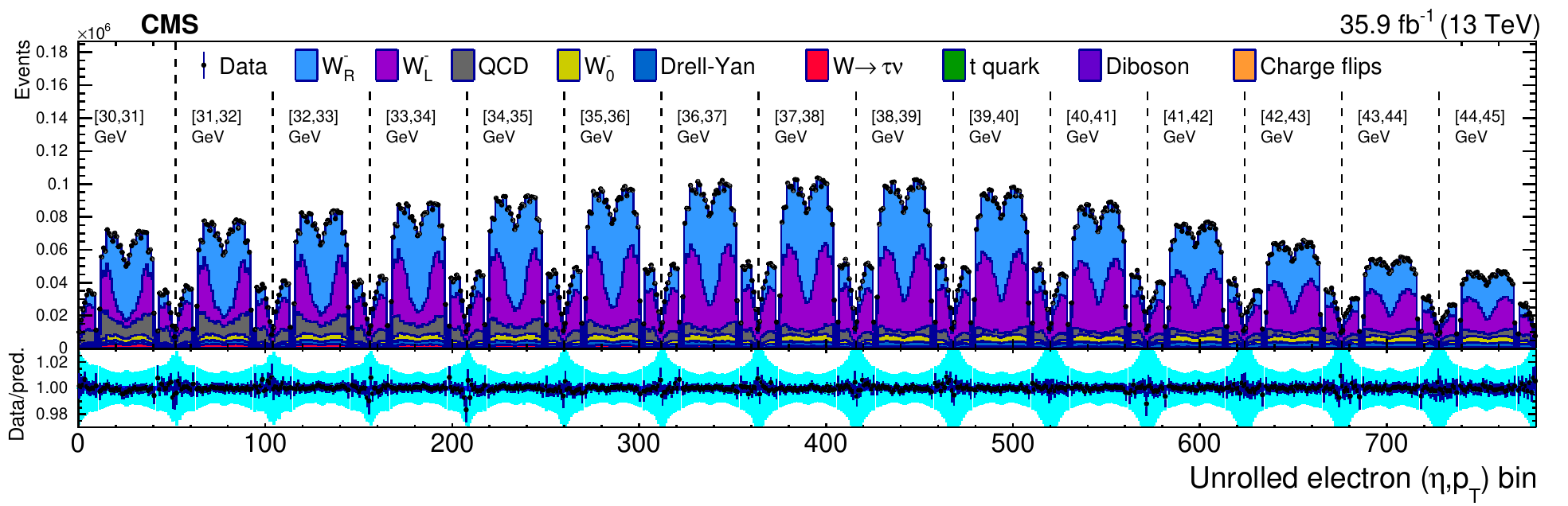}
\caption{Distributions of \etael (upper left), \ptel (upper right), and
  $\text{bin}_\text{unrolled}$ (lower) for $\wminus\to\Pe^-\PAGn$ events
  for observed data superimposed on signal plus background events. The
  signal and background processes are normalized to the result of the
  template fit. The cyan band over the data-to-prediction ratio
  represents the uncertainty in the total yield in each bin after the
  profiling process.
  \label{fig:postfit_elminus}}
\end{figure*}

The template fit to the \etapt distribution is performed on the four
independent channels: \wplusmu, \wminusmu, \wpluse, and \wminuse.
The observed events as a function of lepton $\eta$ and \pt are shown
in Figs.~\ref{fig:postfit_muplus}~(\ref{fig:postfit_muminus}) for the
muon final state and
Figs.~\ref{fig:postfit_elplus}~(\ref{fig:postfit_elminus}) for the
electron final state for the positive (negative) charge.  The upper
distributions in these figures show the 1D projections in \leta and \ptl.
The lower distributions represent the 2D templates ``unrolled'' into one dimension,
such that the integer bin number
$\text{bin}_\text{unrolled}=1+\text{bin}_{\eta}+48(50)\text{bin}_{\pt}$,
with the integers $\text{bin}_\eta\in[0,48(50)]$ and
$\text{bin}_{\pt}\in[0,18(14)]$ for the muon (electron) channel.
In the
projections, the sum in quadrature of the uncertainties in the 2D
distribution is shown, neglecting any correlations. Therefore, these
uncertainties are for illustration purposes only.

\subsection{Cross section measurements}

The $\PW^\pm\to\ell\nu$ cross section measurements are performed in
both the muon and electron channels by using the negative log likelihood
minimization in Eq.~(\ref{eqn:nll}). This provides a cross-check of
experimental consistency of the two decay modes and provides a method
of reducing the impact of the statistical and systematic uncertainties
when combining the measurements in the two channels and accounting for
correlated and uncorrelated uncertainties.

\subsubsection{Combination procedure}

Measurements in different channels are combined by simultaneously
minimizing the likelihood across channels, with common signal
strengths and nuisance parameters as appropriate.  Uncertainties that
are correlated among channels are those corresponding to the
integrated luminosity, the knowledge of specific process cross
sections in the background normalizations when the process is
estimated from simulation, and effects that are common to multiple
processes.  Uncertainties related to the estimate of the QCD
background are considered uncorrelated between muon and electron
channels, since they originate from the closure test of the estimate
in the background-dominated regions, which are independent of each
other. The estimate of the lepton misidentification probability
$\epsilon_\text{pass}$ is also performed independently. The
systematic uncertainty on $\epsilon_\text{pass}$ is~100\%
correlated between the two charges for each lepton flavor.

The statistical uncertainties in the efficiency correction factors are
assumed as uncorrelated among positive and negative charges, and among
the channels, since they are derived from independent samples. The
fully correlated part of the systematic uncertainty in the efficiency
within a channel is assumed uncorrelated between muons and electrons
since the dominant effects from the $\cPZ\to\ell\ell$ line shape and
the background sources are very different.

Most of the theoretical uncertainties are assumed 100\% correlated
among the four channels. They are uncertainties in the boson \pt
spectrum modeling because of \muF and \muR
uncertainties and the uncertainty in the knowledge of
\alpS.  Another large group of nuisance parameters
that are correlated among all the channels represent the effects of
the PDF variations within the \NNPDF set used on both the shape of the
templates used and their normalization. The 60 nuisance parameters
associated with the Hessian representation of the 100 PDF replicas, as
well as the uncertainty in \alpS, are 100\% correlated
among all the four lepton flavor and charge channels.  These 60+1
systematic uncertainties are also fully correlated with the respective
uncertainties considered for the \PZ and $\PW\to\tau\nu$ processes.

\subsubsection{Differential cross sections in \texorpdfstring{$\abs{y_{\PW}}$}{abs[y(W)]} }

The measured \absyw-dependent cross section, for the left- and
right-handed polarizations, is extracted from the fit in 10 bins of
\absyw with a constant width of $\Delta\yw=0.25$ in a
range~$\absyw<2.5$. The cross sections in the two additional
bins,~$2.5<\absyw<2.75$ and $2.75<\absyw<10$, that integrate over the
kinematic region in which the detector acceptance is small, are fixed
to the expectation from
\MGvATNLO with a large 30\% normalization uncertainty.  To
achieve a partial cancellation of uncertainties that are largely
correlated among all \absyw bins, the cross sections are normalized to
the fitted total \PW boson cross section integrating over all the
rapidity bins within the acceptance.  As stated before, the
longitudinally polarized component is fixed to the \MGvATNLO
prediction with a 30\% normalization uncertainty. Therefore, it is not
a freely floating parameter in the fit, and hence only the \wleft and
\wright components are shown in the following.

\begin{figure}[ht!]
\centering
\includegraphics[width=0.49\textwidth]{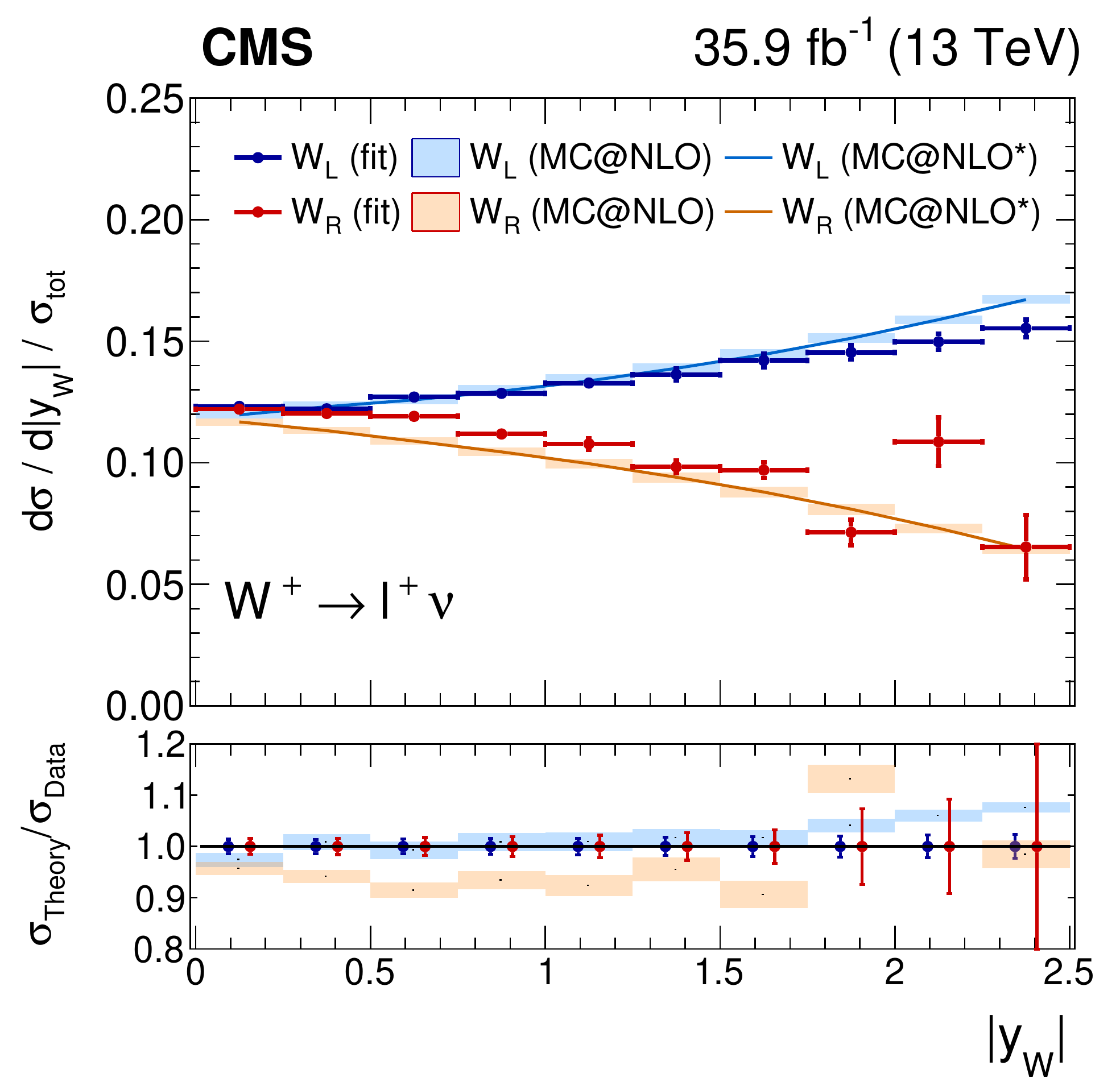}
\includegraphics[width=0.49\textwidth]{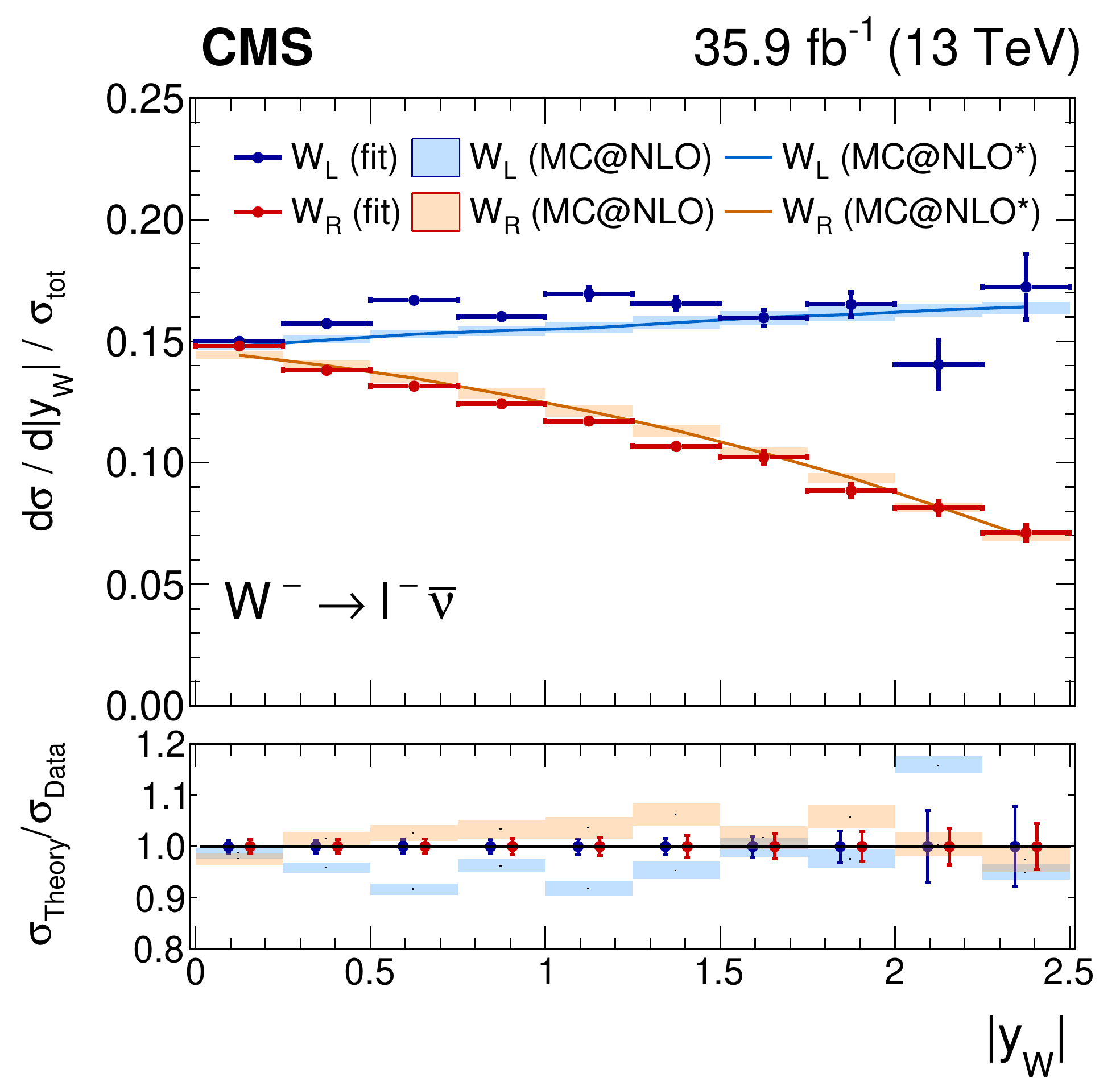}
\caption{Measured normalized $\wplus\to\ell^+\nu$ (\cmsLeft plot) or
  $\wminus\to\ell^-\PAGn$ (\cmsRight plot) cross section as functions of
  \absyw for the left-handed and right-handed helicity states from
  the combination of the muon and electron channels, normalized to the
  total cross section. Also shown is the ratio of the prediction
  from \MGvATNLO to the data. The \MGvATNLO{}$^{*}$ spectrum stands
  for the prediction with the \ptw weighting applied.  The
  lightly filled band corresponds to the expected uncertainty from the
  PDF variations, \muF and \muR scales, and \alpS. \label{fig:rap_pol}
  }
\end{figure}

The measured \PW boson production cross sections, split into the
left- and right-handed helicity states for the combination of the
muon and the electron channels, are presented in
Fig.~\ref{fig:rap_pol}, normalized to the total cross section in the
whole rapidity range. The experimental distributions are compared with
the theoretical prediction from \MGvATNLO. The central value from the
\MGvATNLO prediction, where the \ptw spectrum in simulation is
weighted by the ratio of measured and predicted spectrum for DY
production as described in Sec.~\ref{sec:dataSimulation}, 
is also shown as a line within the error bands and denoted as
\MGvATNLO{}$^{*}$. It is evident that this weighting has a small impact
on the rapidity spectrum, and the alternative expected distributions
are well within the other theoretical uncertainties.  The uncertainty
shown in the theoretical prediction includes the contribution from the
PDFs (\NNPDF set), the envelope of the \muF and
$\mu_\mathrm{R}$ variations, and the \alpS.

\begin{figure}[ht!]
\centering
\includegraphics[width=0.49\textwidth]{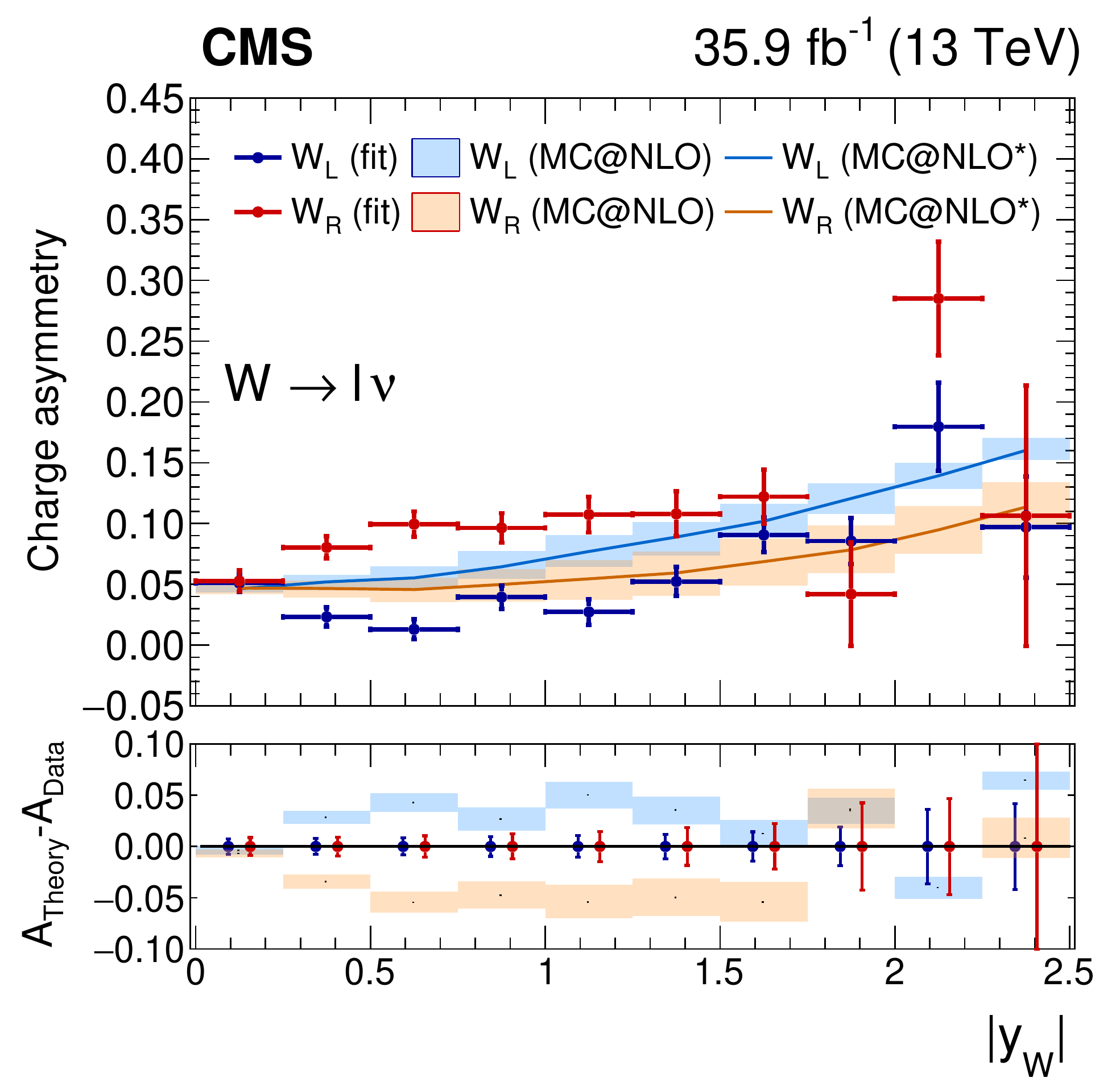}
\caption{Measured \PW boson charge asymmetry as functions of
  \absyw for the left-handed and right-handed helicity states from
  the combination of the muon and electron channels.  Also shown is the
  ratio of the prediction from \MGvATNLO to the
  data. The \MGvATNLO{}$^{*}$ spectrum stands for the prediction with
  the \ptw weighting applied.  The lightly filled band corresponds to
  the expected uncertainty from the PDF variations, \muF and \muR
  scales, and \alpS. \label{fig:rap_pol_asy} }
\end{figure}

The main systematic uncertainty in the signal cross section, the 2.5\%
uncertainty in the integrated luminosity~\cite{CMS:2017sdi}, is fully
correlated across all the rapidity bins, thus it cancels out when
taking the ratio to the total \PW cross section.  The ratio of the
expected normalized cross section using the nominal \MGvATNLO
simulation to the measured one in data is also presented.  As
described in Sec.~\ref{sub:asymmetry}, the fitted \absyw-dependent
cross sections are used to simultaneously derive the differential
charge asymmetry. This is presented in Fig.~\ref{fig:rap_pol_asy}, differentially in \absyw and
polarization.

There are significant correlated uncertainties between neighboring
\PW boson rapidity bins.  The correlations arising only from the
overlap of the signal templates in the \etapt~plane, \ie, of a purely
statistical nature, are in the range 50--80\% for adjacent \PW boson
rapidity bins ($\Delta\absyw=1$), raising with \absyw, about 20\%
for $\Delta\absyw=2$, about 10\% or less for $\Delta\absyw=3$, and
negligible otherwise. An overall correlation sums up to these
statistical correlations, originating from systematic uncertainties
common to all the signal processes, such as the uncertainty in the
integrated luminosity.

The cross section results differential in \PW boson rapidity are
tested for statistical compatibility with a smooth functional shape,
taking these correlations into account.  Monte Carlo
pseudoexperiments show that the results are quantitatively consistent
with smooth third-order polynomial functions of
\absyw. This test is performed simultaneously in both helicity
states, both charges, and all \absyw bins, taking into account the
full covariance matrix of the fit.

\begin{figure}[ht!]
\centering
\includegraphics[width=0.49\textwidth]{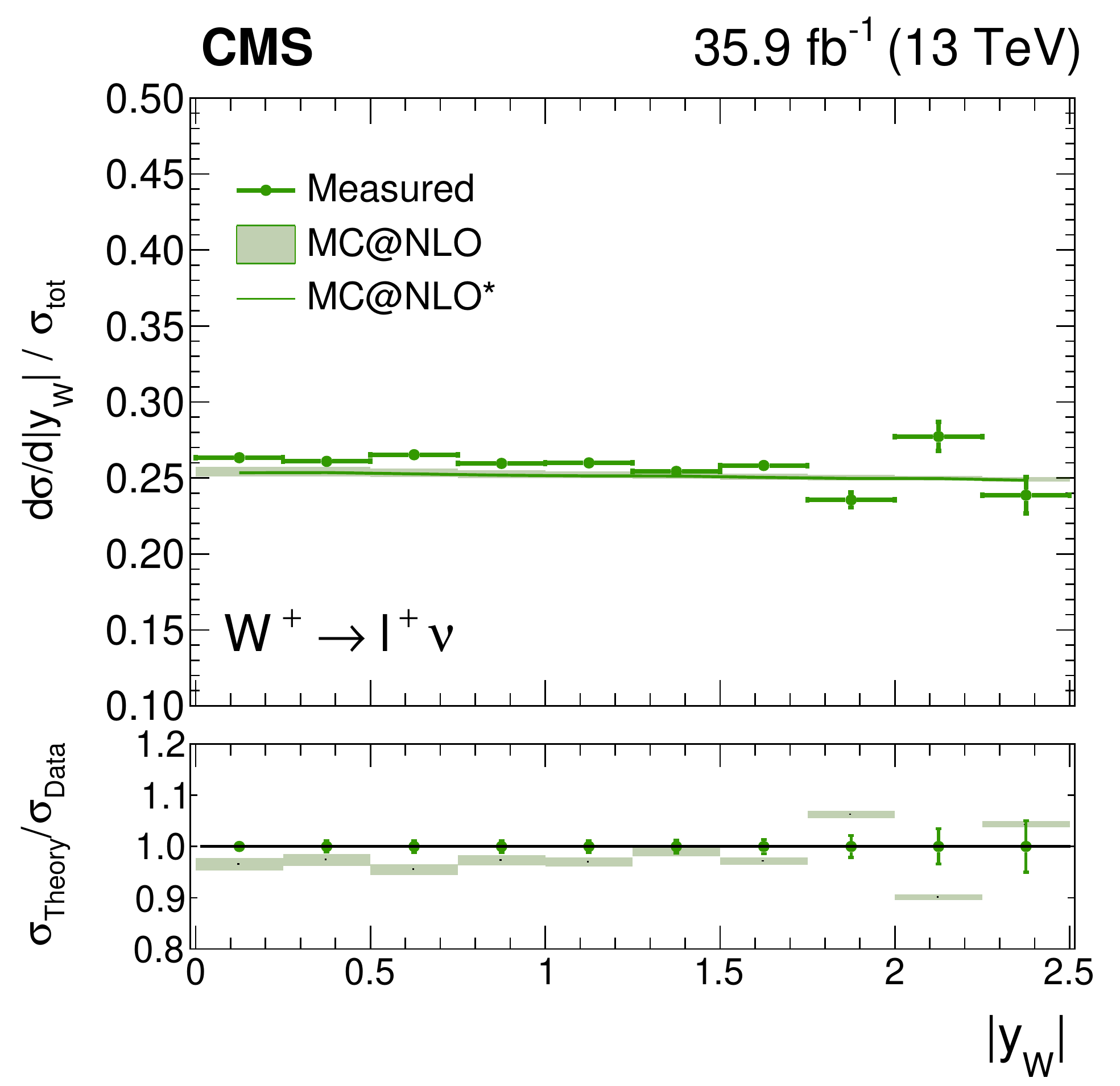}
\includegraphics[width=0.49\textwidth]{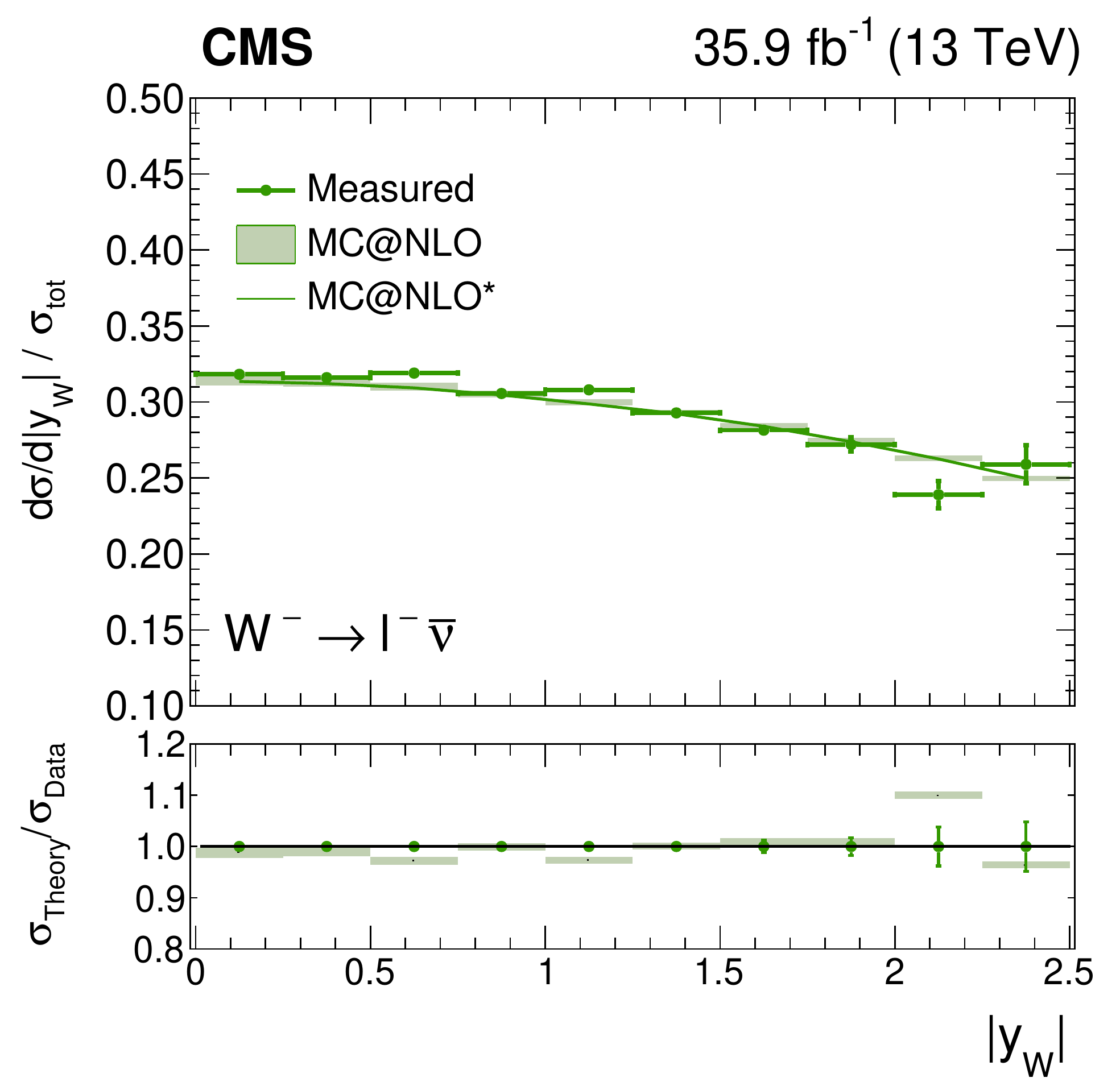}
\caption{Measured normalized $\wplus\to\ell^+\nu$ (\cmsLeft plot) and
  $\wminus\to\ell^-\PAGn$ (\cmsRight plot) cross sections as a function of
  \absyw from the combination of the muon and electron channels,
  normalized to the total cross section, and integrated over the \PW
  polarization states.  Also shown is the ratio of the prediction
  from \MGvATNLO to the data.  The \MGvATNLO{}$^{*}$ spectrum stands
  for the prediction with the \ptw weighting applied. The
  lightly filled band corresponds to the expected uncertainty from the
  PDF variations, \muF and \muR scales,
  and \alpS. \label{fig:rap_unpol}}
\end{figure}

\begin{figure}[ht!]
\centering
\includegraphics[width=0.49\textwidth]{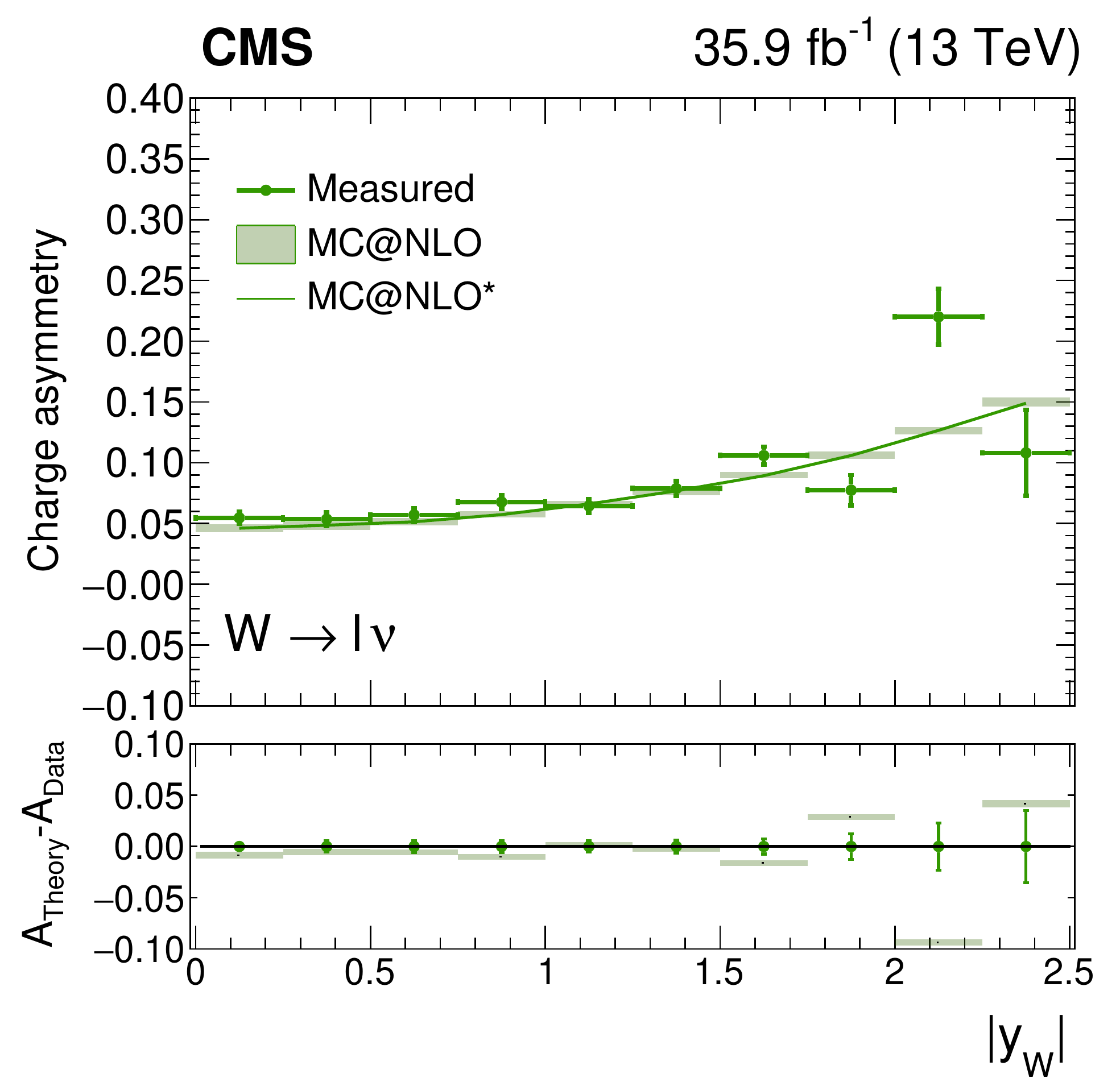}
\caption{Measured \PW charge asymmetry as a function of
  \absyw from the combination of the muon and electron channels,
  integrated over the \PW polarization states. Also shown is the ratio
  of the prediction from \MGvATNLO to the data.  The \MGvATNLO{}$^{*}$
  spectrum stands for the prediction with the \ptw weighting
  applied. The lightly filled band corresponds to the expected
  uncertainty from the PDF variations, \muF and \muR scales,
  and \alpS. \label{fig:rap_unpol_asy}}
\end{figure}

Results are also shown as an unpolarized normalized cross section,
\ie, by summing over all helicity states as a function of \absyw in
Fig.~\ref{fig:rap_unpol}.  The unpolarized charge asymmetry as a
function of \absyw is shown in Fig.~\ref{fig:rap_unpol_asy}.

\begin{figure}[ht!]
\centering
\includegraphics[width=0.49\textwidth]{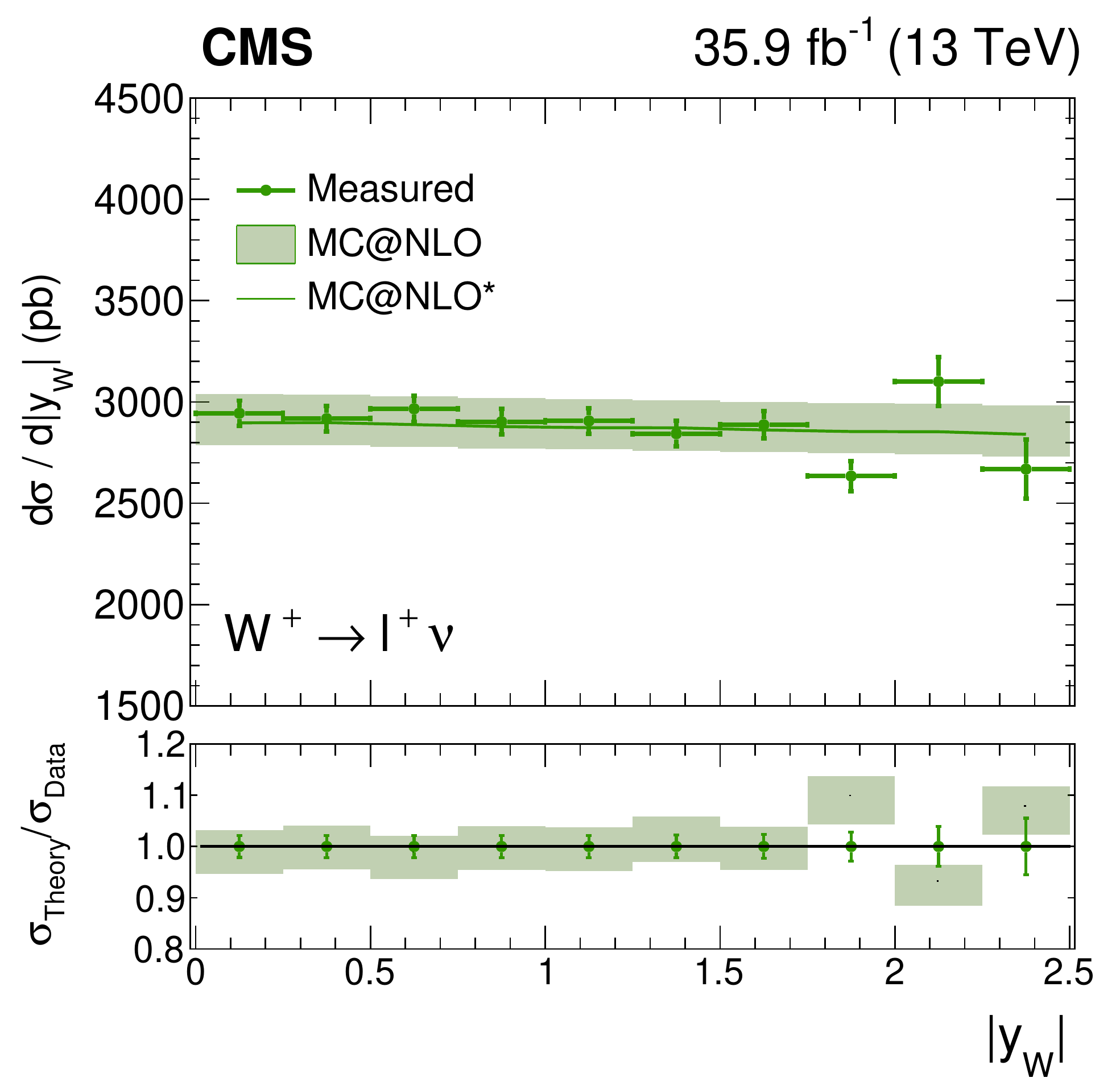}
\includegraphics[width=0.49\textwidth]{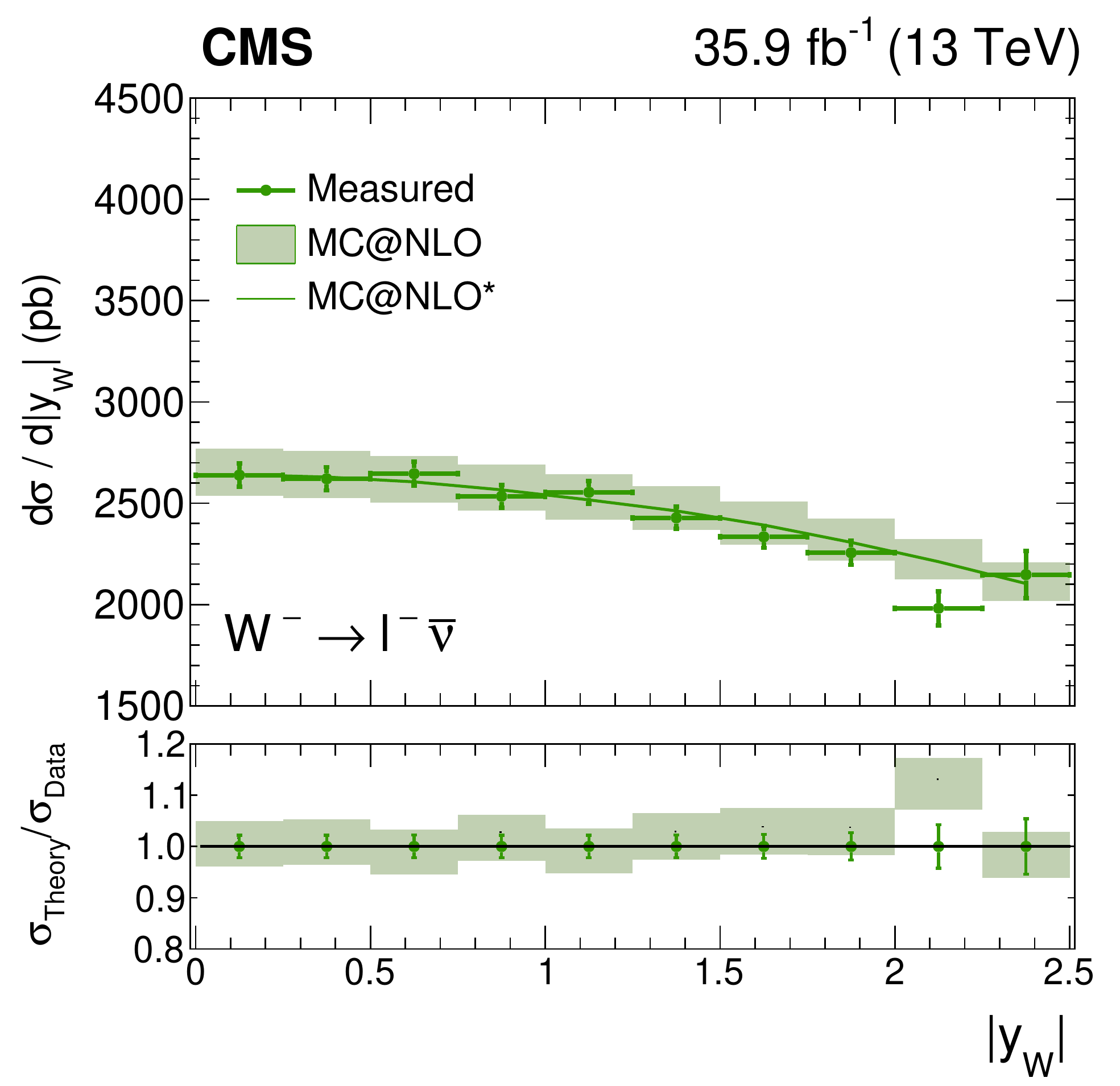}
\caption{Measured absolute $\wplus\to\ell^+\nu$ (\cmsLeft plot) or
  $\wminus\to\ell^-\PAGn$ (\cmsRight plot) cross sections as functions of
  \absyw from the combined flavor fit. The ratio of the prediction
  from \MGvATNLO to the data is also shown. The \MGvATNLO{}$^{*}$
  spectrum stands for the prediction with the \ptw weighting applied.
  The lightly filled band corresponds to the expected uncertainty from
  the PDF variations, \muF and \muR scales,
  and \alpS. \label{fig:rap_unpol_abs}}
\end{figure}

In addition to these normalized and unpolarized cross sections, the
results of the fits are also presented as absolute cross sections in
Fig.~\ref{fig:rap_unpol_abs}, where the absolute unpolarized cross
sections are shown for the combined flavor fit. Generally, good
agreement is observed in the shape of the measured distribution with
respect to the expectation, albeit with an offset of the order of a
few percent.

After the fit with floating cross sections is performed, only a few
nuisance parameters are significantly constrained. Mainly the nuisance
parameters related to the normalization of the nonprompt-lepton
background and its shape in \leta and \ptl are constrained by the
fit. Because of the large data sample, this effect is expected.

\subsubsection{Double-differential cross sections in \texorpdfstring{\ptl}{} and \texorpdfstring{\absleta}{}}

\begin{figure*}[t]
\centering
    \includegraphics[width=0.97\linewidth]{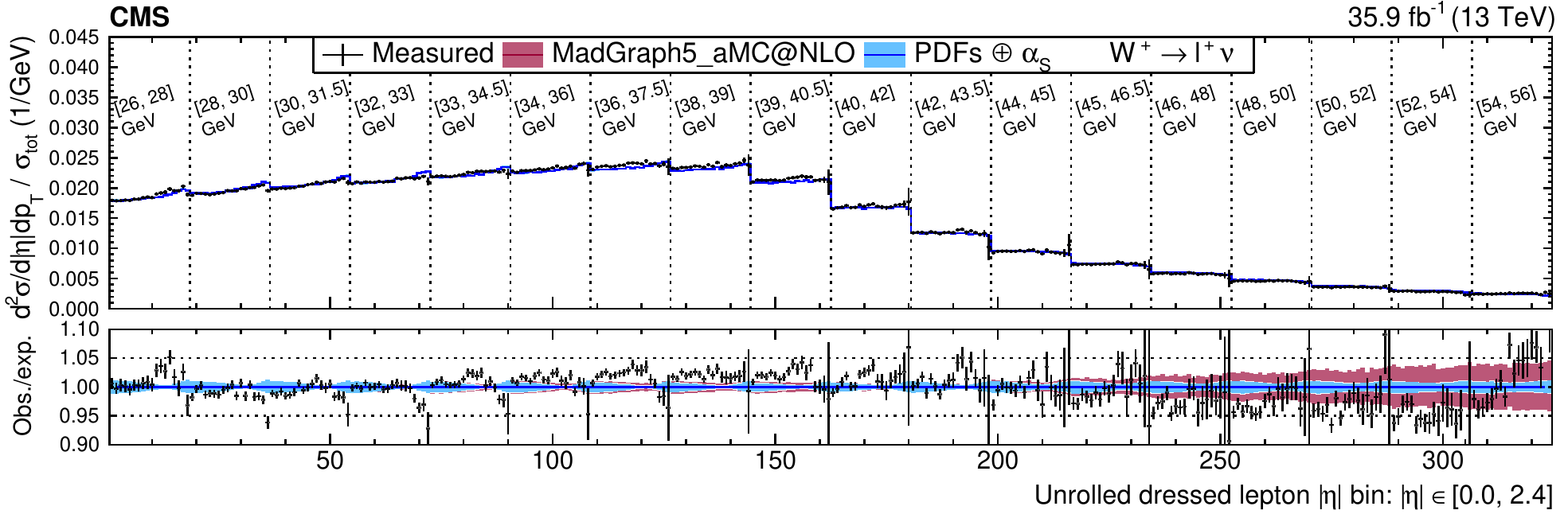} \\
    \includegraphics[width=0.97\linewidth]{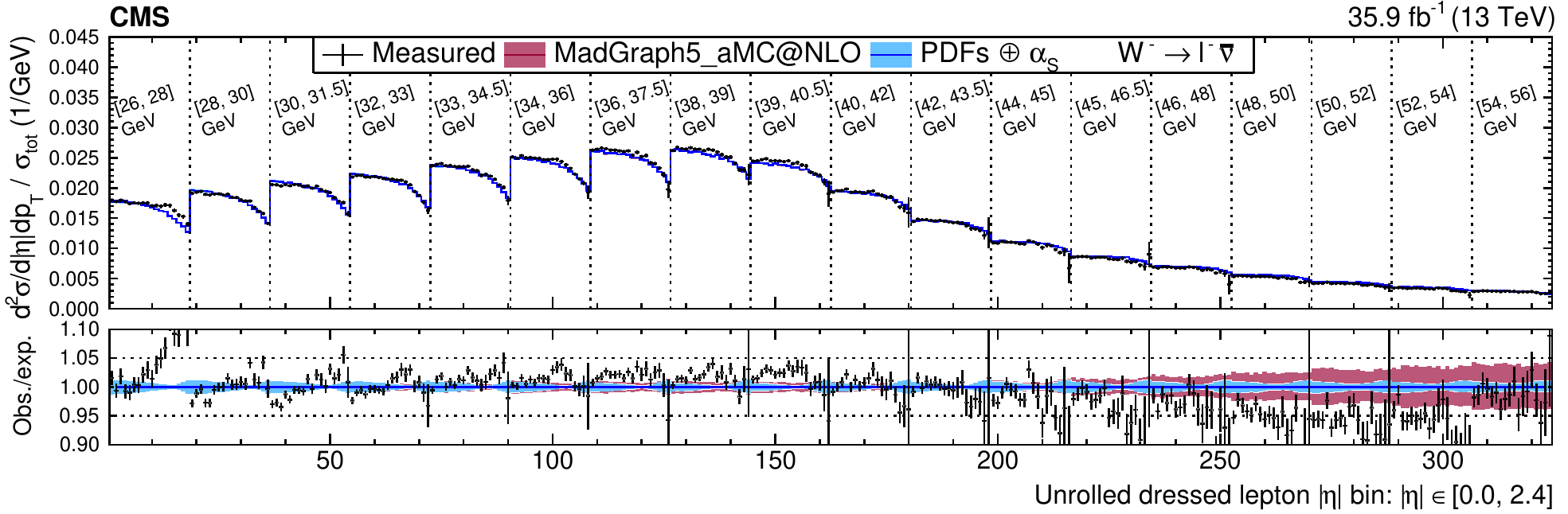}
    \caption{Normalized double-differential cross section 
      as a function of \ptl and \absleta, unrolled in a 1D histogram
      along \absleta~for the positive (negative) charge on the upper
      (lower) plot. The lower panel in each plot shows the ratio
      of the observed and expected cross sections. The colored bands represent the prediction
      from \MGvATNLO with the expected uncertainty from the quadrature
      sum of the PDF$\oplus\alpS$ variations (blue) and the \muF
      and \muR scales (bordeaux).\label{fig:unfoldedDiffCrossSections}}
\end{figure*}

\begin{figure*}[t]
\centering
    \includegraphics[width=0.97\linewidth]{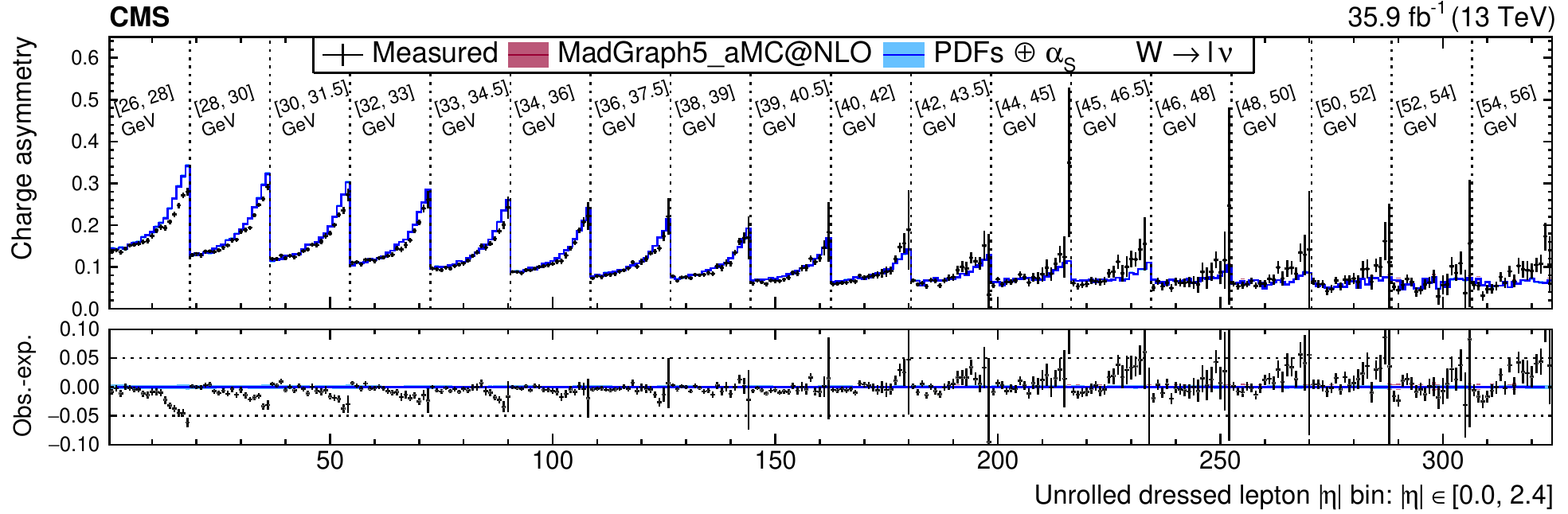}
    \caption{Double-differential \PW boson charge asymmetry as a
      function of \ptl and \absleta, unrolled in a 1D histogram
      along \absleta. The lower panel shows the difference of the
      observed and expected charge asymmetry. The colored bands
      represent the prediction from \MGvATNLO with the expected
      uncertainty from the quadrature sum of the PDF$\oplus\alpS$
      variations (blue) and the \muF and \muR scales
      (bordeaux).\label{fig:unfoldedDiffAsy}}
\end{figure*}

Double-differential cross sections in \ptl and \absleta~are measured
from a fit to the observed data in the \etapt plane. The underlying
generated templates are unfolded to the dressed lepton definition in
18 bins of \ptl and 18 bins of \absleta, as described in
Sec.~\ref{sub:fitDiffXsec}.  These cross sections are shown in
Fig.~\ref{fig:unfoldedDiffCrossSections}, normalized to the total
cross section. These results come from the combination of the muon and
electron final states, divided into two categories of the lepton
charge. From the measured cross sections, the double-differential
charge asymmetry is computed, where the uncertainty is computed from
the full covariance matrix from the fit, and it is shown in
Fig.~\ref{fig:unfoldedDiffAsy}.

The agreement of the measured normalized \PW boson cross sections
and charge asymmetry with the prediction of \MGvATNLO is at the level
of 1\% in the central part of the lepton acceptance ($\absleta<1$).  In
the outer end cap sections of the detector, especially for lower
\ptl, the agreement with the prediction becomes worse.

\begin{figure*}[t]
\centering
    \includegraphics[width=0.97\linewidth]{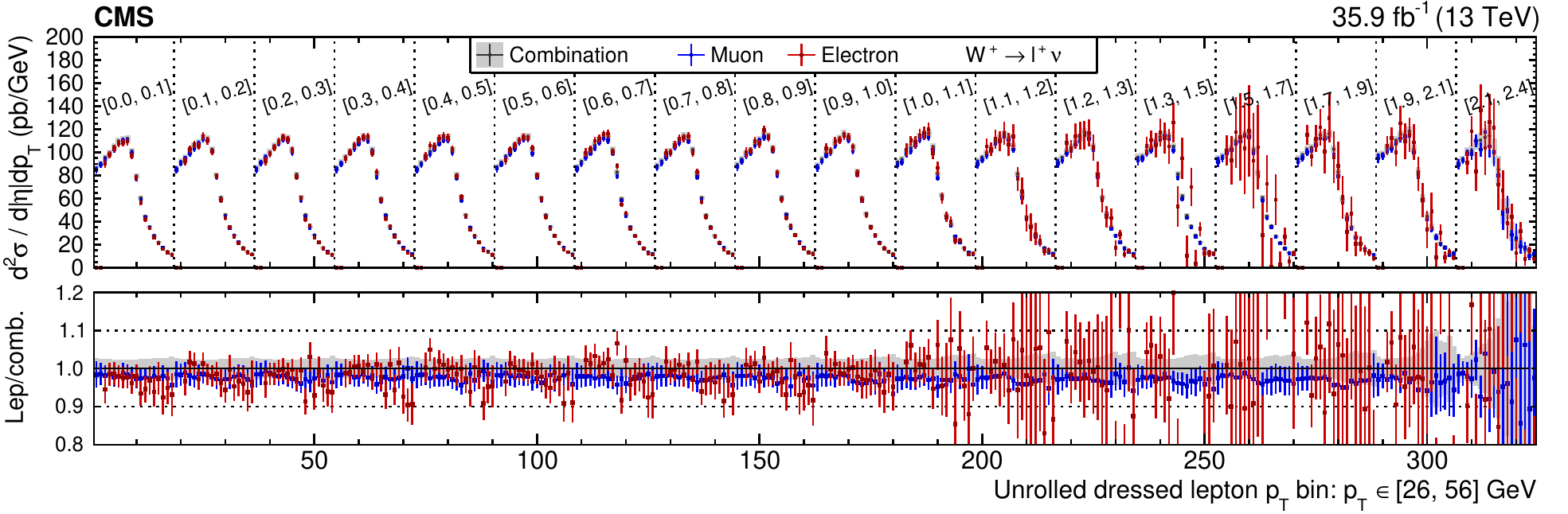} \\
    \includegraphics[width=0.97\linewidth]{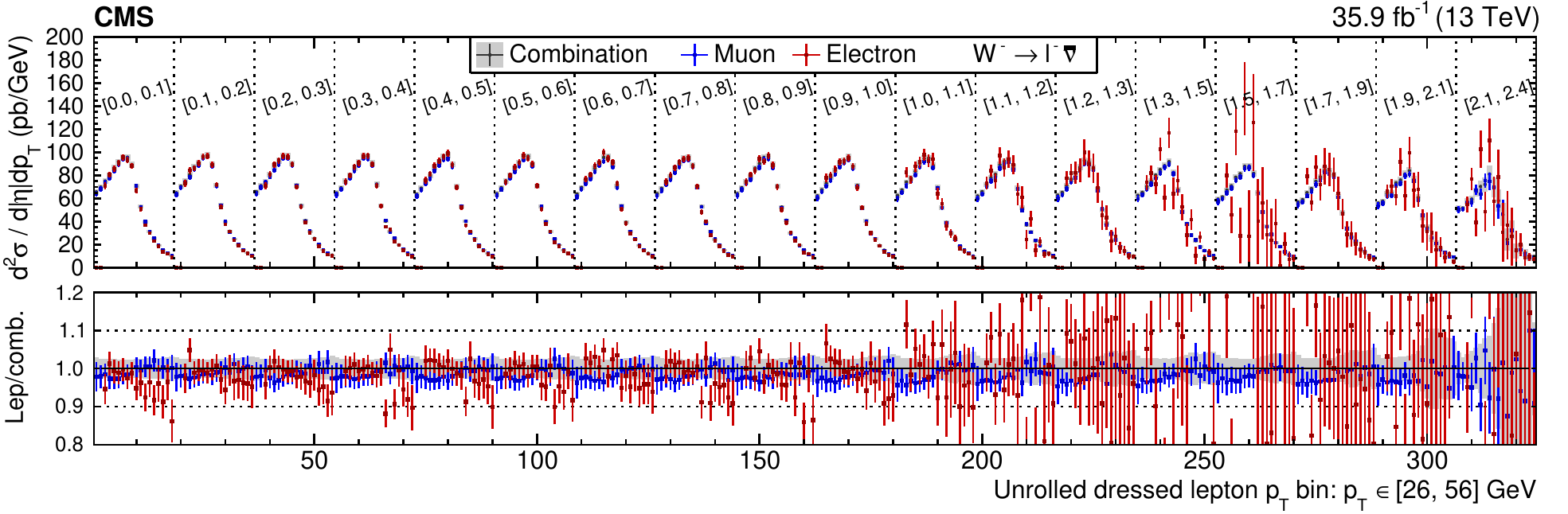} \\
    \caption{Absolute double-differential cross section as a function
      of \ptl and \absleta, unrolled in a 1D histogram along \ptl in bins
      of \absleta~for the positive (negative) charge on upper (lower)
      panel. The combined muon and electron fit is shown in green
      markers, the muon-only fit in blue markers, and the
      electron-only fit in red markers. The error bars correspond to
      the total uncertainty from the respective fits. The filled gray
      band in the lower panel represents the total uncertainty from
      the combined fit.\label{fig:unfoldedDiffFlavor}}
\end{figure*}

Although these normalized cross sections of the combined flavor fit
represent the result with the smallest total uncertainty because of
the cancellation of the fully correlated components, the absolute
cross sections are also of interest. In particular, the agreement of
the absolute cross sections measured in each flavor channel separately highlights the
understanding of the experimental systematic uncertainties, which are 
largely uncorrelated between the two flavors. These plots are
displayed in Fig.~\ref{fig:unfoldedDiffFlavor}, where the measured
absolute cross sections are shown separately for the muon, electron,
and combined fits. Good agreement is found within the uncertainties in
the regions with sufficient event count. Uncertainties become large in
the high-\absleta~region for the electron-only fit, rendering a precise
comparison difficult.

\begin{figure}
\centering
\includegraphics[width=0.49\textwidth]{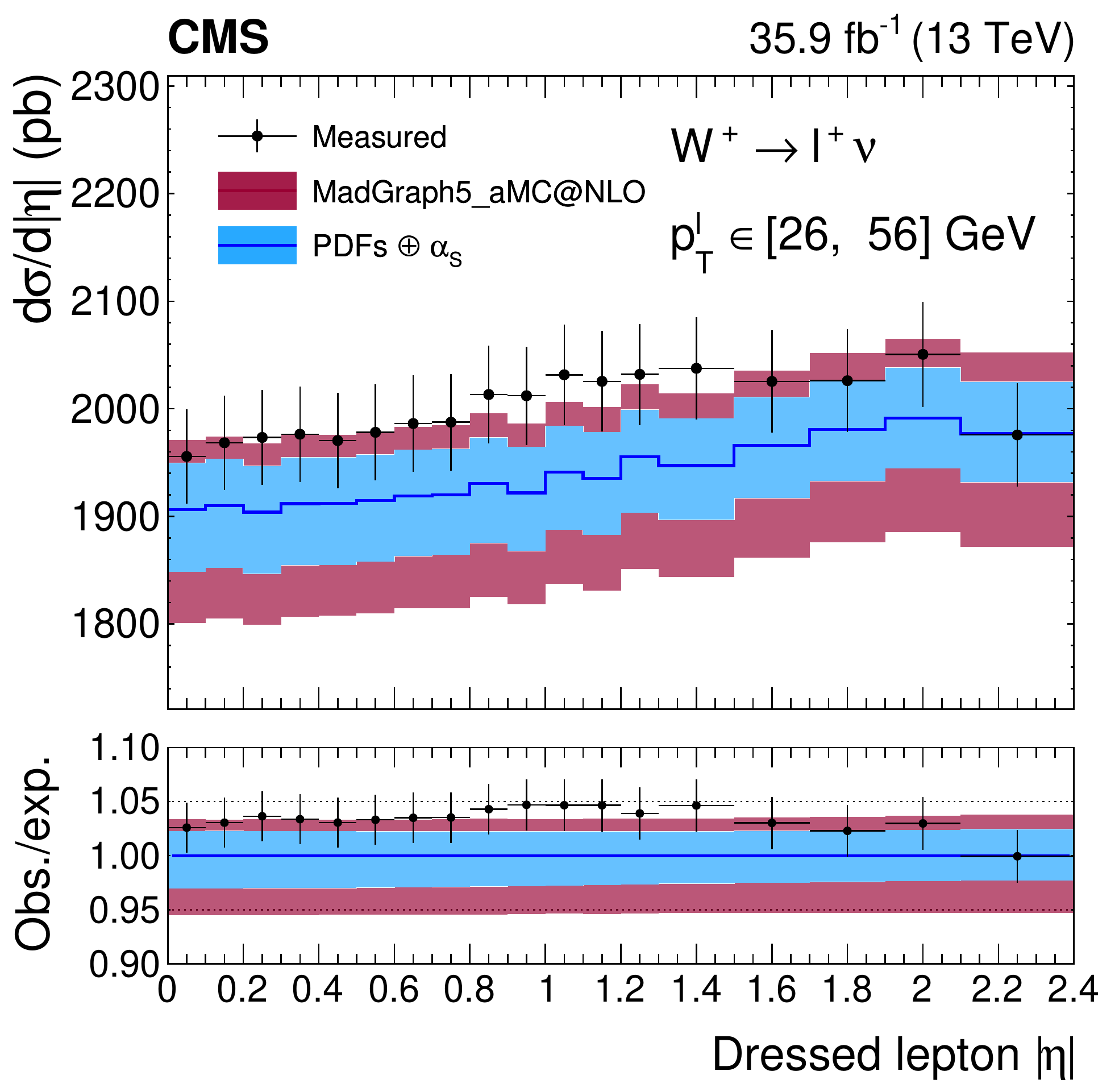}
\includegraphics[width=0.49\textwidth]{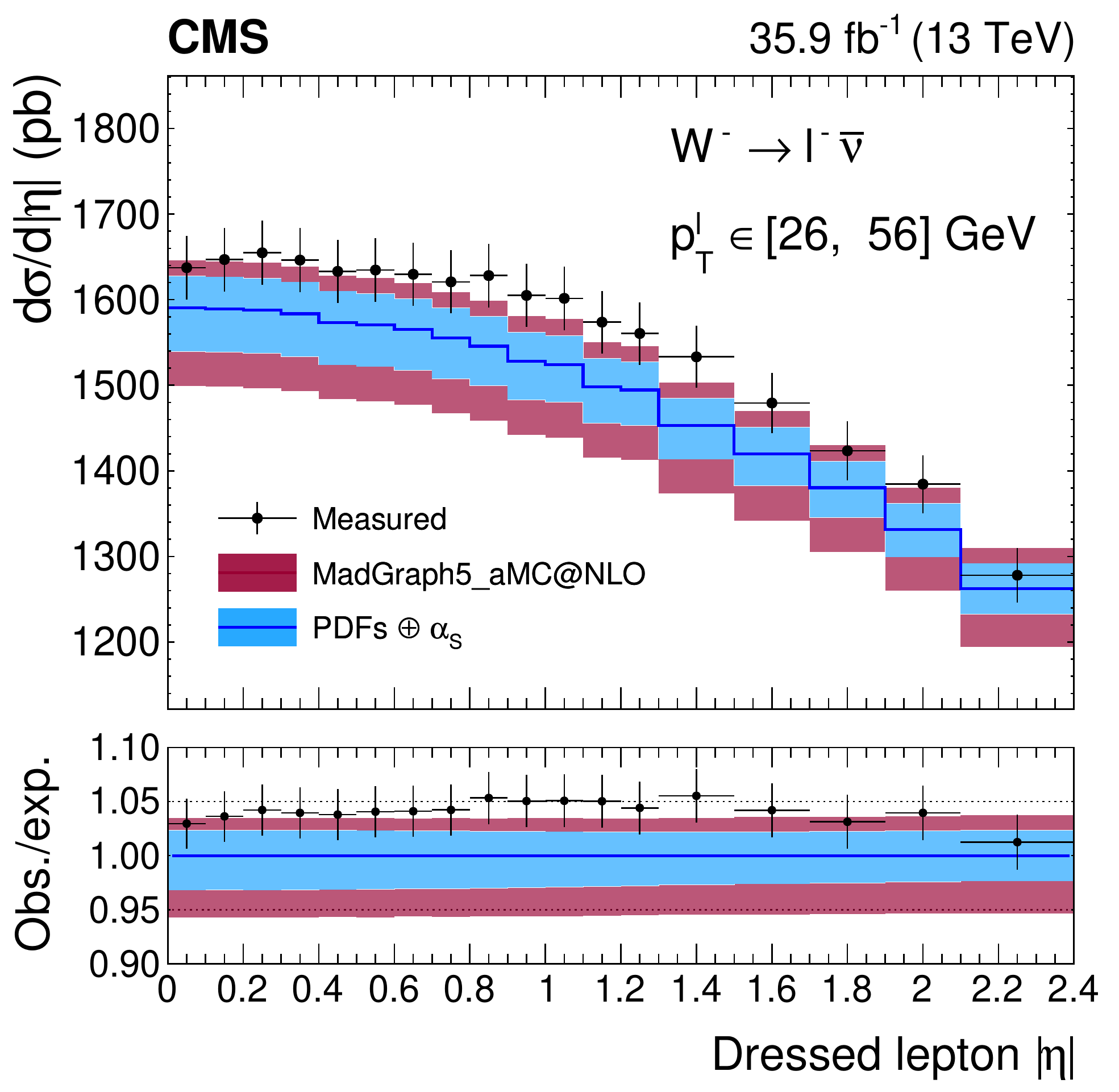}
\caption{Absolute differential cross section as a function of \absleta for the
  \wplusell (\cmsLeft) and \wminusell channel (\cmsRight). The
  measurement is the result of the combination of the muon and
  electron channels. The lower panel in each plot shows the ratio
  of observed and expected cross sections. The
  colored bands represent the prediction from \MGvATNLO with the
  expected uncertainty from the quadrature sum of the PDF$\oplus\alpS$
  variations (blue) and the \muF and \muR scales
  (bordeaux).\label{fig:unfoldedDiffCrossSectionsEta}}
\end{figure}

\begin{figure}
\centering
\includegraphics[width=0.49\textwidth]{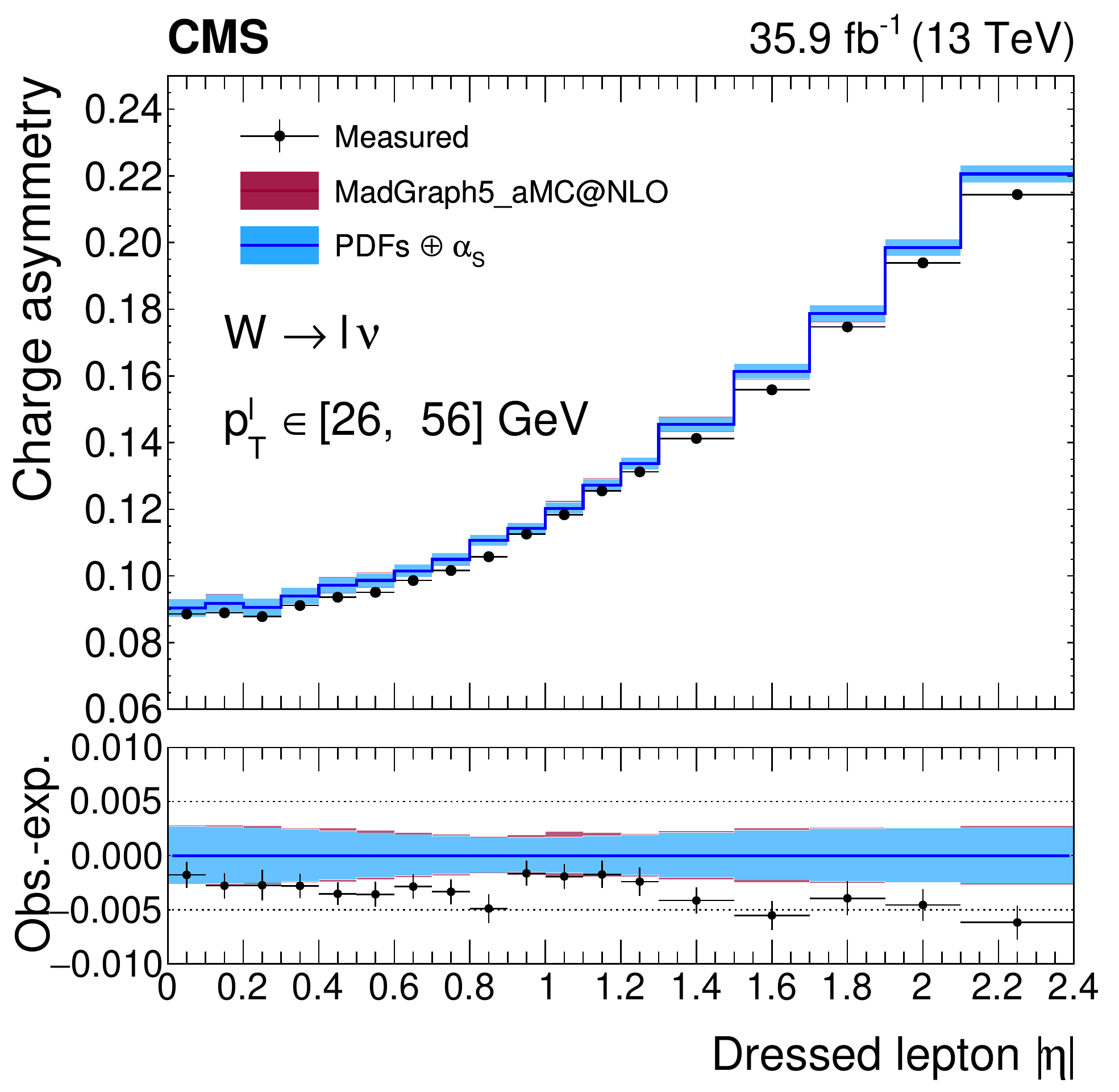}
\caption{Absolute differential \PW boson charge asymmetry as a function of \absleta.  The
  measurement is the result of the combination of the muon and
  electron channels. The lower panel shows the difference of observed
  and expected charge asymmetry. The colored bands represent the
  prediction from \MGvATNLO with the expected uncertainty from the
  quadrature sum of the PDF$\oplus\alpS$ variations (blue) and
  the \muF and \muR scales
  (bordeaux).\label{fig:unfoldedDiffAsyEta}}
\end{figure}

From the results of this fit, the single-differential cross section is
measured by integrating in one of the two dimensions, as a function of
the other variable.  Along with these cross sections, the charge
asymmetry differential in one dimension is extracted. This approach
has the added value, with respect to a single-differential
measurement, that it is independent of the modeling of the lepton
kinematics in the variable that is integrated over.  The resulting
absolute cross sections for the combination of the two lepton flavors
is shown as a function of \leta for both \wplus and \wminus in
Fig.~\ref{fig:unfoldedDiffCrossSectionsEta}. The corresponding \PW
charge asymmetry is shown in Fig.~\ref{fig:unfoldedDiffAsyEta}. This
result can be directly compared with previous measurements of the \PW
boson differential cross section and charge asymmetry as functions of
\leta performed at 7 and 8\TeV by the CMS and ATLAS
Collaborations~\cite{Chatrchyan:2011jz,Aaboud:2016btc}.

\begin{figure}
\centering
\includegraphics[width=0.97\linewidth]{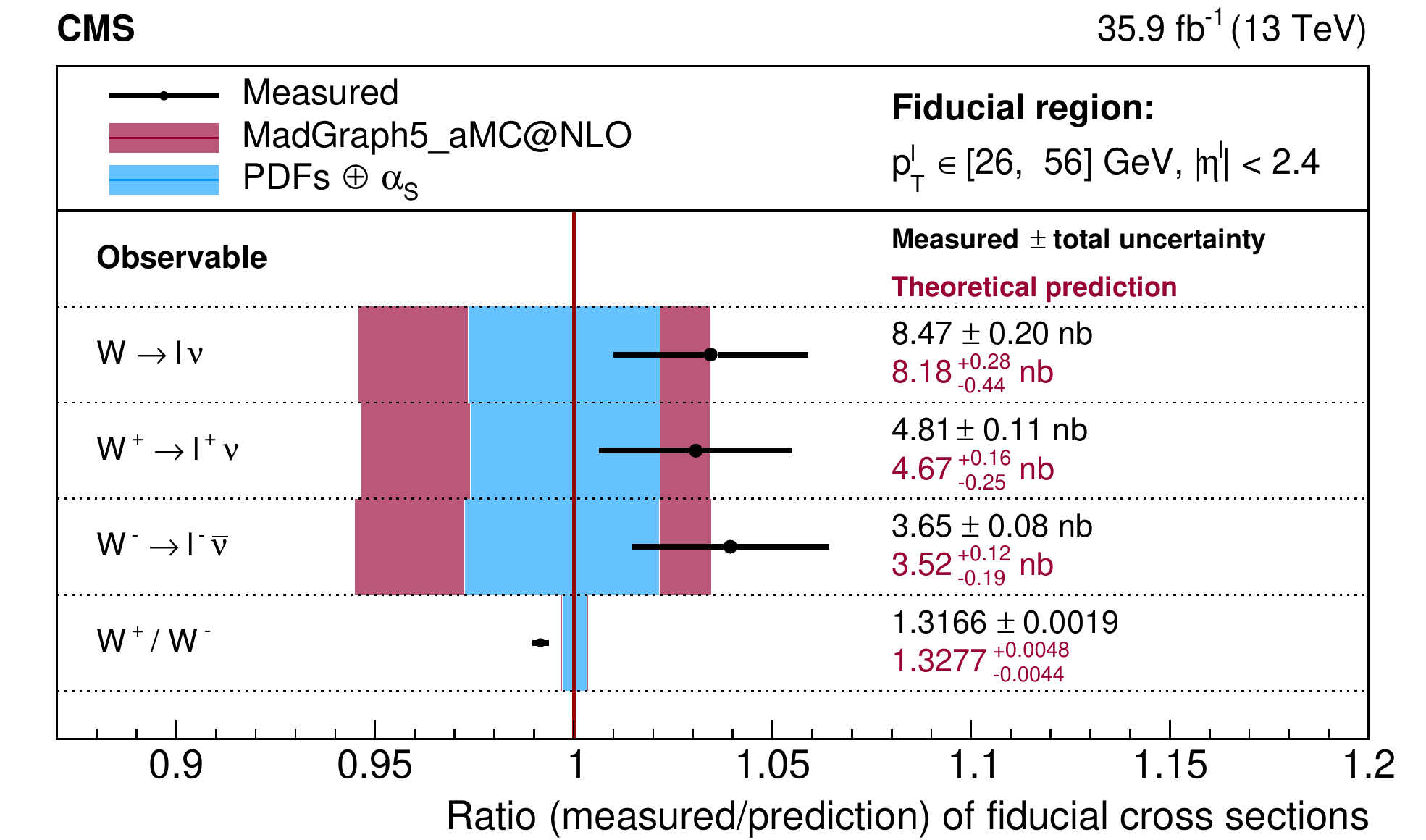} \\
\caption{Ratio of the measured over predicted absolute inclusive cross
  section in the fiducial region $26<\ptl<56\GeV$ and $\absleta<2.5$,
  charge-integrated, charge-dependent, and the ratio for \wplus and
  \wminus.  The measurement is the result of the combination of the
  muon and electron channels. The colored bands represent the
  prediction from \MGvATNLO with the expected uncertainty from the
  quadrature sum of the PDF$\oplus\alpS$ variations
  (blue) and the \muF and \muR scales
  (bordeaux).\label{fig:inclusive_xsec}}
\end{figure}

As a further summary of this fit, the total \PW boson production
cross section, integrated over the fiducial region, $26<\ptl<56\GeV$
and $\absleta<2.4$, is measured.  The fiducial charge-integrated
cross section is $8.47\pm0.10\unit{nb}$, which agrees well with the
NLO prediction. The values for each charge, and their ratio to the
theoretical prediction, are also shown in
Fig.~\ref{fig:inclusive_xsec}, as well as the ratio of the two charges
to the prediction from \MGvATNLO.

\subsection{Constraining the PDF nuisances through likelihood profiling}

When the cross section parameters in the likelihood function of
Eq.~(\ref{eqn:nll}) are fixed to their expected values ($\mu_p = 1$)
within their uncertainties, the fit has the statistical power to
constrain the PDF nuisance parameters.  This procedure corresponds to
the PDF profiling method described in Ref.~\cite{Aaboud:2016btc}, with
associated caveats about the interpretation of constraints far from
the initial predictions.  The constraints in this case are derived
directly from the detector-level measurements rather than passing
through an intermediate step of unfolded cross sections.

The input PDF and MC predictions are both accurate to NLO in QCD, with
the MC prediction implicitly including resummation corrections through
the parton shower.  The theoretical uncertainties included in this
procedure for missing higher orders in QCD correspond to the full
model used for the measurement as described in
Sec.~\ref{sec:theounc}.  This is in contrast to typical global PDF
fits or QCD analyses that are performed at NNLO accuracy, though at
fixed order without resummation, and with the inclusion of missing
higher order uncertainties only in dedicated studies at NLO so far
\cite{AbdulKhalek:2019bux,AbdulKhalek:2019ihb}.

For each variation, the fit input value (prefit) is trivially
represented by a parameter with mean zero and width one. The expected
postfit values of these parameters all have mean zero, but a reduced
uncertainty after the likelihood profiling procedure, \ie, width
smaller than unity. Finally, the points representing the observed
postfit values of the parameters may have a mean different from zero,
indicating a pull of the associated systematic uncertainty, and a width
smaller than 1.

\begin{figure*}[t]
\centering
\includegraphics[width=0.97\textwidth]{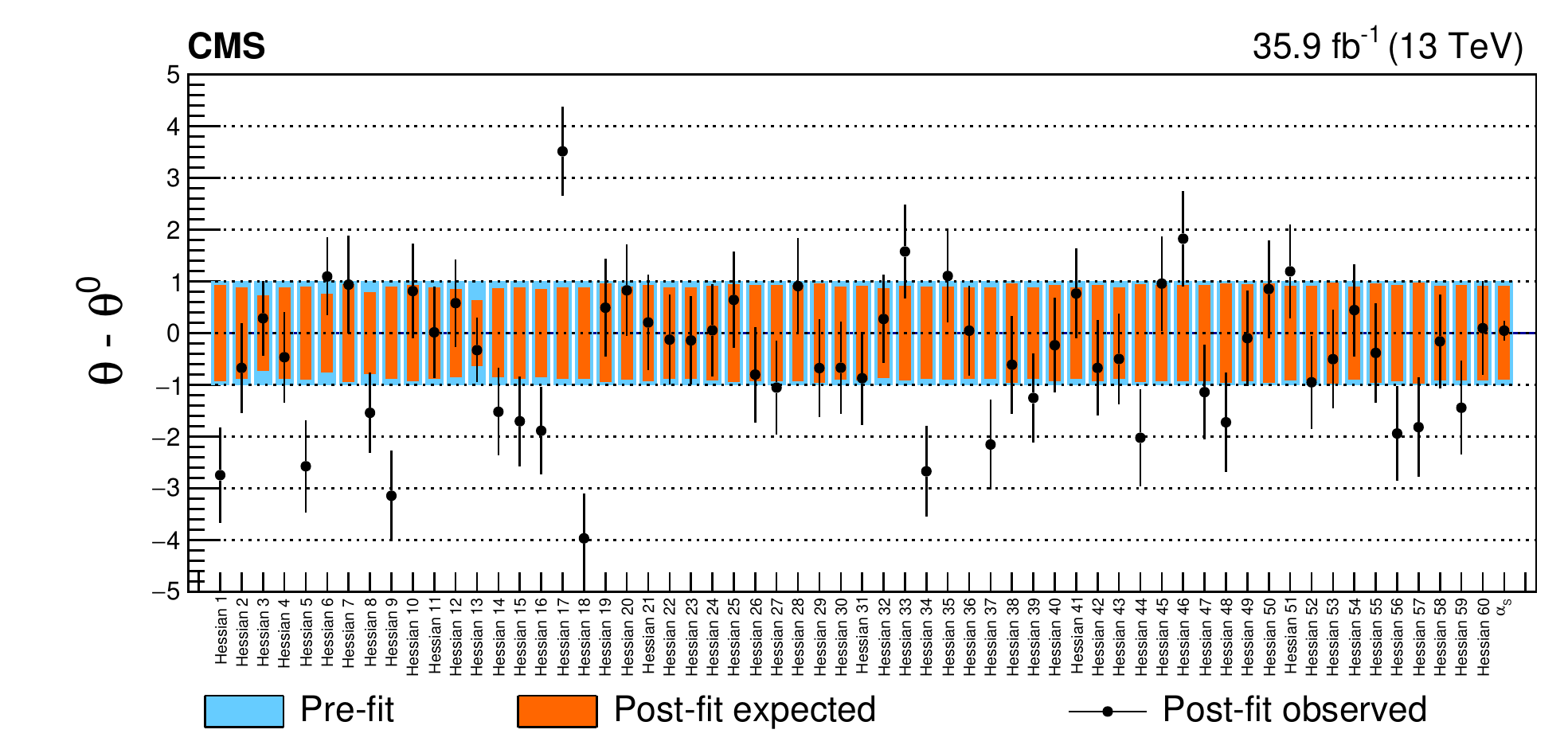} \\
\caption{Pulls and constraints of the 60 Hessian variations of the
  \NNPDF PDF set, and of the \alpS parameter, from the combined
  fit of muon and electron channels. The underlying fit is performed
  by fixing the \PW boson cross sections to their expectation in all
  helicity and charge processes. The cyan band represents the input
  values (which all have zero mean and width one), the orange bands
  show the postfit expected values, and black points represent the
  observed pulls and constraint values.
\label{fig:pdfConstraints}}
\end{figure*}

Such a result can be obtained in both the helicity and the
double-differential cross section fits, and they indeed provide a
consistent set of PDF nuisance parameter values.  The ones reported in
this section, shown in Fig.~\ref{fig:pdfConstraints}, come from the
former fit. These parameters correspond to the 60 orthogonalized
Hessian PDF variations corresponding to the \NNPDF replicas, plus
\alpS. All of the variants, \ie, prefit, postfit expected, and
postfit observed, are shown. Postfit constraints of $\simeq$70\% of
the prefit values are observed in some of the PDF nuisance
parameters, Whereas the mean constraint is closer to $\simeq$90\%.
The postfit nuisance parameter values, with respect to the prefit
values and uncertainties, give a $\chi^2$ value of 117 for 61 degrees
of freedom.  This suggests that the PDF set used here at NLO QCD plus
parton shower accuracy may not be sufficient to describe the data.  It
is possible that NNLO QCD accuracy combined with additional
developments in fitting methodology incorporated in more recent PDF
fits may improve the situation, and this can be studied in detail on
the basis of the unfolded cross sections measured here.

\subsection{Additional plots}
\label{sub:additionals}
Additional plots on the helicity and rapidity analysis are presented in Appendix~\ref{appsub:helicity}, and
additional plots on the two-dimensional cross sections are presented in Appendix~\ref{appsub:2ddiff}.

\ifthenelse{\boolean{cms@external}}{}{\clearpage}
\section{Summary}
\label{sec:conclusions}

The differential \PW boson cross sections
as functions of the \PW boson rapidity, \absyw, and for the two
charges separately, $\wplus \to \ell^+\nu$ and $\wminus \to
\ell^-\PAGn$, are measured in the \PW boson 
helicity states.  Double-differential cross sections of the \PW
boson are measured as a function of the charged-lepton transverse
momentum \ptl and absolute pseudorapidity \absleta. For both \wplus
and \wminus bosons, the differential charge asymmetry is also
extracted.

The measurement is based on data taken in proton-proton collisions at
the LHC at a center-of-mass energy of $\sqrt{s}=13 \TeV$,
corresponding to an integrated luminosity of \lint.  Differential
cross sections are presented, both absolute and normalized to the
total production cross section within a given acceptance.  For the
helicity measurement, the range $\absyw<2.5$ is presented, whereas for
the double-differential cross section the range $\absleta<2.4$ and
$26<\ptl<56\GeV$ is used.  The measurement is performed using both
the muon and electron channels, combined together considering all
sources of correlated and uncorrelated uncertainties.

The precision in the measurement as a function of \absyw, using a
combination of the two channels, is about 2\% in central \absyw bins
and 5 to 20\%, depending on the charge-polarization combination, in
the outermost acceptance bins.  The precision of the
double-differential cross section, relative to the total, is about 1\%
in the central part of the detector of $\absleta<1$ and better than
2.5\% up to $\absleta<2$ for each of the two \PW boson charges.

Charge asymmetries are also measured, differentially in \absyw and
polarization, as well as in \ptl and \absleta. The uncertainties in
these asymmetries range from 0.1\% in high-acceptance bins to roughly
2.5\% in regions of phase space with lower detector acceptance.
Furthermore, fiducial cross sections are presented by integrating the
two-dimensional differential cross sections over the full acceptance
of the analysis.

The measurement of the \PW boson polarized cross sections as
functions of \absyw is used to constrain the parameters related to
parton distribution functions in a simultaneous fit of the two
channels and the two \PW boson charges.  The constraints are derived
at the detector level on 60 uncorrelated eigenvalues of the \NNPDF set
of PDFs within the \MGvATNLO event generator, and show a total
constraint down to $\simeq$70\% of the prefit uncertainties for
certain variations of the PDF nuisance parameters.

\begin{acknowledgments}
  
\hyphenation{Bundes-ministerium Forschungs-gemeinschaft Forschungs-zentren Rachada-pisek} We congratulate our colleagues in the CERN accelerator departments for the excellent performance of the LHC and thank the technical and administrative staffs at CERN and at other CMS institutes for their contributions to the success of the CMS effort. In addition, we gratefully acknowledge the computing centers and personnel of the Worldwide LHC Computing Grid for delivering, so effectively, the computing infrastructure essential to our analyses. Finally, we acknowledge the enduring support for the construction and operation of the LHC and the CMS detector provided by the following funding agencies: the Austrian Federal Ministry of Education, Science and Research and the Austrian Science Fund; the Belgian Fonds de la Recherche Scientifique, and Fonds voor Wetenschappelijk Onderzoek; the Brazilian Funding Agencies (CNPq, CAPES, FAPERJ, FAPERGS, and FAPESP); the Bulgarian Ministry of Education and Science; CERN; the Chinese Academy of Sciences, Ministry of Science and Technology, and National Natural Science Foundation of China; the Colombian Funding Agency (COLCIENCIAS); the Croatian Ministry of Science, Education and Sport, and the Croatian Science Foundation; the Research and Innovation Foundation, Cyprus; the Secretariat for Higher Education, Science, Technology and Innovation, Ecuador; the Ministry of Education and Research, Estonian Research Council via PRG780, PRG803 and PRG445 and European Regional Development Fund, Estonia; the Academy of Finland, Finnish Ministry of Education and Culture, and Helsinki Institute of Physics; the Institut National de Physique Nucl\'eaire et de Physique des Particules~/~CNRS, and Commissariat \`a l'\'Energie Atomique et aux \'Energies Alternatives~/~CEA, France; the Bundesministerium f\"ur Bildung und Forschung, the Deutsche Forschungsgemeinschaft (DFG) under Germany's Excellence Strategy -- EXC 2121 ``Quantum Universe" -- 390833306, and Helmholtz-Gemeinschaft Deutscher Forschungszentren, Germany; the General Secretariat for Research and Technology, Greece; the National Research, Development and Innovation Fund, Hungary; the Department of Atomic Energy and the Department of Science and Technology, India; the Institute for Studies in Theoretical Physics and Mathematics, Iran; the Science Foundation, Ireland; the Istituto Nazionale di Fisica Nucleare, Italy; the Ministry of Science, ICT and Future Planning, and National Research Foundation (NRF), Republic of Korea; the Ministry of Education and Science of the Republic of Latvia; the Lithuanian Academy of Sciences; the Ministry of Education, and University of Malaya (Malaysia); the Ministry of Science of Montenegro; the Mexican Funding Agencies (BUAP, CINVESTAV, CONACYT, LNS, SEP, and UASLP-FAI); the Ministry of Business, Innovation and Employment, New Zealand; the Pakistan Atomic Energy Commission; the Ministry of Science and Higher Education and the National Science Center, Poland; the Funda\c{c}\~ao para a Ci\^encia e a Tecnologia, Portugal; JINR, Dubna; the Ministry of Education and Science of the Russian Federation, the Federal Agency of Atomic Energy of the Russian Federation, Russian Academy of Sciences, the Russian Foundation for Basic Research, and the National Research Center ``Kurchatov Institute"; the Ministry of Education, Science and Technological Development of Serbia; the Secretar\'{\i}a de Estado de Investigaci\'on, Desarrollo e Innovaci\'on, Programa Consolider-Ingenio 2010, Plan Estatal de Investigaci\'on Cient\'{\i}fica y T\'ecnica y de Innovaci\'on 2017--2020, research project IDI-2018-000174 del Principado de Asturias, and Fondo Europeo de Desarrollo Regional, Spain; the Ministry of Science, Technology and Research, Sri Lanka; the Swiss Funding Agencies (ETH Board, ETH Zurich, PSI, SNF, UniZH, Canton Zurich, and SER); the Ministry of Science and Technology, Taipei; the Thailand Center of Excellence in Physics, the Institute for the Promotion of Teaching Science and Technology of Thailand, Special Task Force for Activating Research and the National Science and Technology Development Agency of Thailand; the Scientific and Technical Research Council of Turkey, and Turkish Atomic Energy Authority; the National Academy of Sciences of Ukraine; the Science and Technology Facilities Council, UK; the US Department of Energy, and the US National Science Foundation.

Individuals have received support from the Marie Curie program and the European Research Council and Horizon 2020 Grant, Contracts No.\ 675440, 752730, and 765710 (European Union); the Leventis Foundation; the A.P.\ Sloan Foundation; the Alexander von Humboldt Foundation; the Belgian Federal Science Policy Office; the Fonds pour la Formation \`a la Recherche dans l'Industrie et dans l'Agriculture (FRIA-Belgium); the Agentschap voor Innovatie door Wetenschap en Technologie (IWT-Belgium); the F.R.S.-FNRS and FWO (Belgium) under the ``Excellence of Science -- EOS" -- be.h project n.\ 30820817; the Beijing Municipal Science \& Technology Commission, Grant No. Z191100007219010; the Ministry of Education, Youth and Sports (MEYS) of the Czech Republic; the Lend\"ulet (``Momentum") Program and the J\'anos Bolyai Research Scholarship of the Hungarian Academy of Sciences, the New National Excellence Program \'UNKP, the NKFIA research Grants No. 123842, 123959, 124845, 124850, 125105, 128713, 128786, and 129058 (Hungary); the Council of Scientific and Industrial Research, India; the HOMING PLUS program of the Foundation for Polish Science, cofinanced from European Union, Regional Development Fund, the Mobility Plus program of the Ministry of Science and Higher Education, the National Science Center (Poland), Contracts No. Harmonia 2014/14/M/ST2/00428, Opus 2014/13/B/ST2/02543, 2014/15/B/ST2/03998, and 2015/19/B/ST2/02861, Sonata-bis 2012/07/E/ST2/01406; the National Priorities Research Program by Qatar National Research Fund; the Ministry of Science and Higher Education, project No. 02.a03.21.0005 (Russia); the Tomsk Polytechnic University Competitiveness Enhancement Program and ``Nauka" Project FSWW-2020-0008 (Russia); the Programa de Excelencia Mar\'{i}a de Maeztu, and the Programa Severo Ochoa del Principado de Asturias; the Thalis and Aristeia programs cofinanced by EU-ESF, and the Greek NSRF; the Rachadapisek Sompot Fund for Postdoctoral Fellowship, Chulalongkorn University, and the Chulalongkorn Academic into Its 2nd Century Project Advancement Project (Thailand); the Kavli Foundation; the Nvidia Corporation; the SuperMicro Corporation; the Welch Foundation, Contract No. C-1845; and the Weston Havens Foundation (USA).
\end{acknowledgments}

\bibliography{auto_generated}
\ifthenelse{\boolean{cms@external}}{}{\clearpage}
\appendix
\numberwithin{figure}{section}
\section{Additional material}
\label{app:additionalMaterial}

This Appendix includes the plots and figures that shall be provided
as additional material in addition to those already featured in the
body of the text.

\subsection{Helicity and rapidity analysis}
\label{appsub:helicity}

\begin{figure}[h!tbp]
\centering
\includegraphics[width=0.49\textwidth]{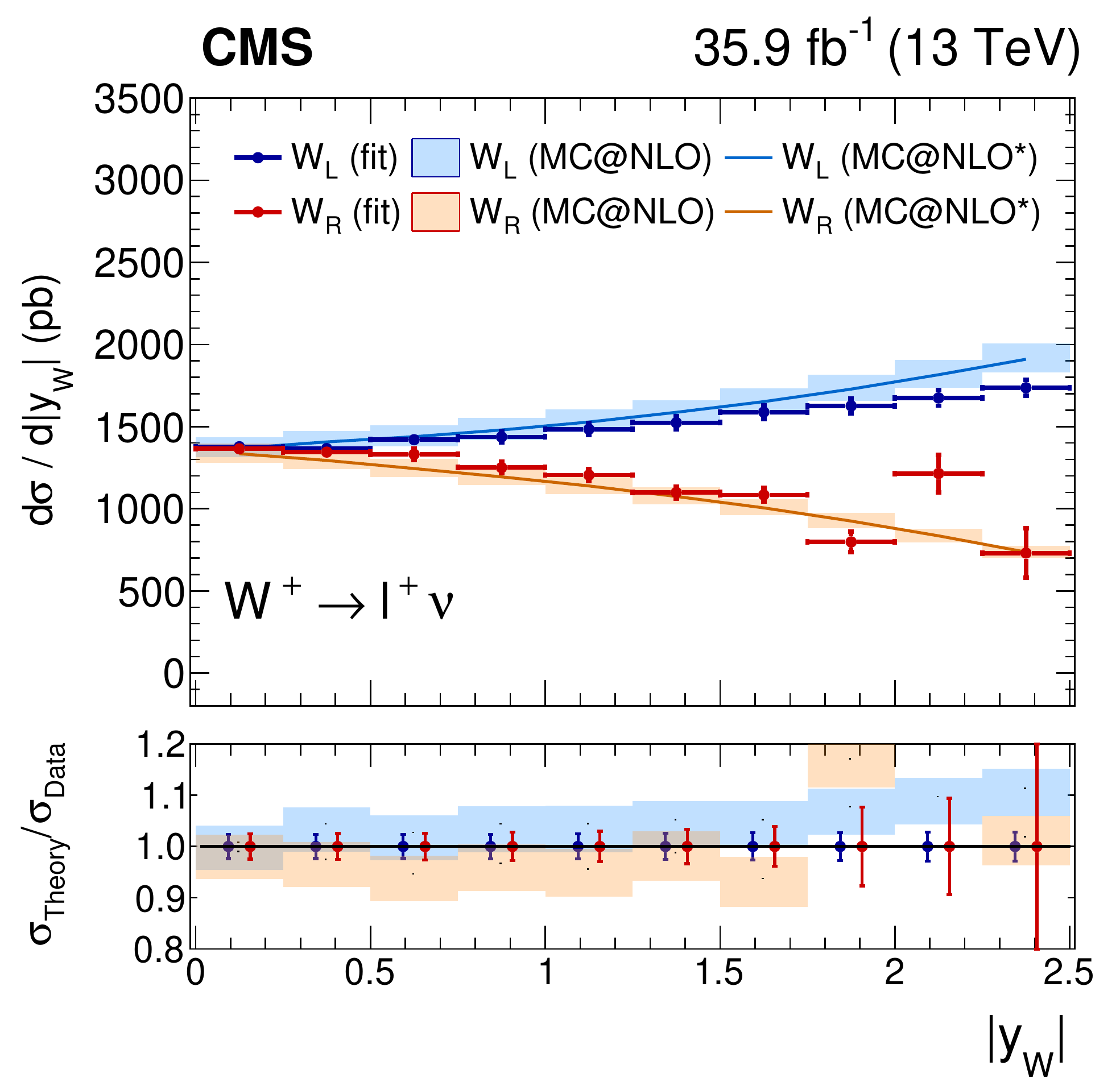}
\includegraphics[width=0.49\textwidth]{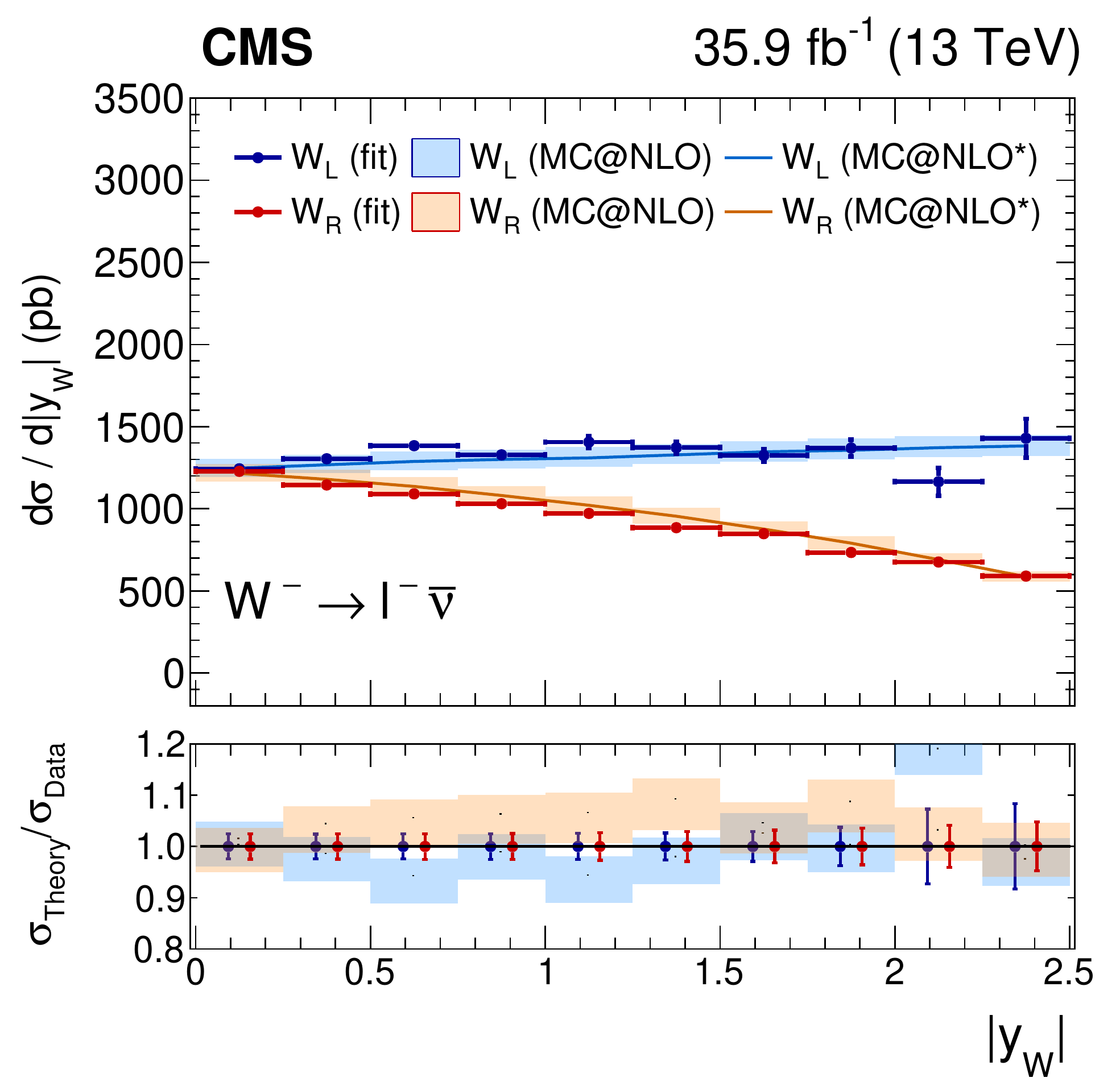}
\caption{Measured absolute $\wplus\to\ell^+\nu$ (\cmsLeft) and
  $\wminus\to\ell^-\PAGn$ (\cmsRight) cross section as a function of
  \absyw for the left-handed and right-handed helicity states from
  the combination of muon and electron channels. The
  ratio of the prediction from \MGvATNLO to the data is also shown. The
  lightly-filled band corresponds to the expected uncertainty from the
  PDF variations, \muF and \muR scales, and
  \alpS. \label{fig:xsecAbsPol}}
\end{figure}

Figure~\ref{fig:xsecAbsPol} shows the absolute polarized
cross sections as functions of \absyw from the combined muon and
electron fit for both charges of the \PW boson.

\begin{figure}[h!tbp]
\centering
\includegraphics[width=0.49\textwidth]{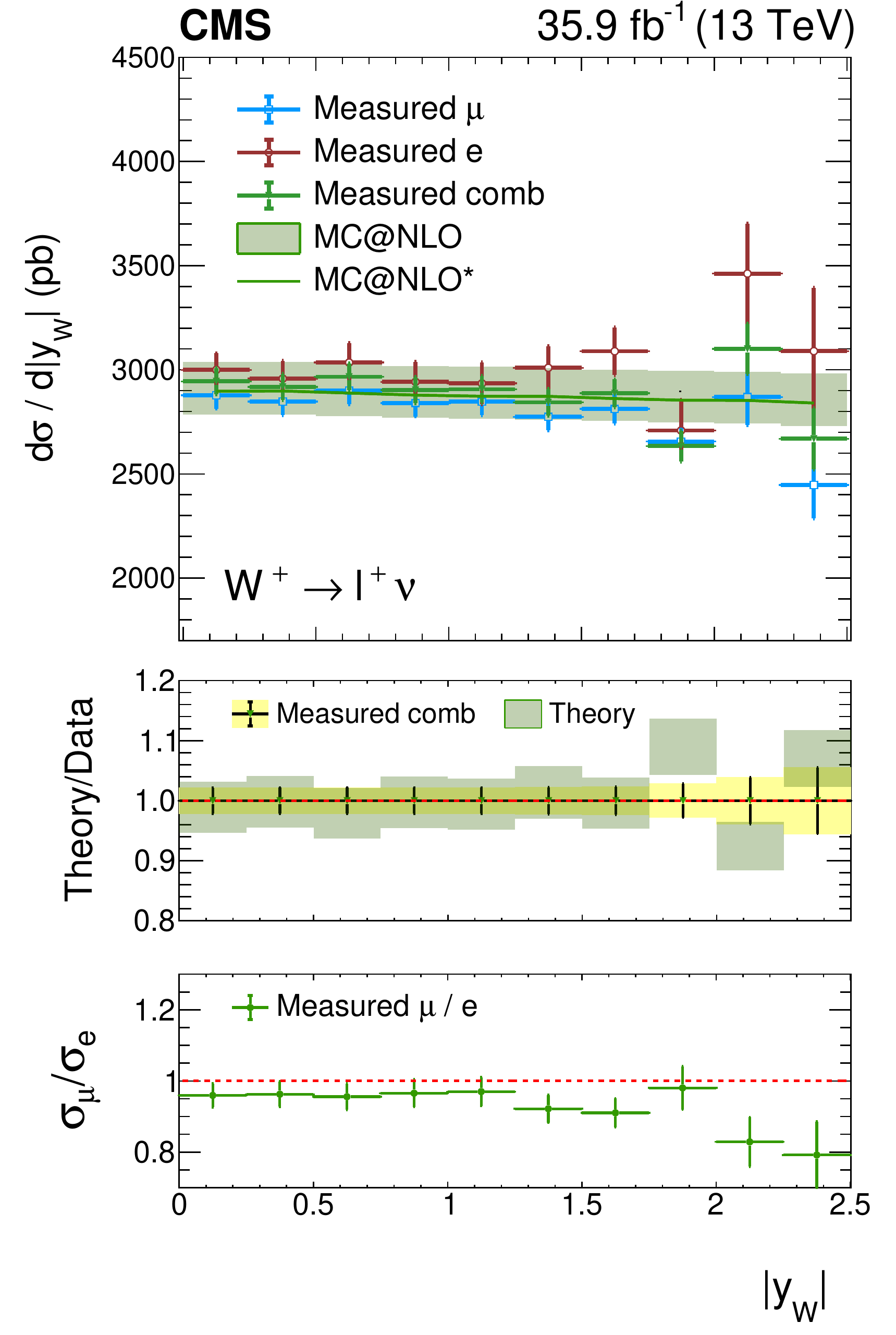}
\caption{Measured absolute $\wplus\to\ell^+\nu$ cross section as a function of
  \absyw from three distinct fits: the combination of muon and
  electron channels (green), the muon-only fit (blue), and the
  electron-only fit (red). The ratio of the prediction from \MGvATNLO
  to the data is also shown. The lightly-filled band corresponds to
  the expected uncertainty from the PDF variations, \muF
  and \muR scales, and
  \alpS. \label{fig:rap_unpol_abs_comp_plus}}
\end{figure}

\begin{figure}[h!tbp]
\centering
\includegraphics[width=0.49\textwidth]{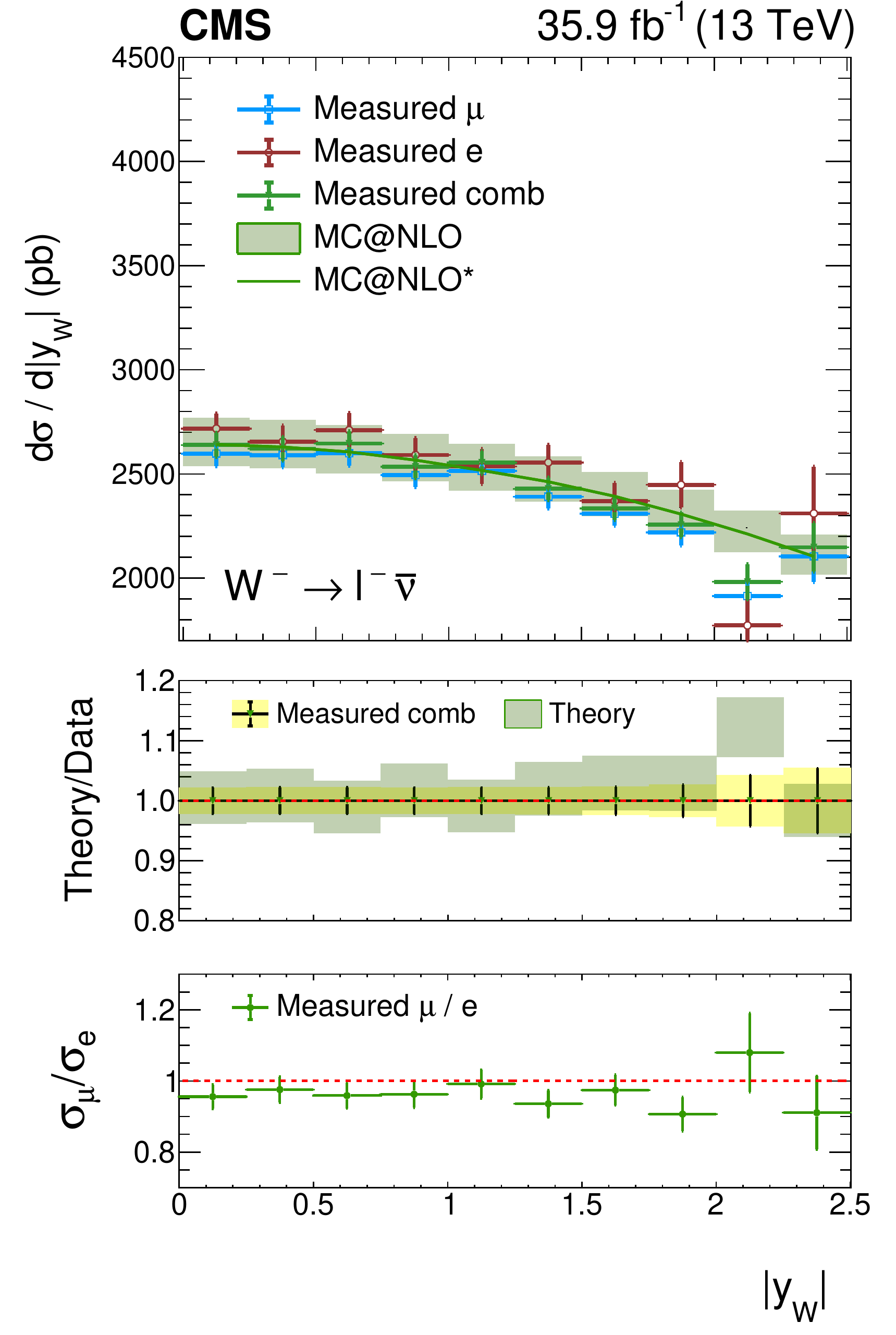}
\caption{Measured absolute $\wminus\to\ell^-\PAGn$ cross section as a
  function of \absyw from three distinct fits: the combination of
  muon and electron channels (green), the muon-only fit (blue), and
  the electron-only fit (red). The ratio of the prediction
  from \MGvATNLO to the data is also shown. The lightly-filled band
  corresponds to the expected uncertainty from the PDF variations,
  \muF and \muR scales, and
  \alpS. \label{fig:rap_unpol_abs_comp_minus}}
\end{figure}

Figures~\ref{fig:rap_unpol_abs_comp_plus}
and~\ref{fig:rap_unpol_abs_comp_minus} show again the absolute
unpolarized cross sections as functions of the \PW boson rapidity
for the positively and negatively charged \PW bosons. These figures,
however, also show the comparison between the two lepton flavors,
\ie, performing the fits separately once in the muon-only, once in
the electron-only, and once in the flavor combination shows the
experimental agreement of the different flavor channels. We show that
the single-flavor fits agree within their uncertainty with each other,
as well as with the combined-flavor fit. The correlation structure of
the three different fits cannot be trivially displayed in the ratios
of the flavors shown in the lower panels of
Figs.~\ref{fig:rap_unpol_abs_comp_plus}
and~\ref{fig:rap_unpol_abs_comp_minus}.

\begin{figure}[h!tbp]
\centering
\includegraphics[width=0.49\textwidth]{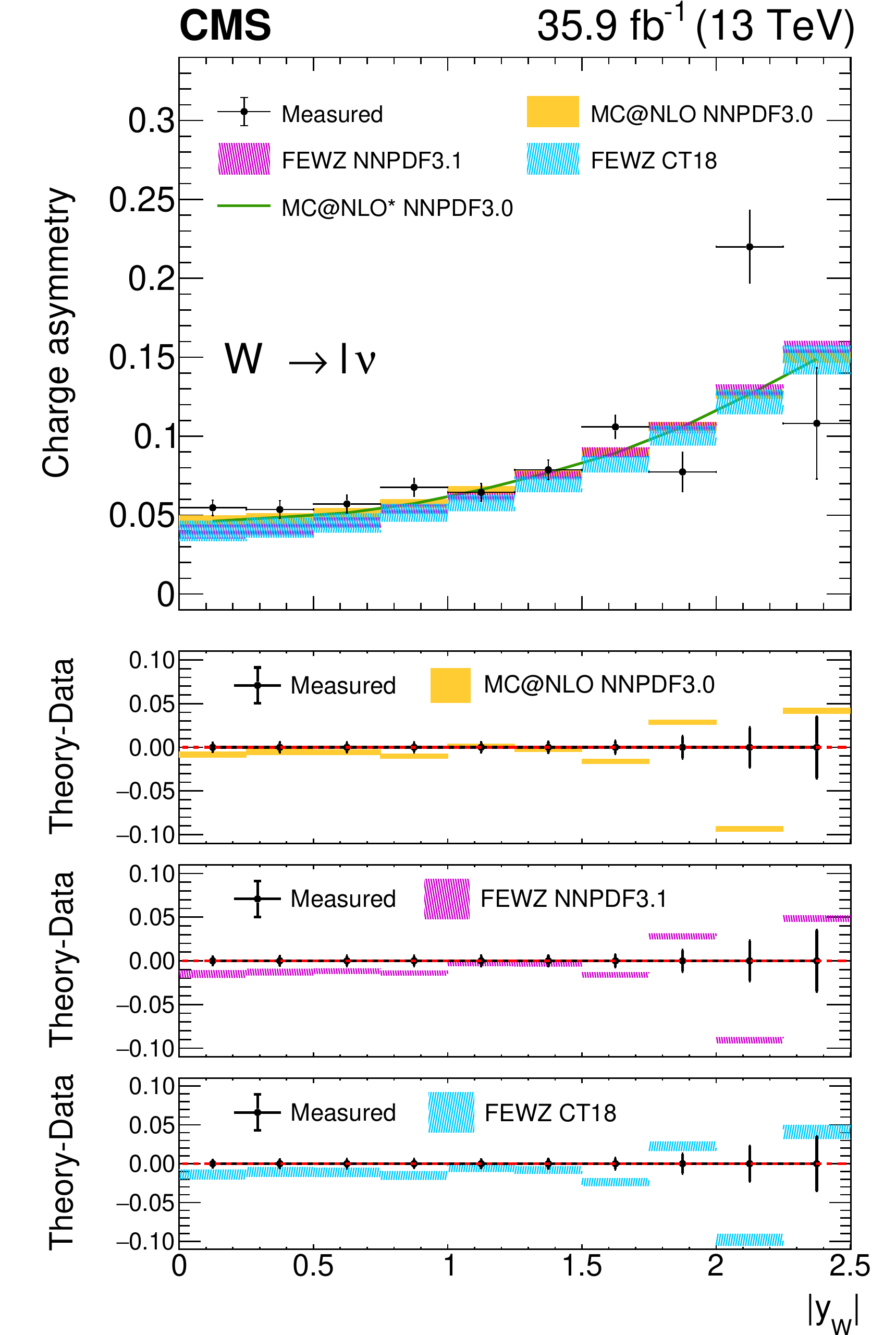}
\caption{Measured \PW boson charge asymmetry as a function of
  \absyw from the combination of the muon and electron channels
  (black dots), compared with different theoretical predictions.  The
  yellow band represents the default generator used in this analysis,
  \MGvATNLO with \NNPDF PDF set, the pink band represents the \FEWZ
  generator with \textsc{NNPDF3.1} PDF set, and the cyan band
  represents the \FEWZ generator with CT18 PDF
  set. The uncertainty bands of the prediction include PDF uncertainties only, which are dominant with respect to \alpS or QCD scale variations for this quantity.\label{fig:asyfewz}}
\end{figure}

The comparison of the measured unpolarized \PW boson charge
asymmetry as a function of \absyw with the prediction from another
matrix-element generator, \FEWZ 2.0~\cite{Gavin:2010az}, is shown in
Fig.~\ref{fig:asyfewz}. The calculation is coupled with either the
\textsc{NNPDF3.1} NNLO PDF set or the CT18~\cite{Hou:2019efy}
NNLO PDF set.

\begin{figure}[h!tbp]
\centering
\includegraphics[width=0.49\textwidth]{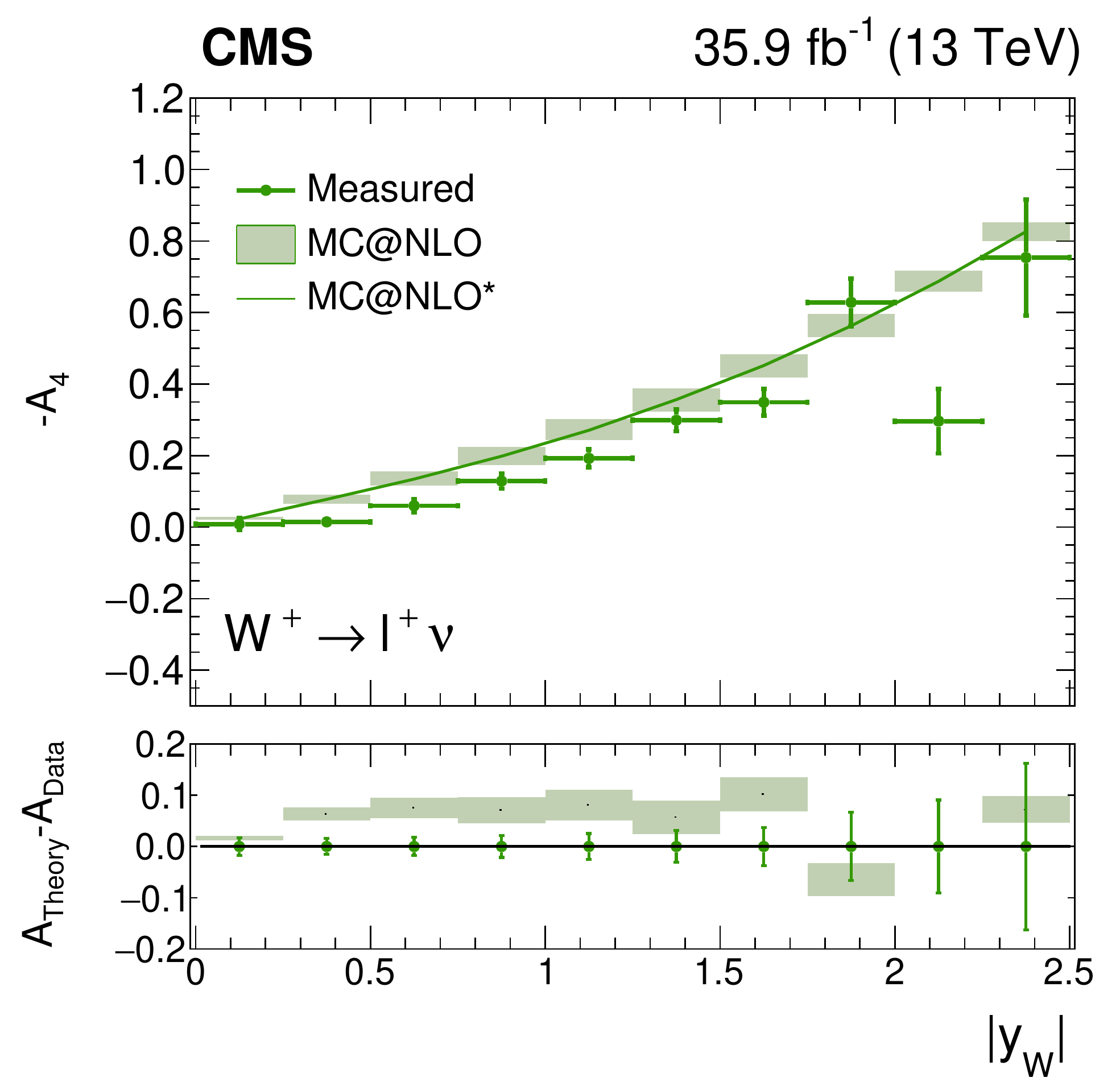}
\caption{Measured $A_4$ coefficient for $\wplus\to\ell^+\nu$ extracted
  from the fit of the polarized cross sections to the combined muon
  and electron channel fit. Note that $A_4$ is negative in this case,
  and the plotted quantity is $-A_4$. The difference between the
  prediction from \MGvATNLO and the measured values is also shown. The
  lightly-filled band corresponds to the expected uncertainty from the
  PDF variations, \muF and \muR scales, and
  \alpS.  \label{fig:xsecA4plus}}
\end{figure}

\begin{figure}[h!tbp]
\centering
\includegraphics[width=0.49\textwidth]{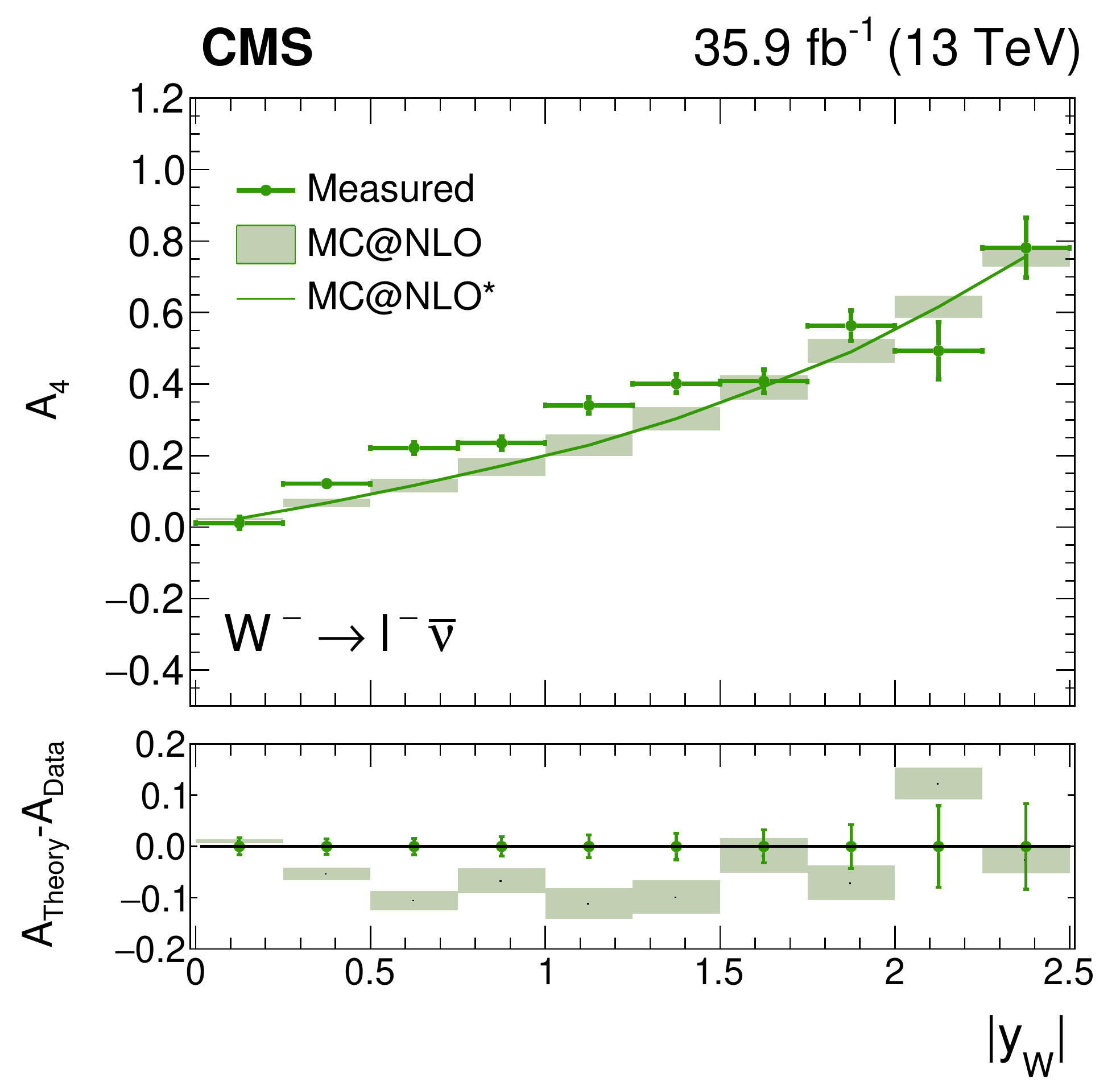}
\caption{Measured $A_4$ coefficient for $\wminus\to\ell^-\nu$ extracted
  from the fit of the polarized cross sections to the combined muon
  and electron channel fit. The difference between the
  prediction from \MGvATNLO and the measured values is also shown. The
  lightly-filled band corresponds to the expected uncertainty from the
  PDF variations, \muF and \muR scales, and
  \alpS.  \label{fig:xsecA4minus}}
\end{figure}

Figures~\ref{fig:xsecA4plus} and~\ref{fig:xsecA4minus} show the
distribution of the $A_4$ coefficient extracted as a function of
\absyw from the combined fit to the muon and electron channels for
the positively and negatively charged \PW bosons, respectively.

\begin{figure}[h!tbp]
\centering
\includegraphics[width=0.49\textwidth]{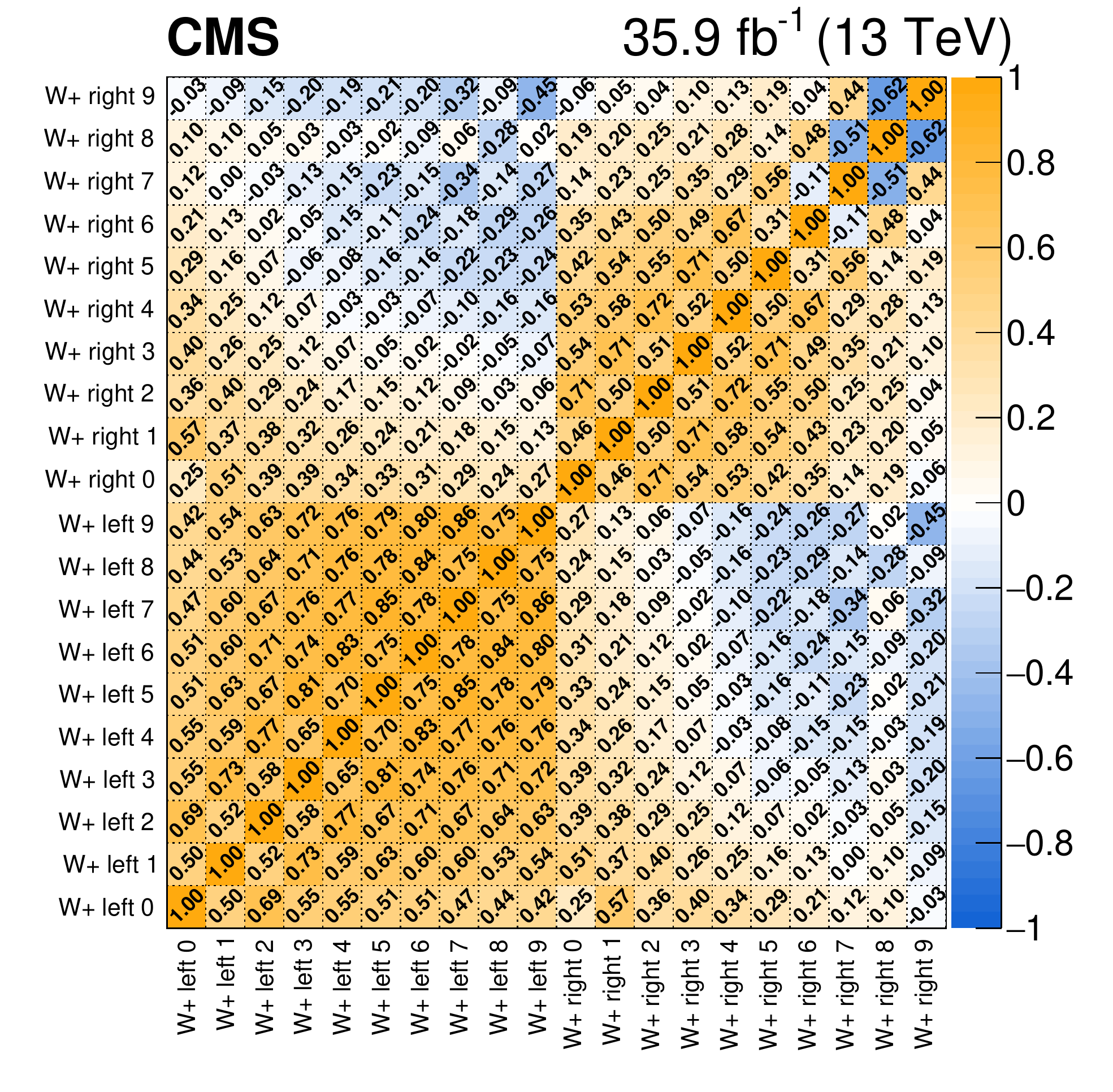}
\includegraphics[width=0.49\textwidth]{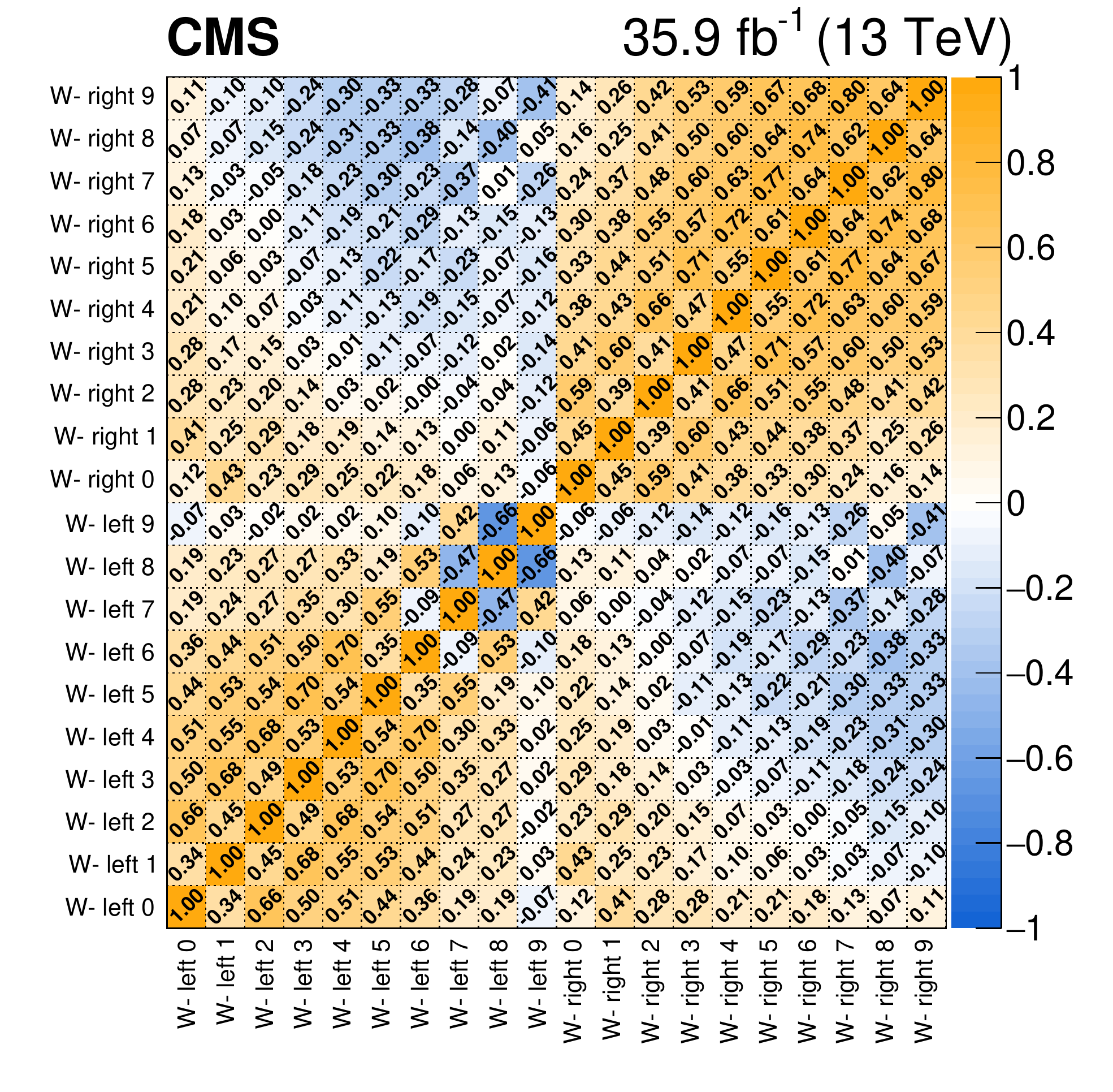}
\caption{Correlation coefficients between the helicity-dependent
  signal cross sections for $\wplus\to\ell^+\nu$ (\cmsLeft) and
  $\wminus\to\ell^-\PAGn$ (\cmsRight) extracted from the fit to the combined
  muon and electron channel fit.\label{fig:correlationsPOIs}}
\end{figure}

Figure~\ref{fig:correlationsPOIs} shows the correlation coefficients
between the different signal processes split into their helicity
components from the combined muon and electron channel fit for the two
charges of the \PW boson. The numbering corresponds to the bins in
\absyw of width 0.25 starting at zero. It is worthwhile to note here
that the correlations of neighboring bins in rapidity are large,
especially for each helicity. There are also nontrivial
correlations across the helicity states.

\begin{figure}[h!tbp]
\centering
\includegraphics[width=0.49\textwidth]{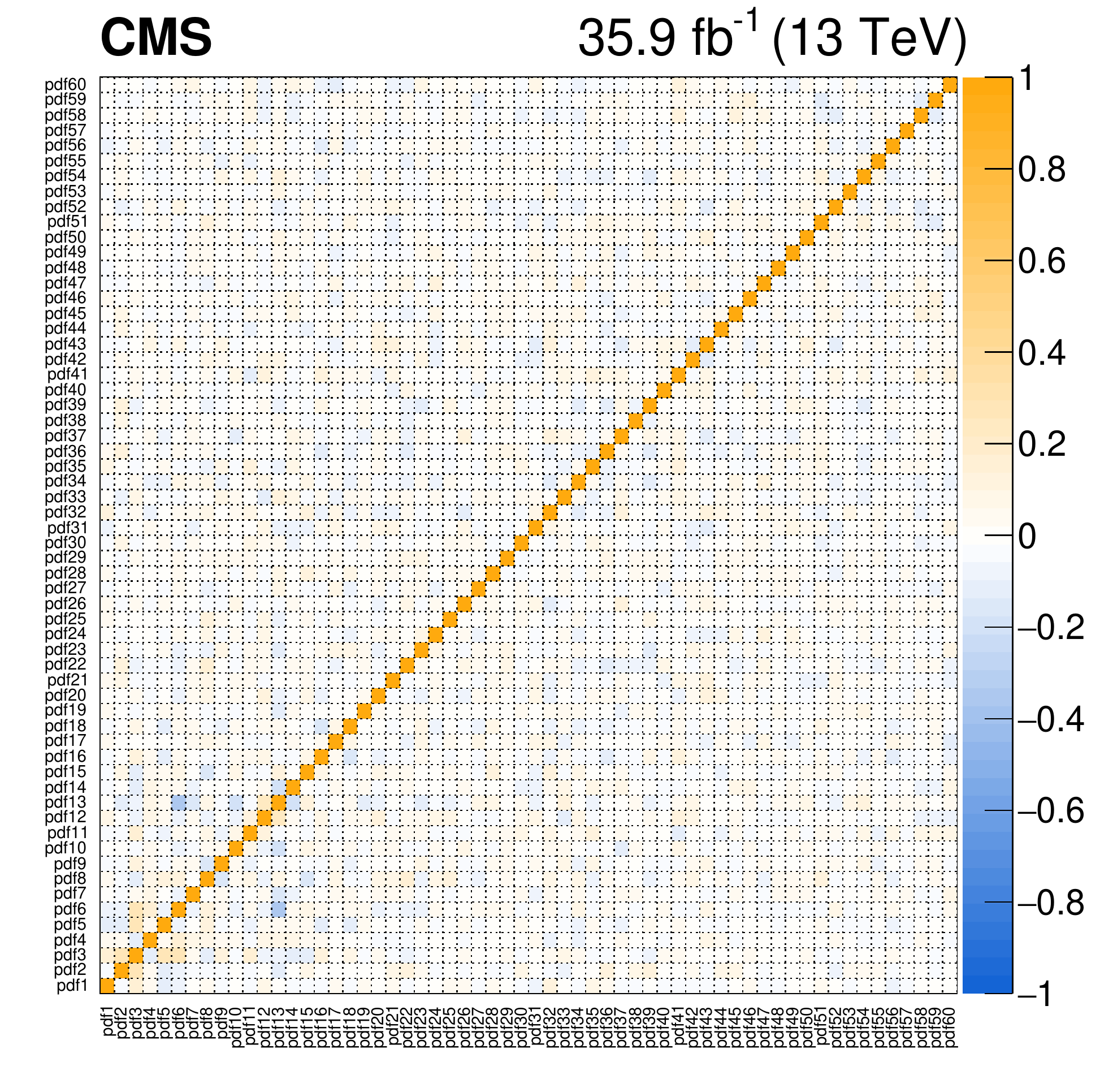}
\caption{Correlation coefficients between the 60 PDF nuisance
  parameters extracted from the fit to the combined muon and electron
  channel fit. The underlying fit is performed by fixing the \PW boson cross sections to their expectation in all helicity and charge processes.\label{fig:correlationsPDFs}}
\end{figure}

Figure~\ref{fig:correlationsPDFs} shows the correlation coefficients
between the different PDF nuisance parameters in the combined muon and
electron channel fit. The numbering of the PDF nuisances derives from
the conversion of the NNPDF3.0 replicas to 60 orthogonal Hessian
nuisance parameters and carries no physical meaning.

\begin{figure*}[h!tbp]
\centering
\includegraphics[width=0.97\linewidth]{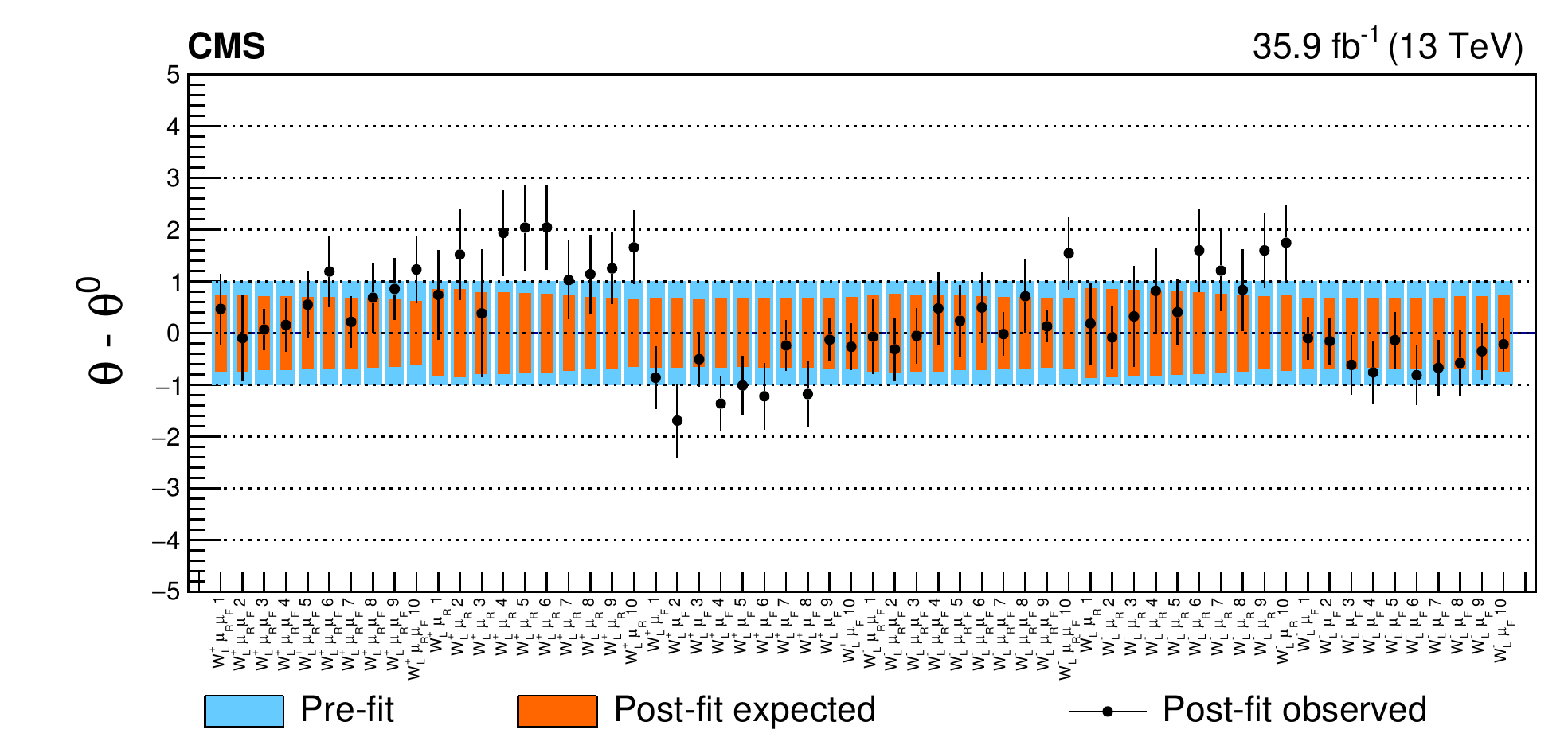} \\
\includegraphics[width=0.97\linewidth]{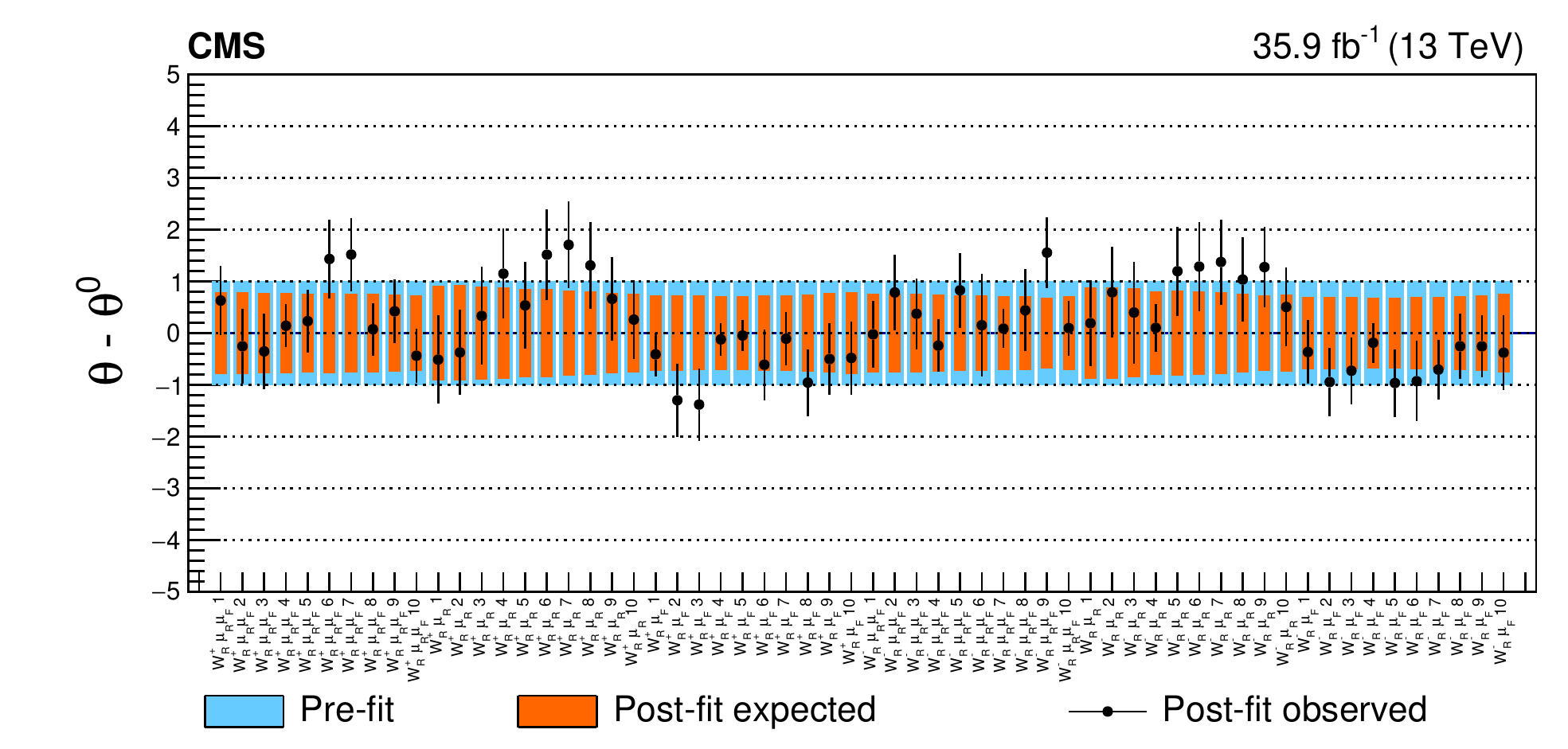} \\
\caption{Post-fit pulls and constraints of the nuisance parameters
  associated with the \muF and \muR scale
  systematic uncertainties. The numbering refers to bins in the \ptw
  spectrum in increasing order. The nuisance parameters applied to the
  ``left'' polarization are shown on the upper panel while the ones
  associated with the ``right'' polarization are shown on the lower
  panel.\label{fig:constraintsScales}}
\end{figure*}

Figure~\ref{fig:constraintsScales} shows the post-fit pulls and their
post-fit constraints of the nuisance parameters associated with the
\muF and \muR scale systematic
uncertainties. The numbering corresponds to the bins in the \ptw
spectrum in increasing order. The numbers result from the combined fit
to the muon and electron channels.

\begin{figure}[h!tbp]
\centering
\includegraphics[width=0.97\linewidth]{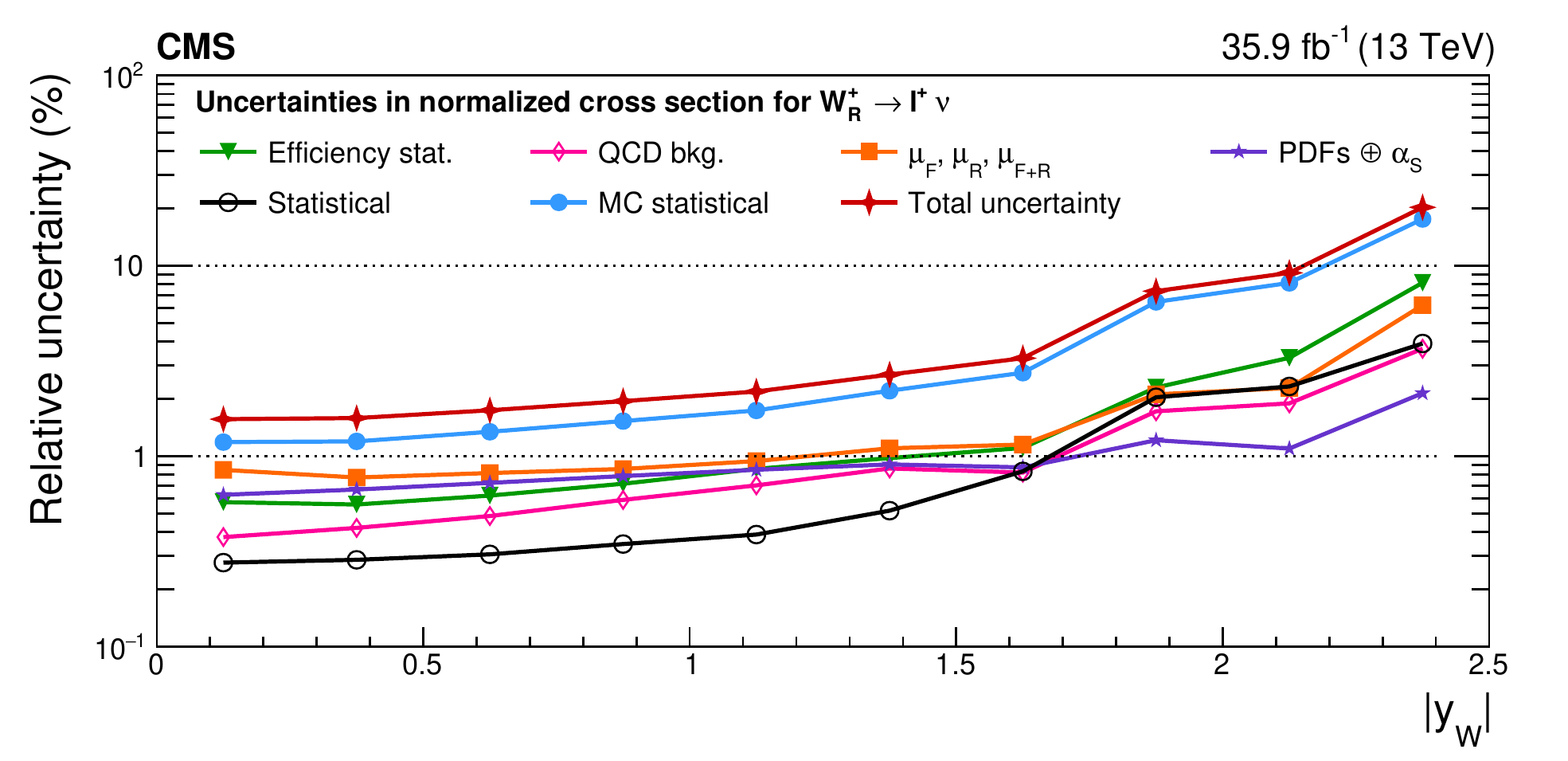} \\
\includegraphics[width=0.97\linewidth]{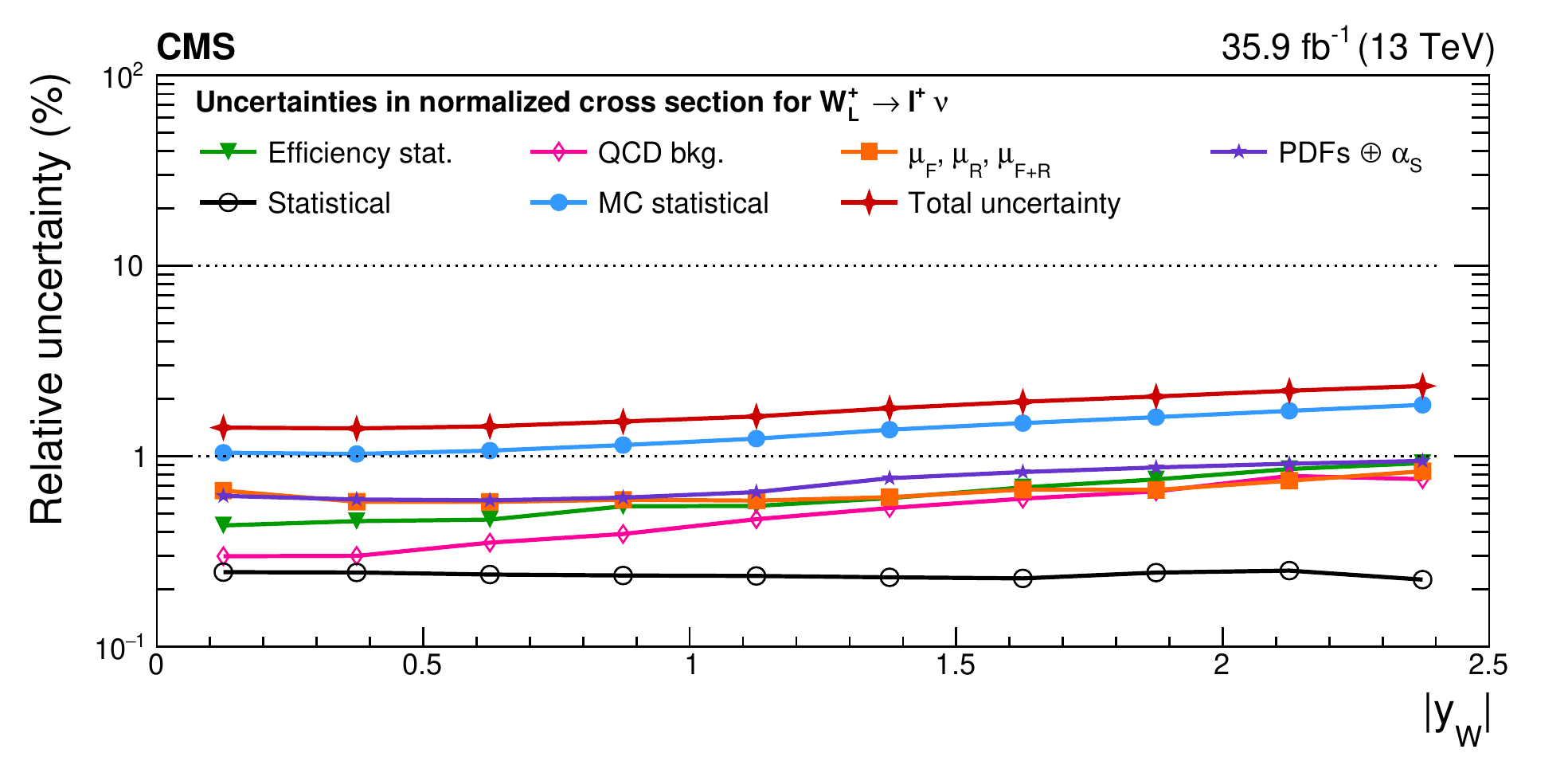} \\
\includegraphics[width=0.97\linewidth]{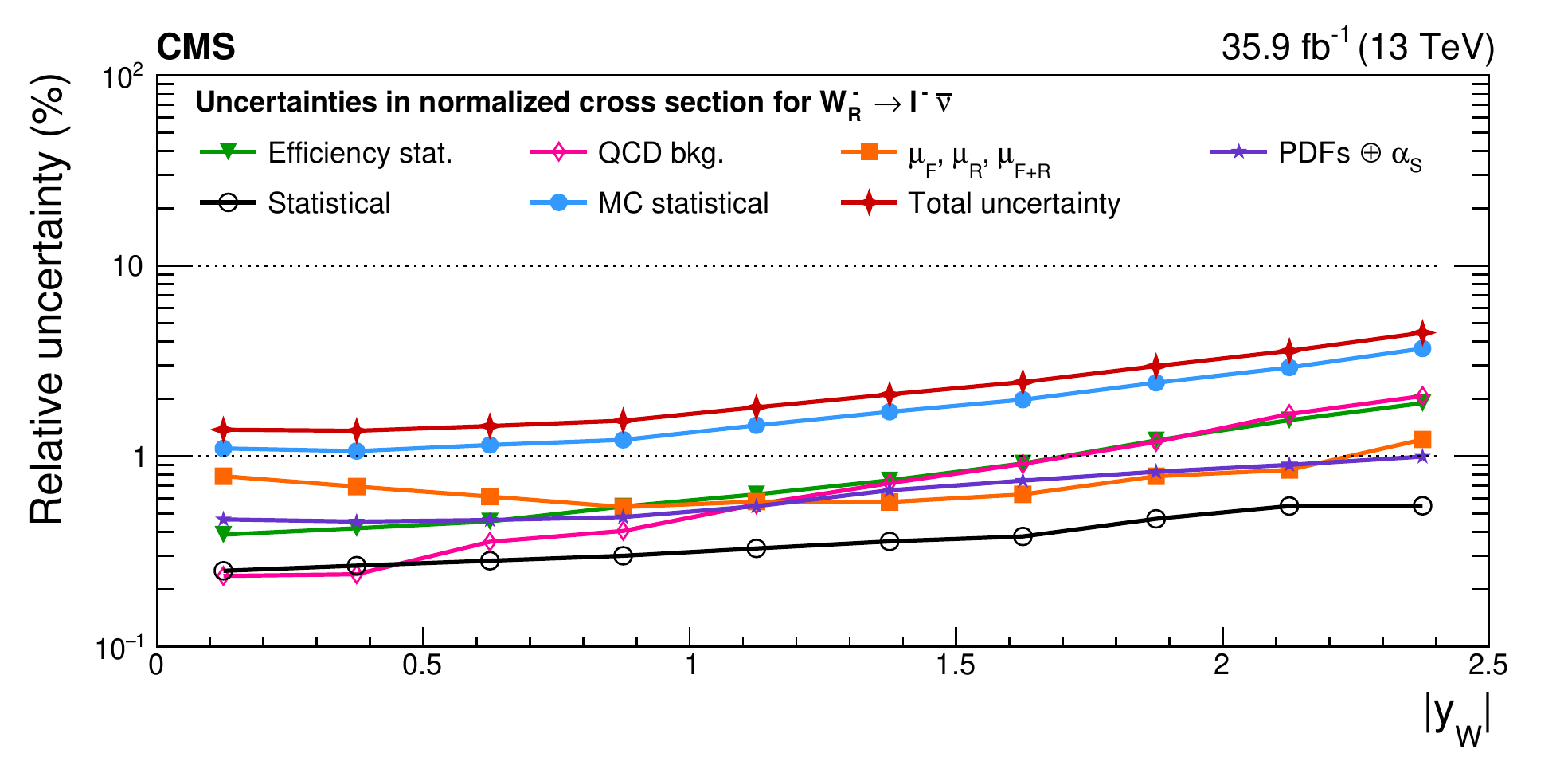} \\
\caption{Remaining impacts on the normalized polarized cross sections
  as functions of the \PW boson rapidity. Shown are the impacts of
  the nuisance groups for \wplusr (upper), \wplusl (middle), and
  \wminusr (lower) bosons in the helicity fit. The groups of uncertainties
  subleading to the ones shown are suppressed
  for simplicity. \label{fig:impactsNormPol}}
\end{figure}

Figure~\ref{fig:impactsNormPol} shows the impacts of the nuisance parameter groups
on the normalized polarized cross sections for \wplusr, \wplusl, and \wminusr.

\begin{figure}[h!tbp]
\centering
\includegraphics[width=0.97\linewidth]{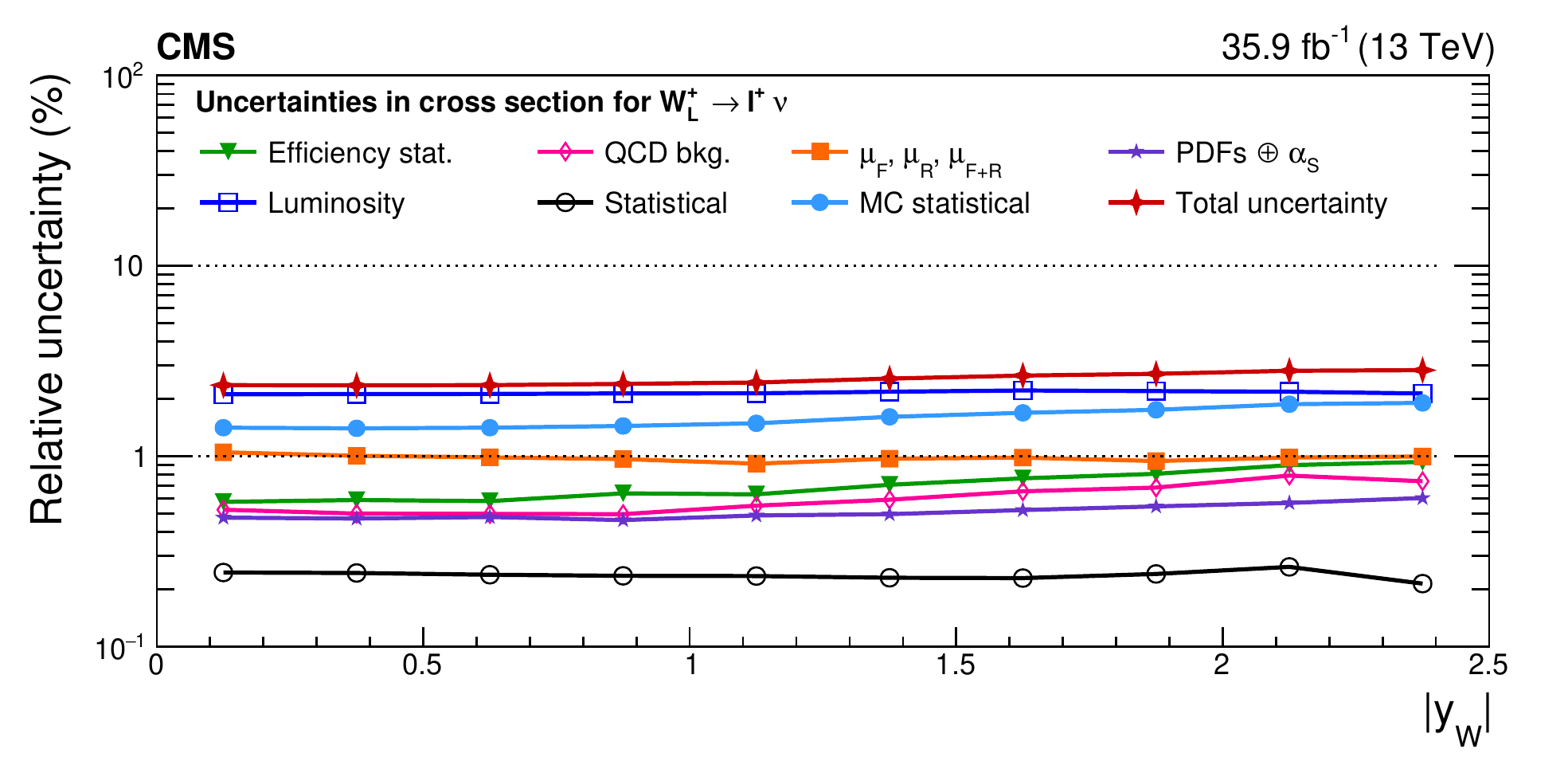} \\
\includegraphics[width=0.97\linewidth]{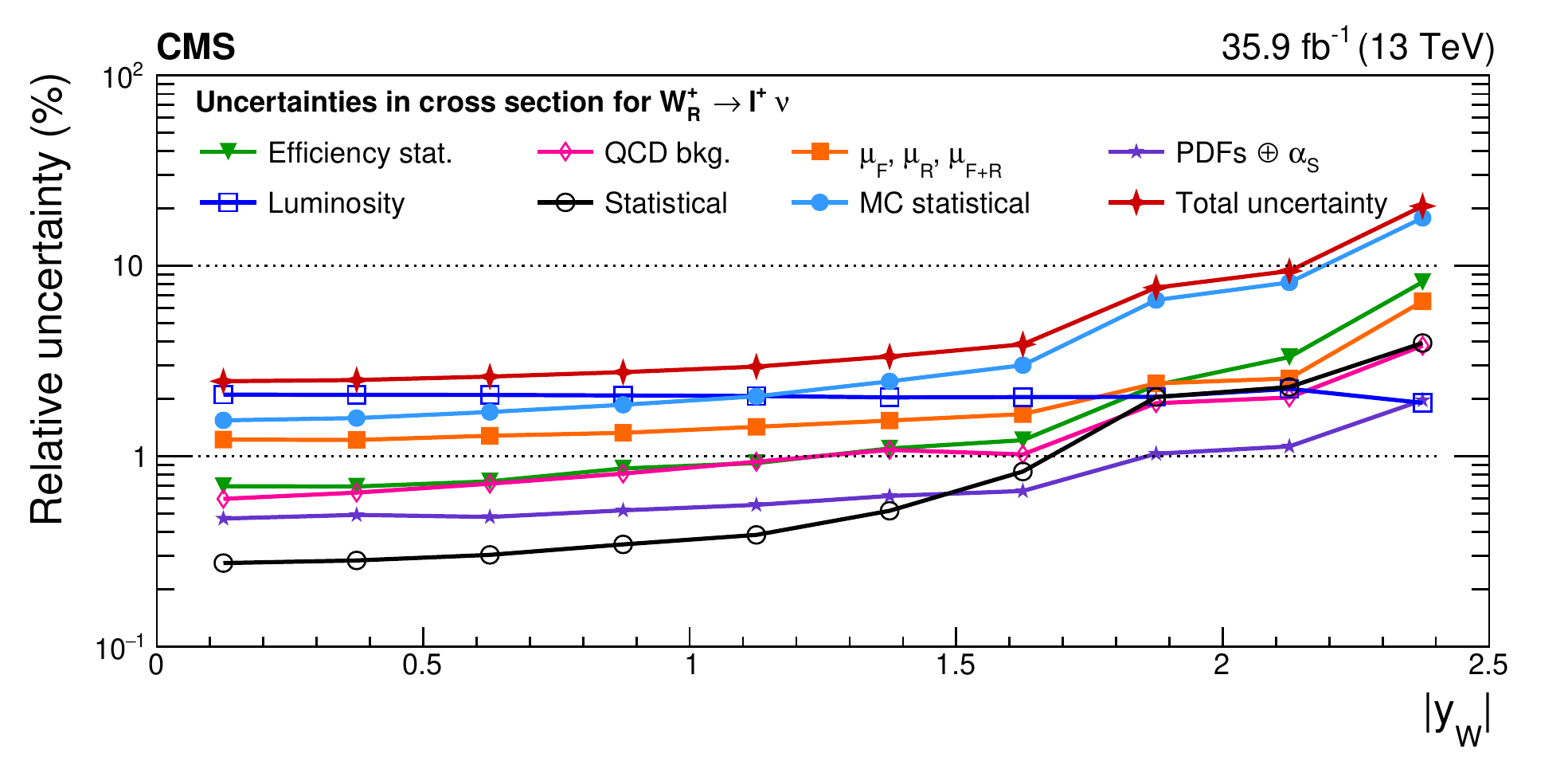} \\
\caption{Impacts on the absolute polarized cross sections as functions
  of the \PW boson rapidity. Shown are the impacts of the nuisance
  groups for \wplusl (upper) and \wplusr (lower) in the helicity
  fit. The groups of uncertainties subleading to the ones
  shown are suppressed for simplicity. \label{fig:impactsAbsPolPlus}}
\end{figure}

\begin{figure}[h!tbp]
\centering
\includegraphics[width=0.97\linewidth]{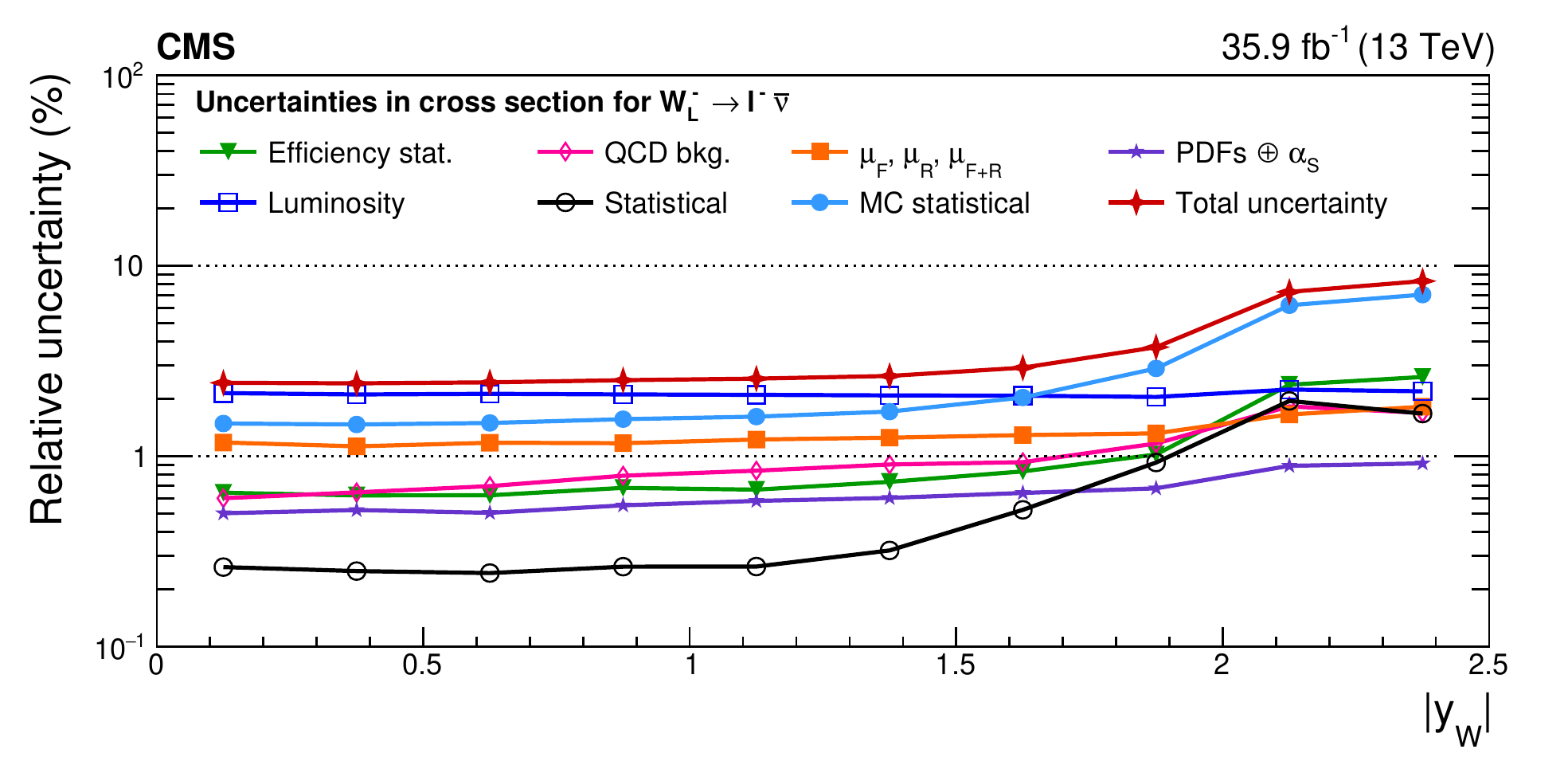} \\
\includegraphics[width=0.97\linewidth]{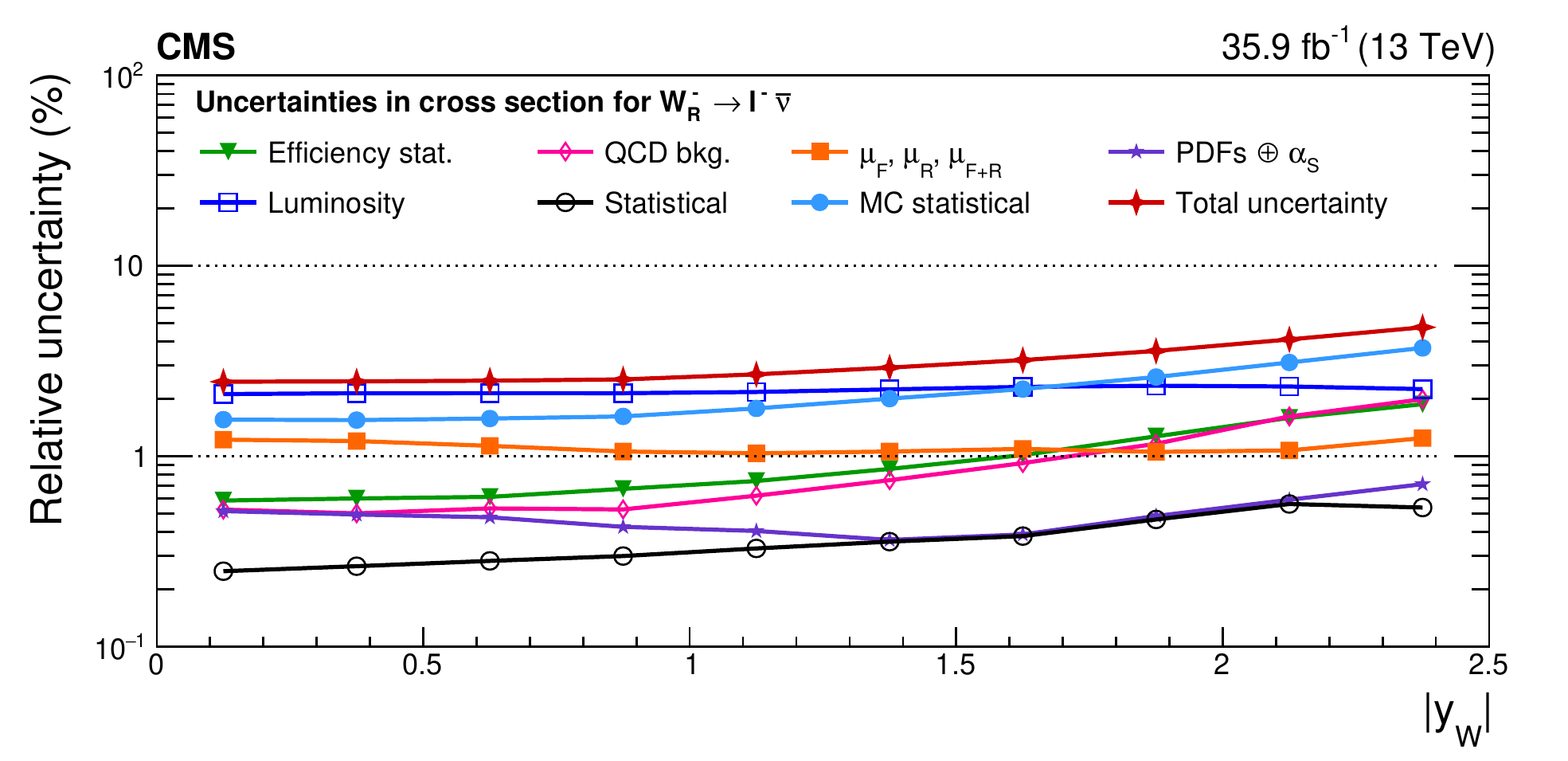} \\
\caption{Impacts on the absolute polarized cross sections as functions
  of the \PW boson rapidity. Shown are the impacts of the nuisance
  groups for \wminusl (upper) and \wminusr (lower) bosons in the
  helicity fit. The groups of uncertainties subleading to
  the ones shown are suppressed for
  simplicity. \label{fig:impactsAbsPolMinus}}
\end{figure}

Figure~\ref{fig:impactsAbsPolPlus} shows the impacts of the nuisance parameter groups on the absolute polarized
cross sections for left-, and right-handed \PW bosons of positive charge. Figure~\ref{fig:impactsAbsPolMinus}
shows the same impacts for negatively charged \PW bosons.

\begin{figure}[h!tbp]
\centering
\includegraphics[width=0.97\linewidth]{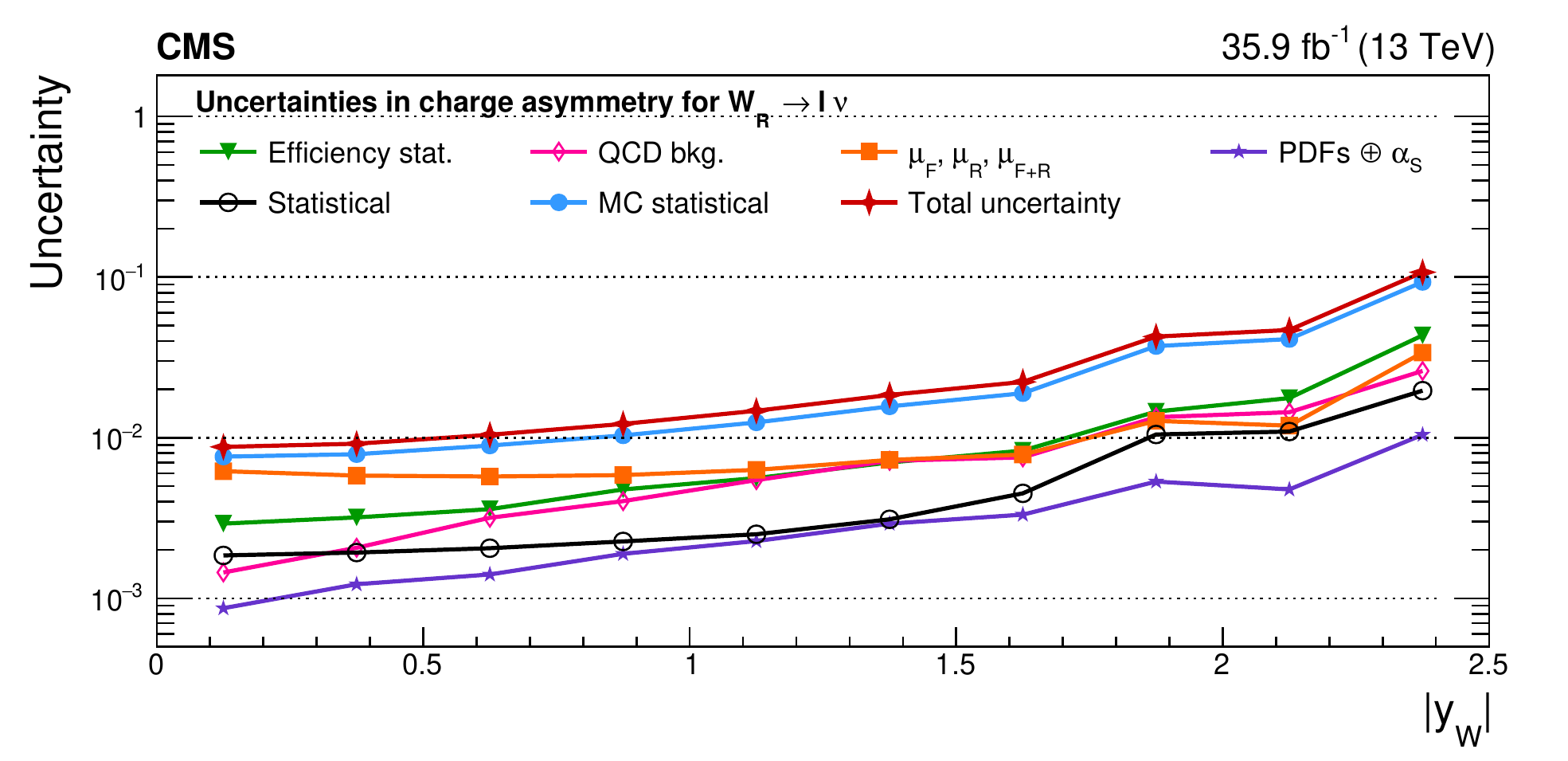}
\caption{Impacts on the charge asymmetry as functions of the \PW
  boson rapidity for \wright bosons in the helicity fit. The groups
  of uncertainties subleading to the ones shown are
  suppressed for simplicity. \label{fig:impactAsymWright}}
\end{figure}

Figure~\ref{fig:impactAsymWright} shows the impacts of the nuisance parameter groups
on the charge asymmetry for \wright bosons.

\begin{figure}[h!tbp]
\centering
\includegraphics[width=0.97\linewidth]{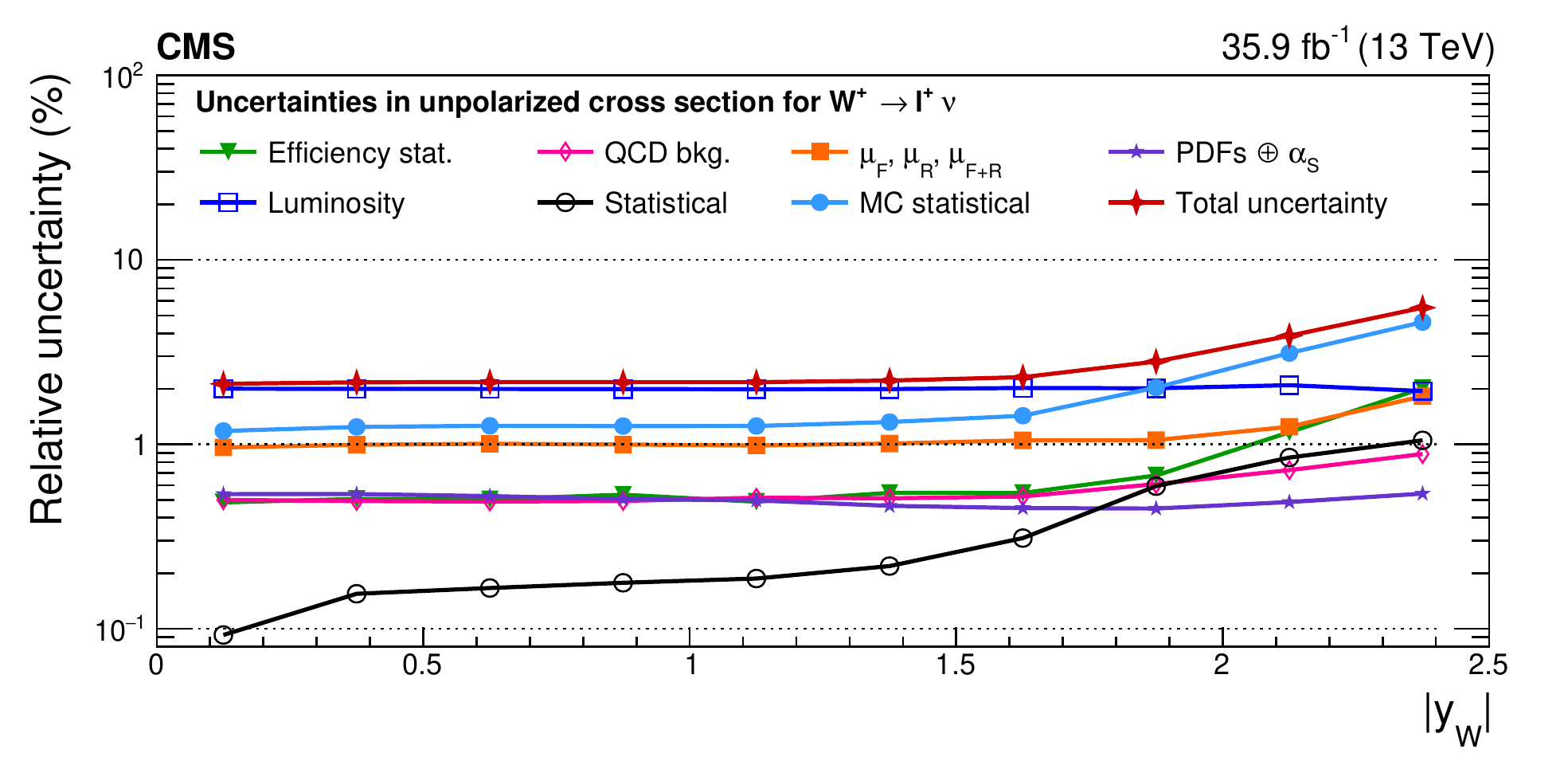} \\
\includegraphics[width=0.97\linewidth]{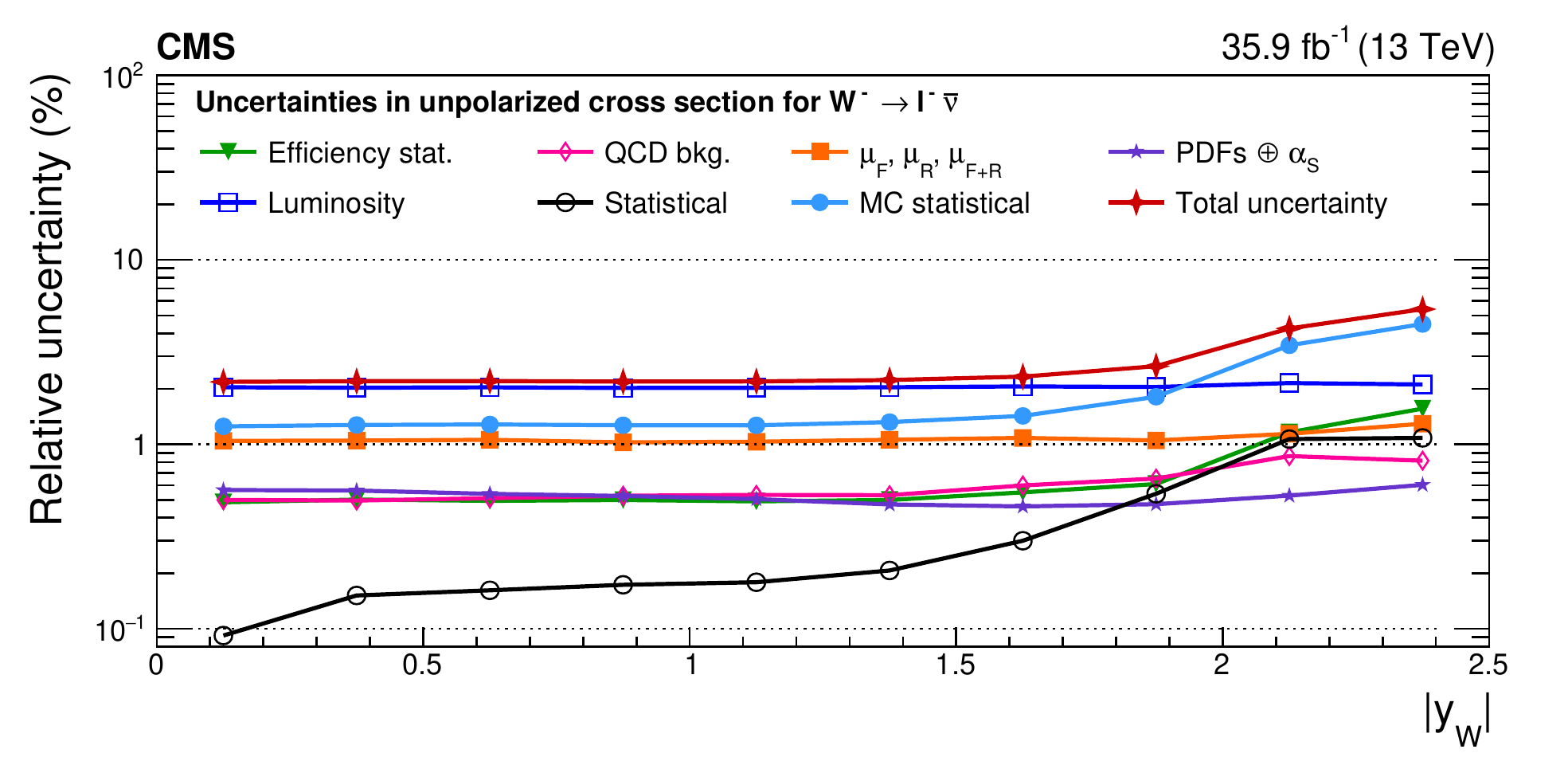} \\
\includegraphics[width=0.97\linewidth]{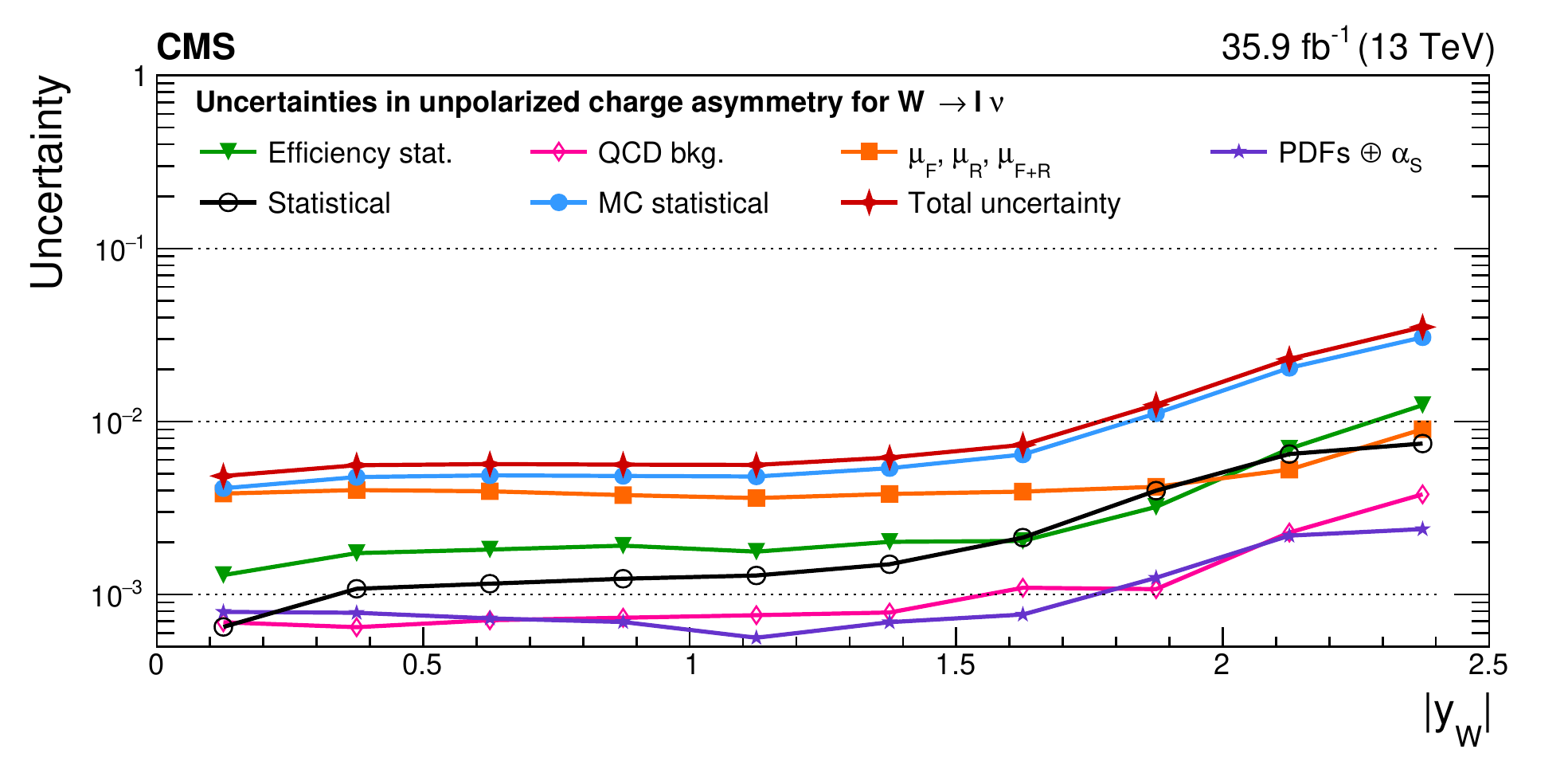} \\
\caption{Impacts on the unpolarized absolute cross sections as
  functions of the \PW boson rapidity for \wplus (upper),
  \wminus (middle), and the unpolarized charge asymmetry (lower)
  bosons in the helicity fit. The groups of uncertainties subleading 
  to the ones shown are suppressed for
  simplicity. \label{fig:impactsUnpol}}
\end{figure}

\begin{figure}[h!tbp]
\centering
\includegraphics[width=0.97\linewidth]{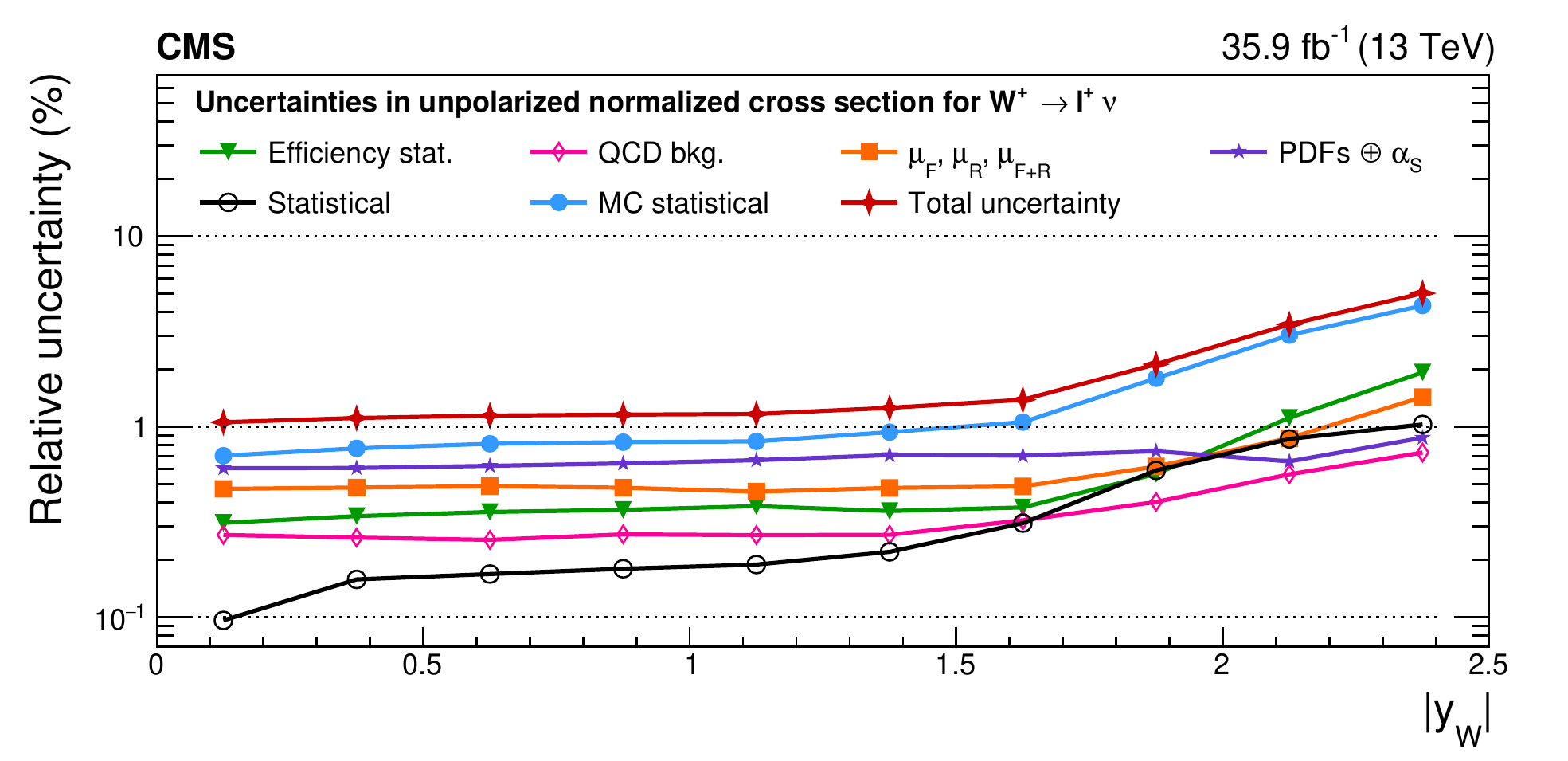} \\
\includegraphics[width=0.97\linewidth]{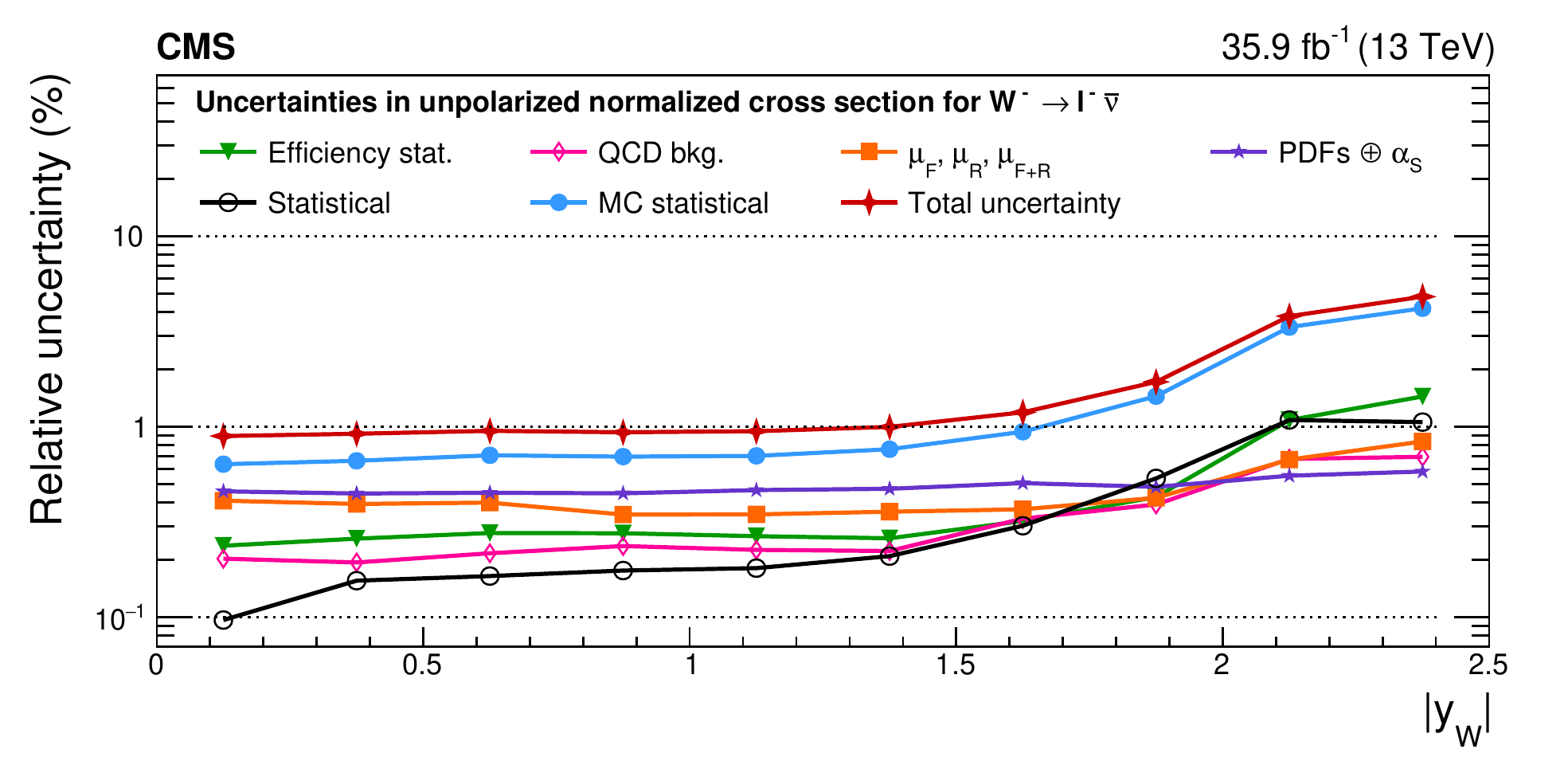} \\
\caption{Impacts on the unpolarized normalized cross sections as
  functions of the \PW boson rapidity for \wplus (upper) and
  \wminus (lower) bosons in the helicity fit. The groups of
  uncertainties subleading to the ones shown are suppressed
  for simplicity. \label{fig:impactsUnpolNorm}}
\end{figure}

Figure~\ref{fig:impactsUnpol} shows the impacts of the nuisance parameter groups
on the unpolarized absolute \PW boson cross sections as a function of \absyw for both 
charges and the charge asymmetry. Figure~\ref{fig:impactsUnpolNorm} shows the same impacts
for the normalized \PW production cross sections.

\begin{figure}[h!tbp]
\centering
\includegraphics[width=0.97\linewidth]{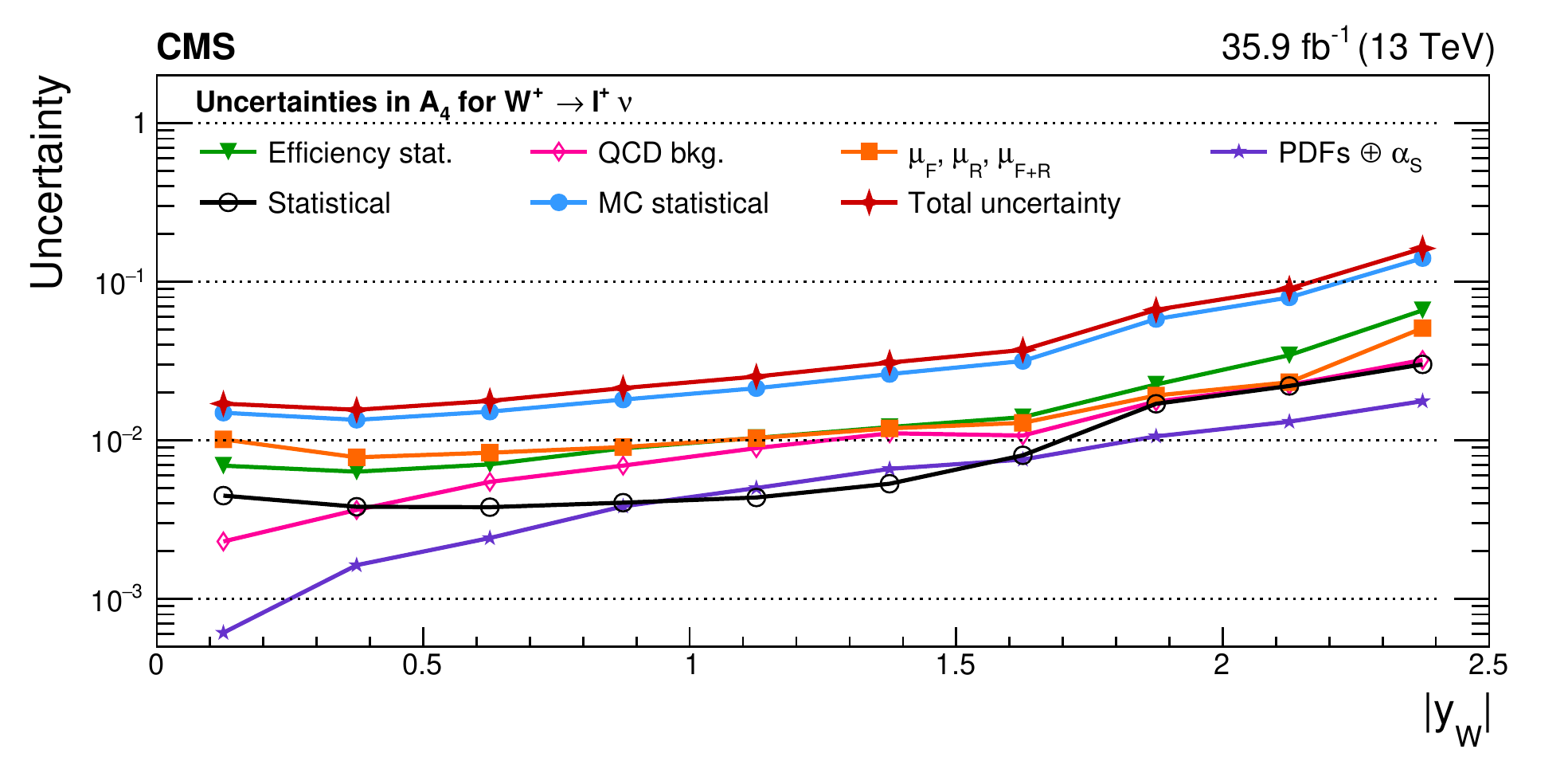} \\
\includegraphics[width=0.97\linewidth]{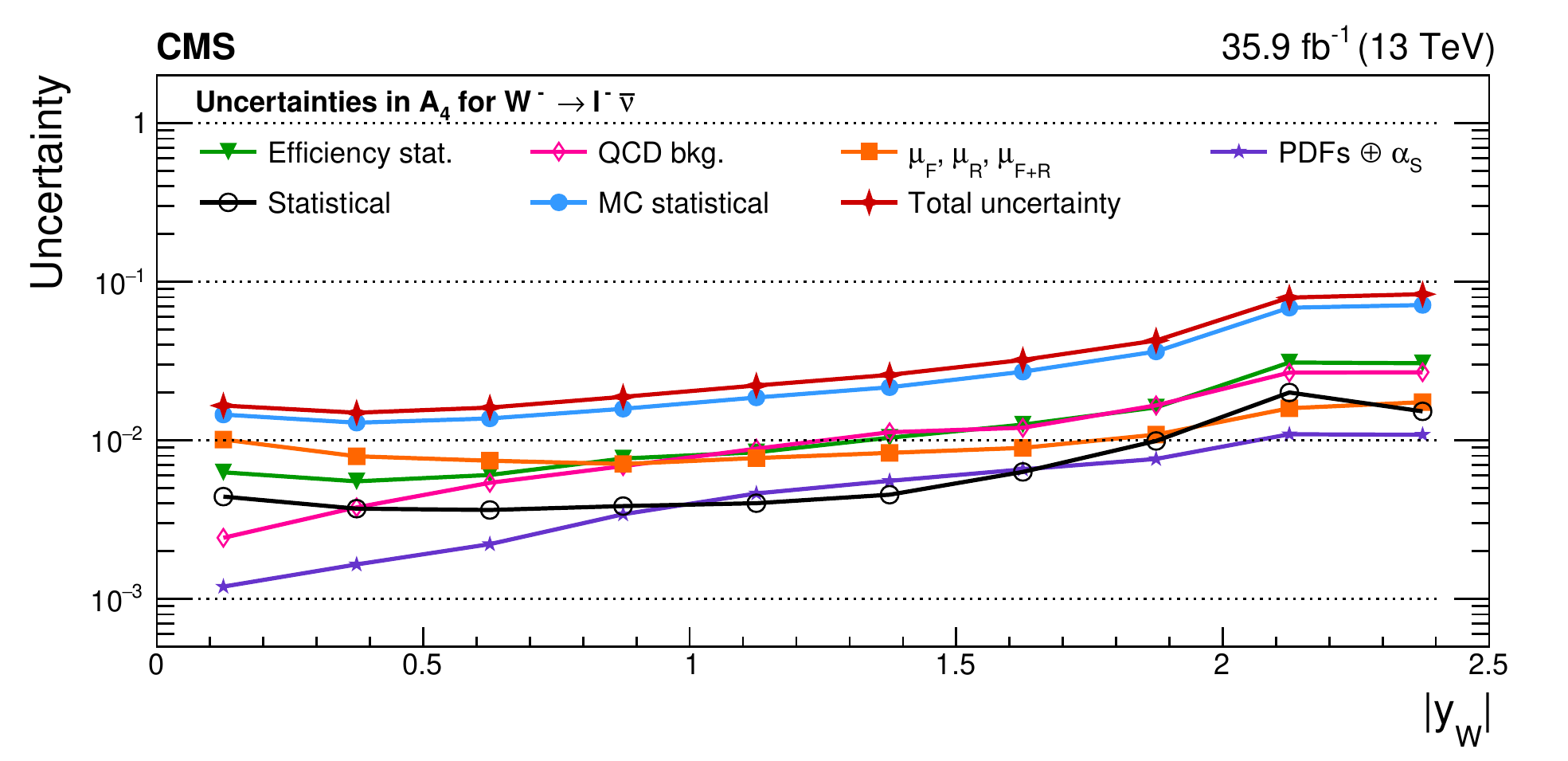} \\
\caption{Impacts on the A$_4$ coefficient as functions of the \PW
  boson rapidity for \wplus (upper) and \wminus (lower) bosons in
  the helicity fit. The groups of uncertainties subleading to
  the ones shown are suppressed for
  simplicity. \label{fig:impactsA4}}
\end{figure}

Figure~\ref{fig:impactsA4} shows the impacts of the nuisance parameter groups on the A$_4$
coefficient as a function of \absyw for both charges of the \PW boson.

\subsection{2D differential cross section}
\label{appsub:2ddiff}

\begin{figure*}[t!hbp]
\centering
\includegraphics[width=1.0\linewidth]{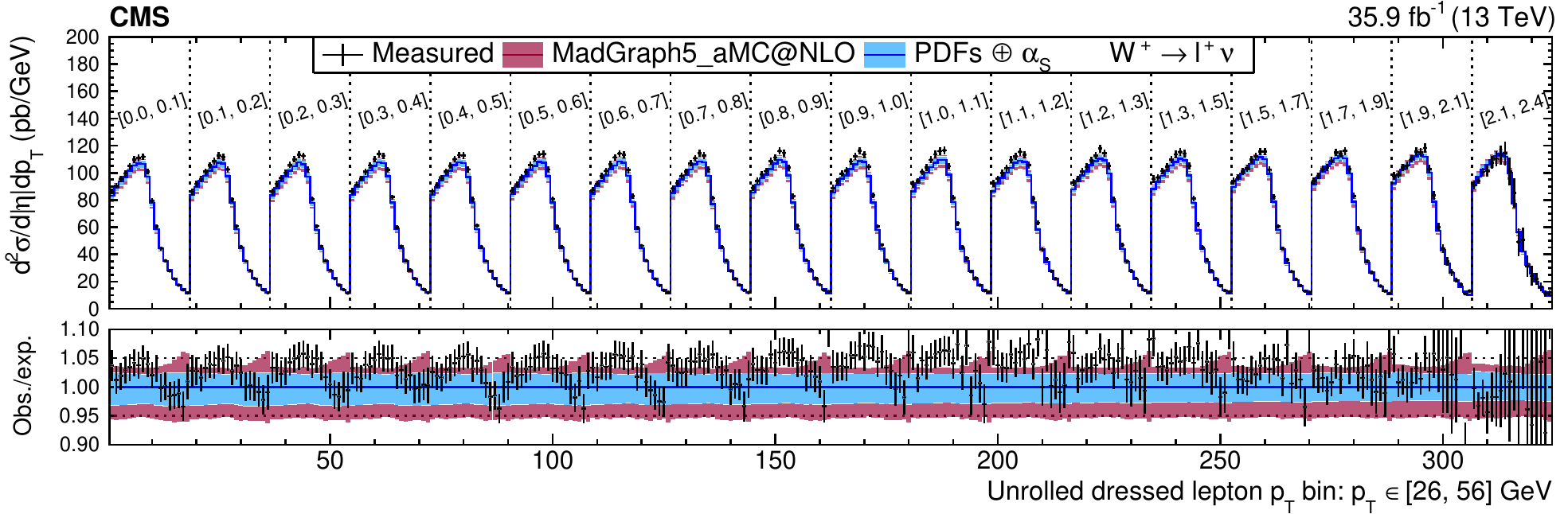} \\
\includegraphics[width=1.0\linewidth]{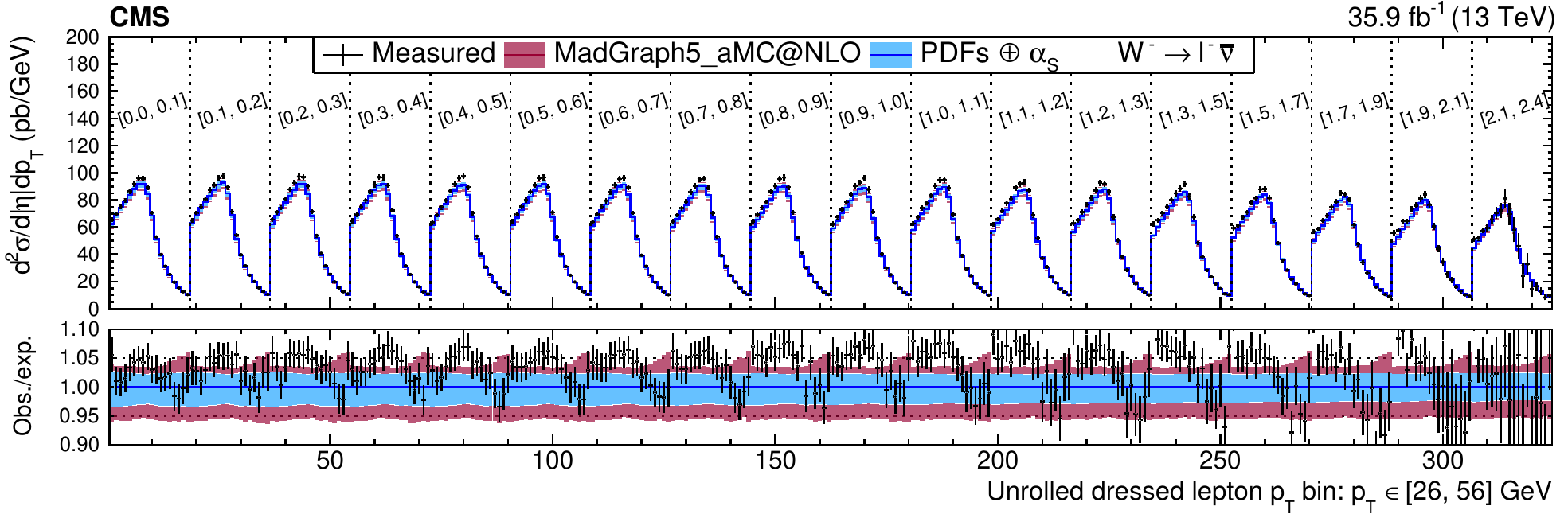} \\
\caption{Unrolled cross sections for the combined muon and electron
  channel fit unrolled along \ptl in bins of \absleta for \wplus
  (upper) and \wminus (lower) bosons. The colored bands represent the prediction from \MGvATNLO with the expected uncertainty from the quadrature sum of the PDF$\oplus\alpS$ variations (blue) and the \muF and \muR scales (bordeaux).\label{fig:combinedOnly_unrolled_abs}}
\end{figure*}

\begin{figure}[h!tbp]
\centering
\includegraphics[width=0.49\textwidth]{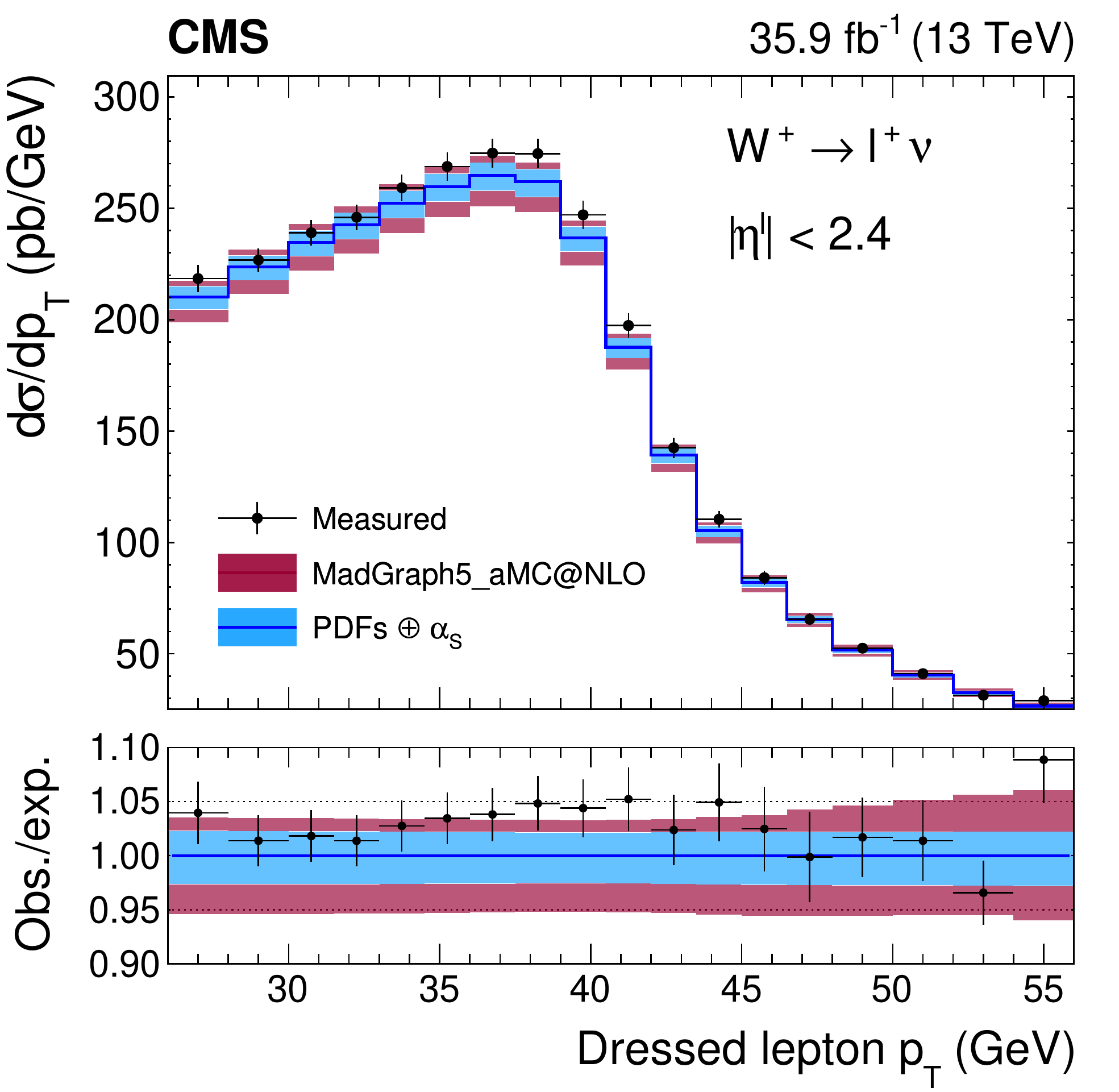}
\includegraphics[width=0.49\textwidth]{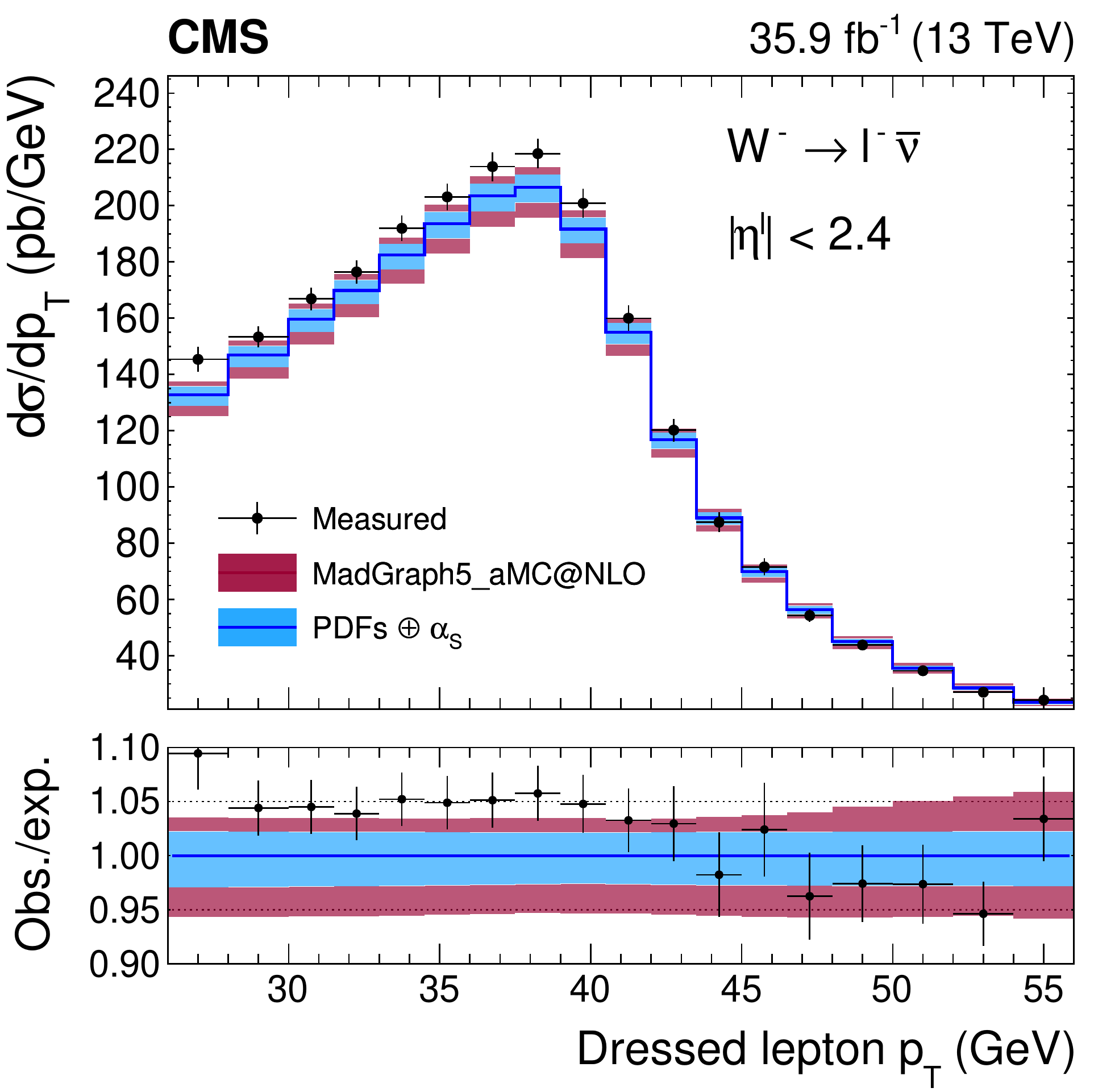}
\caption{Absolute cross sections as functions of \ptl, integrated over
  \absleta for \wplus (\cmsLeft) and \wminus (\cmsRight) bosons.  The colored bands represent the prediction from \MGvATNLO with the expected uncertainty from the quadrature sum of the PDF$\oplus\alpS$ variations (blue) and the \muF and \muR scales (bordeaux).\label{fig:absolutePt}}
\end{figure}

\begin{figure}[h!tbp]
\centering
\includegraphics[width=0.49\textwidth]{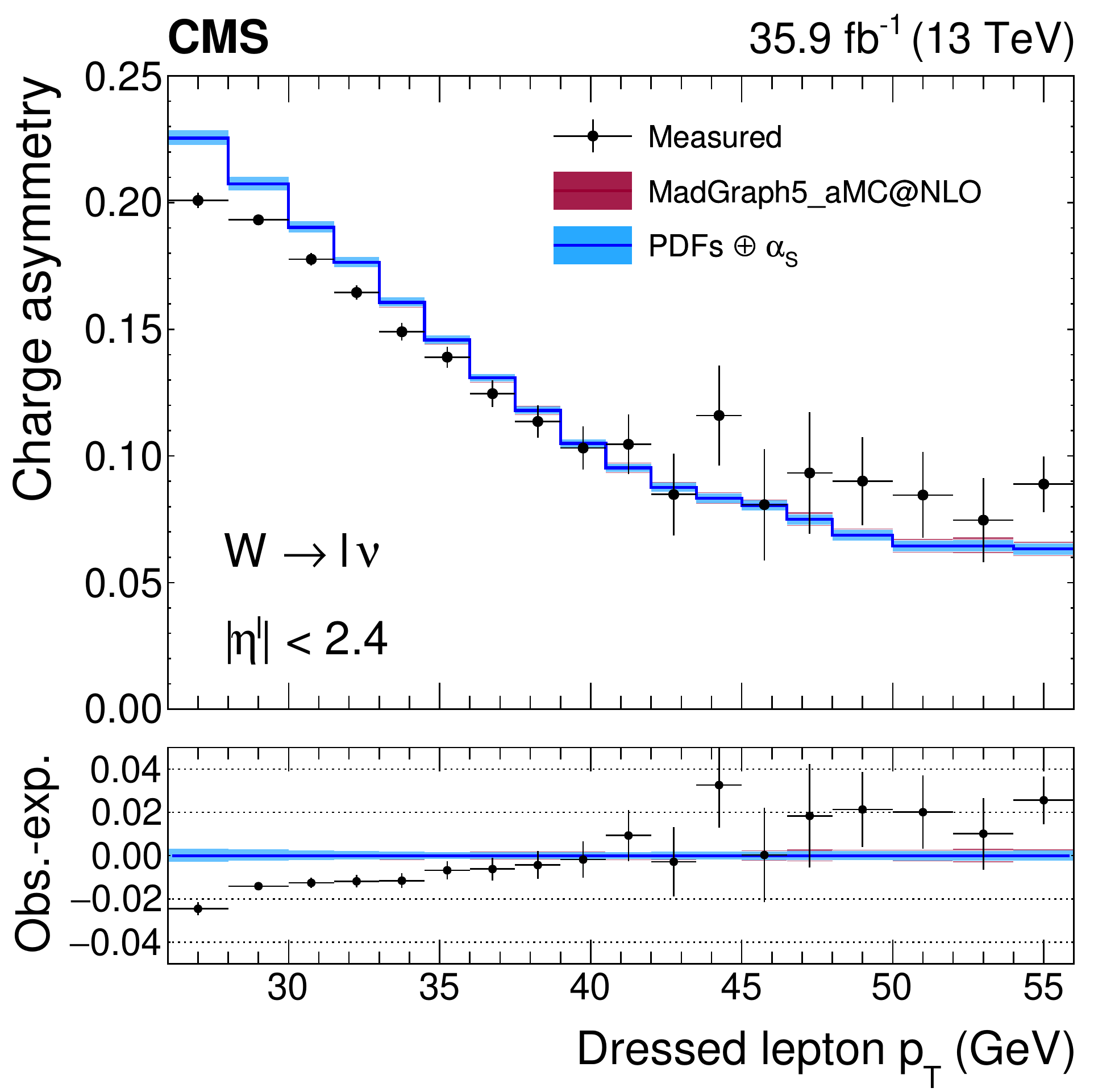} \\
\caption{\PW charge asymmetry as functions of \ptl, integrated
  over \absleta.  The colored bands represent the prediction from \MGvATNLO with the expected uncertainty from the quadrature sum of the PDF$\oplus\alpS$ variations (blue) and the \muF and \muR scales (bordeaux). \label{fig:absolutePtAsy}}
\end{figure}

Figure~\ref{fig:combinedOnly_unrolled_abs} shows the absolute cross sections for the combined 
muon and electron channel fit unrolled along \ptl, in bins of \absleta for both charges
of the \PW boson. Figure~\ref{fig:absolutePt} and \ref{fig:absolutePtAsy} shows these same absolute cross sections and charge asymmetry, but
integrated over all the bins in \absleta.

\begin{figure}[h!tbp]
\centering
\includegraphics[width=0.49\textwidth]{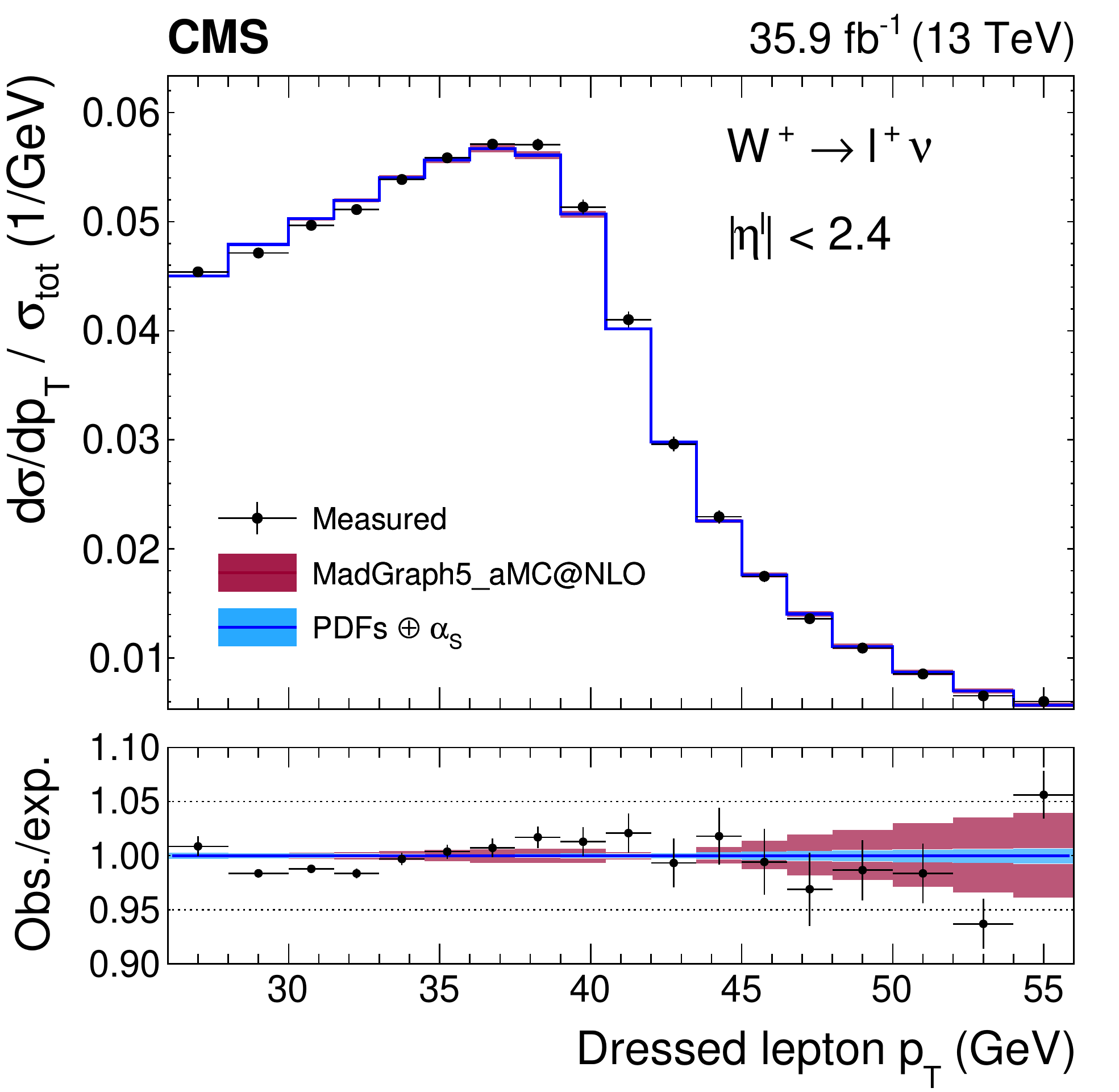}
\includegraphics[width=0.49\textwidth]{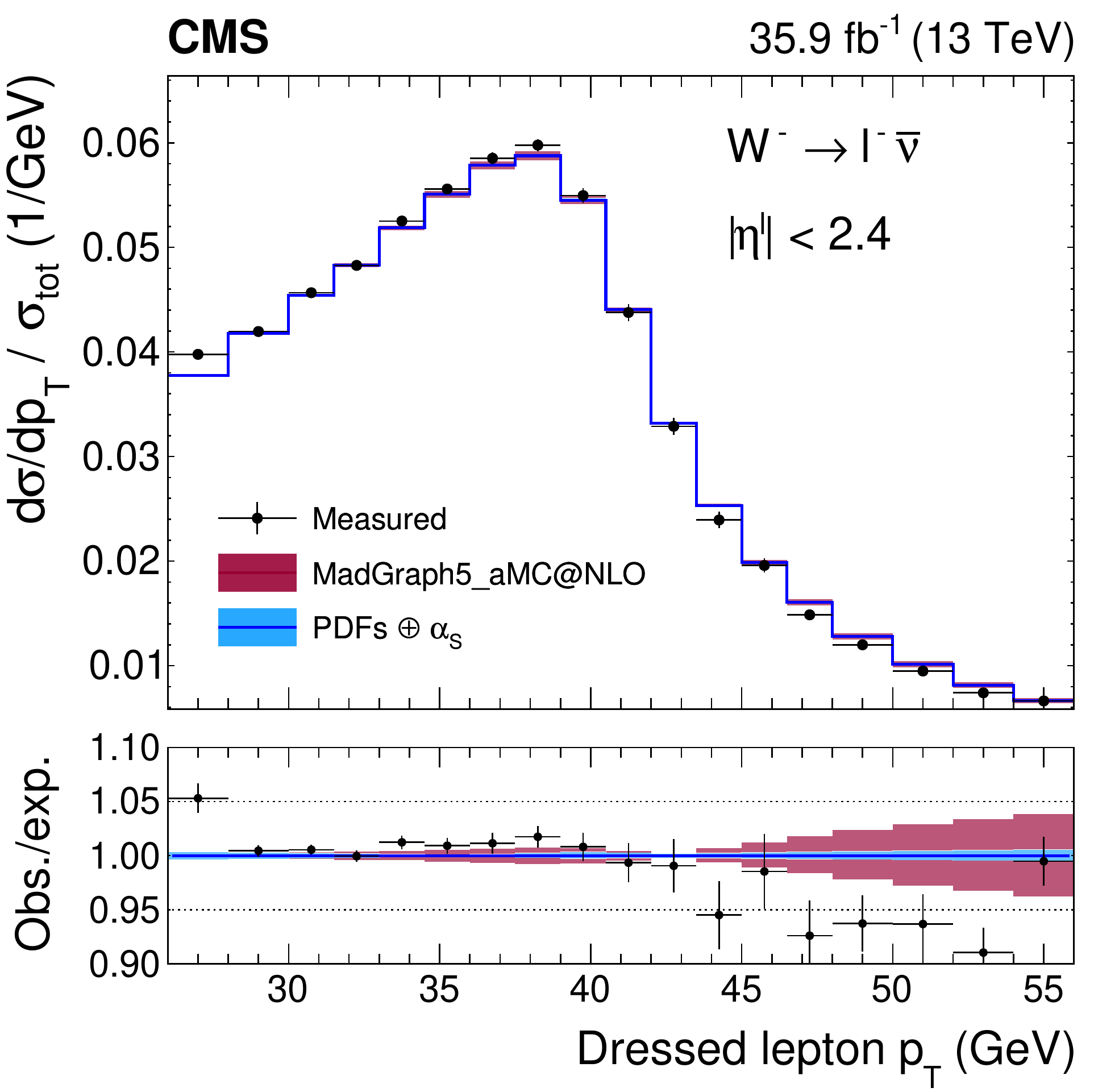}
\caption{Normalized cross sections as functions of \ptl, integrated
  over \absleta for \wplus (\cmsLeft) and \wminus (\cmsRight) bosons.  The colored bands represent the prediction from \MGvATNLO with the expected uncertainty from the quadrature sum of the PDF$\oplus\alpS$ variations (blue) and the \muF and \muR scales (bordeaux). \label{fig:normalizedPt}}
\end{figure}

\begin{figure}[h!tbp]
\centering
\includegraphics[width=0.49\textwidth]{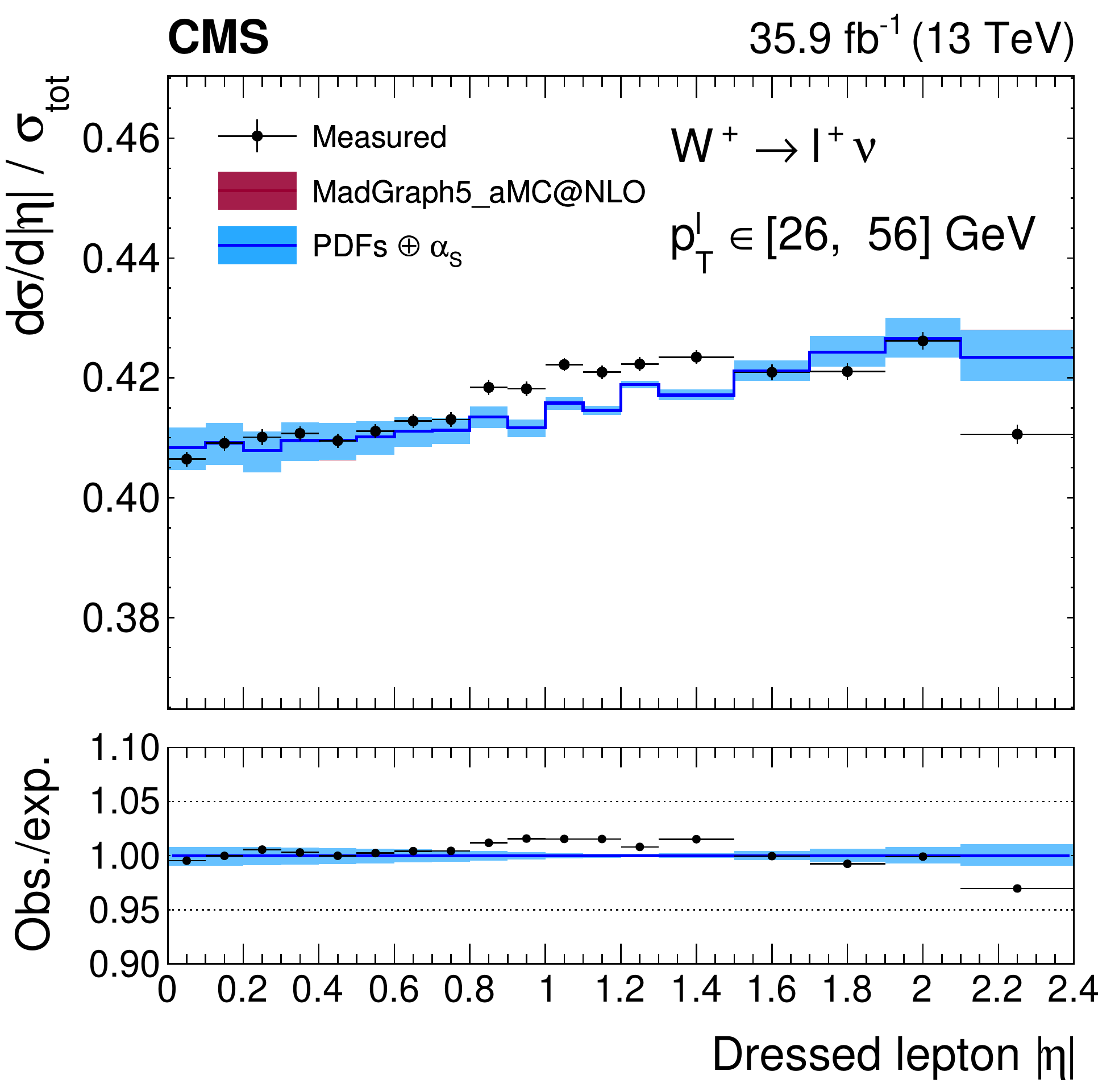}
\includegraphics[width=0.49\textwidth]{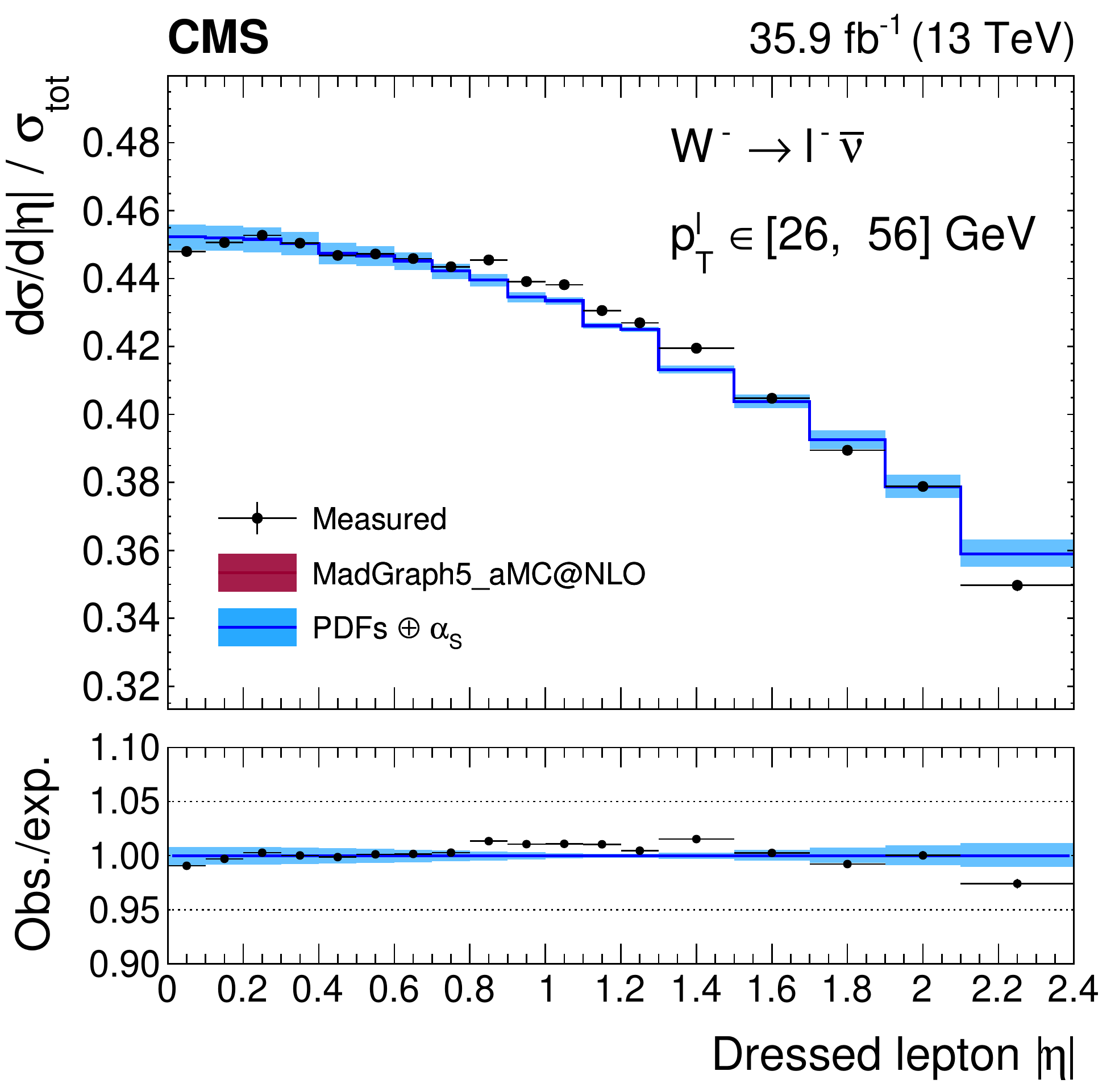}
\caption{Normalized cross sections as functions of \absleta,
  integrated over \ptl for \wplus (\cmsLeft) and \wminus (\cmsRight) bosons. The colored bands represent the prediction from \MGvATNLO with the expected uncertainty from the quadrature sum of the PDF$\oplus\alpS$ variations (blue) and the \muF and \muR scales (bordeaux). The uncertainty band is almost entirely dominated
  by the PDF$\oplus\alpS$ variations, while the missing higher order QCD
  uncertainties almost perfectly cancel and are therefore invisible.\label{fig:combinedFit_normalized_eta}}
\end{figure}

Figures~\ref{fig:normalizedPt} and~\ref{fig:combinedFit_normalized_eta} show the normalized differential
cross sections for both charges of the \PW boson as a function of \ptl and \absleta, respectively. The charge asymmetry 
as a function of \ptl is also shown in Fig.~\ref{fig:normalizedPt}.

\begin{figure}[h!tbp]
\centering
\includegraphics[width=0.97\linewidth]{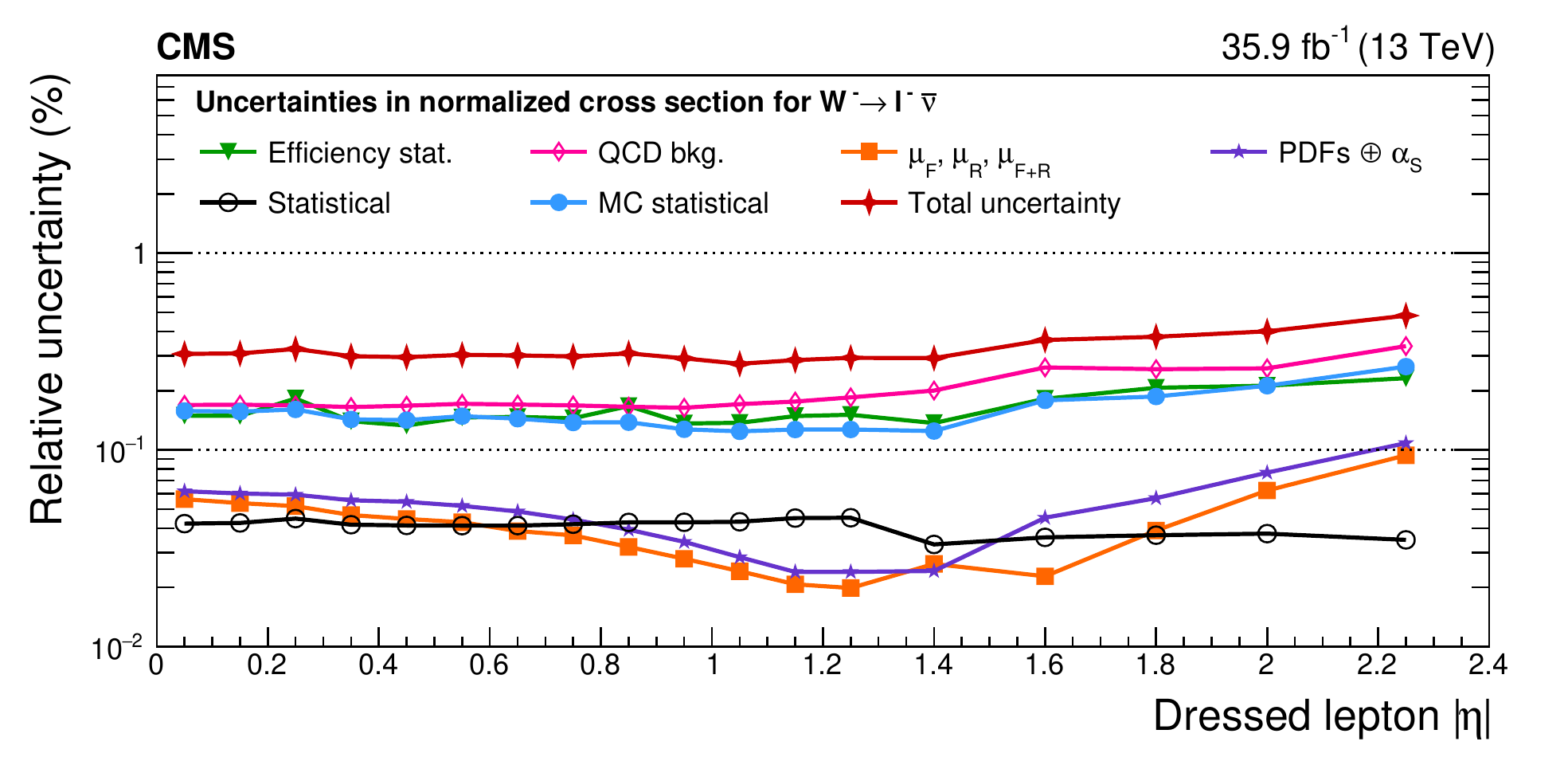} \\
\caption{Remaining impacts of the nuisance groups on the normalized
  cross sections as functions of \absleta, integrated in \ptl, for
  \wminus bosons in the double-differential cross section fit. The
  groups of uncertainties subleading to the ones shown are
  suppressed for simplicity. \label{fig:impactsNormEtaWminus}}
\end{figure}

\begin{figure}[h!tbp]
\centering
\includegraphics[width=0.97\linewidth]{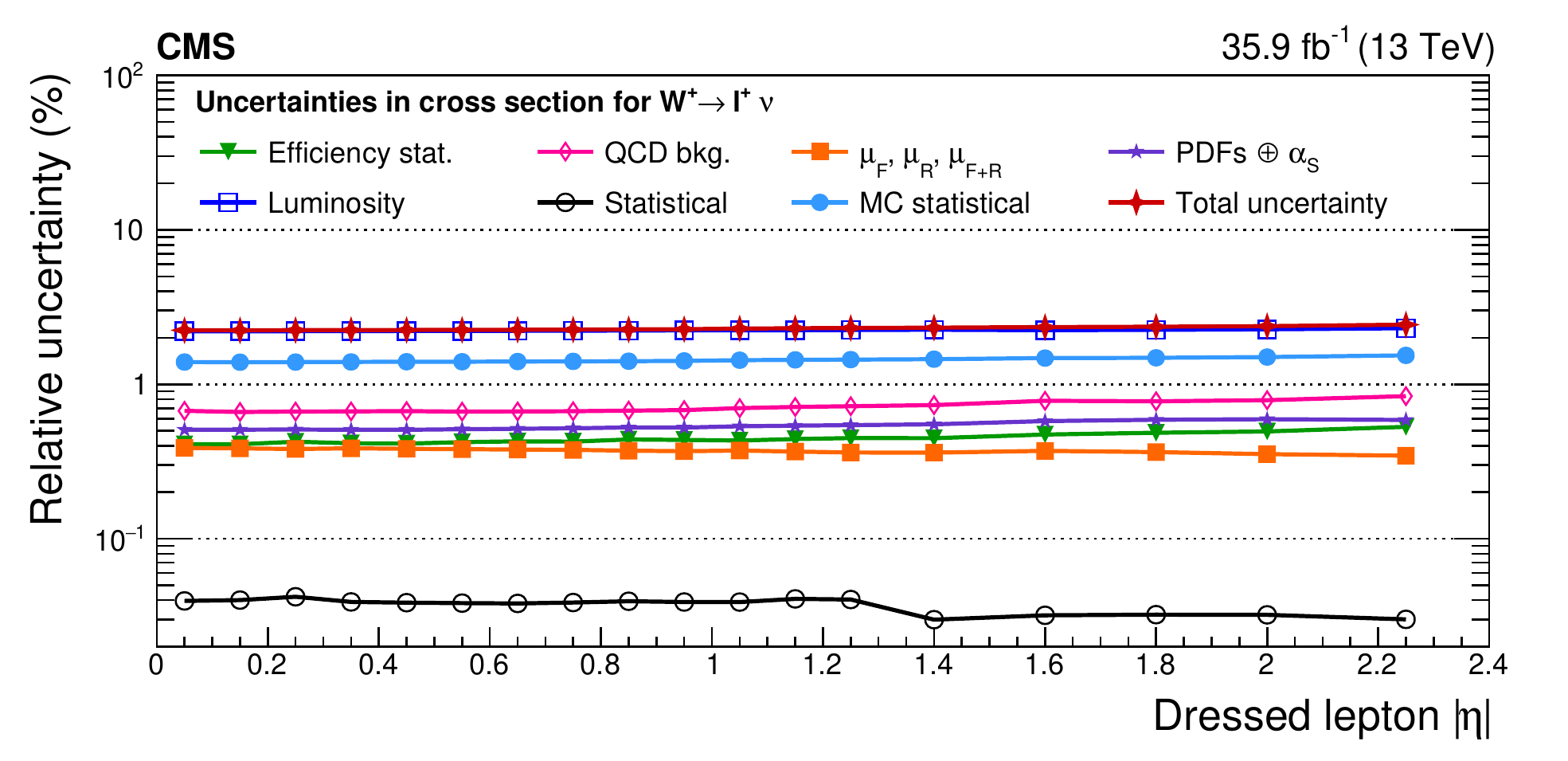} \\
\includegraphics[width=0.97\linewidth]{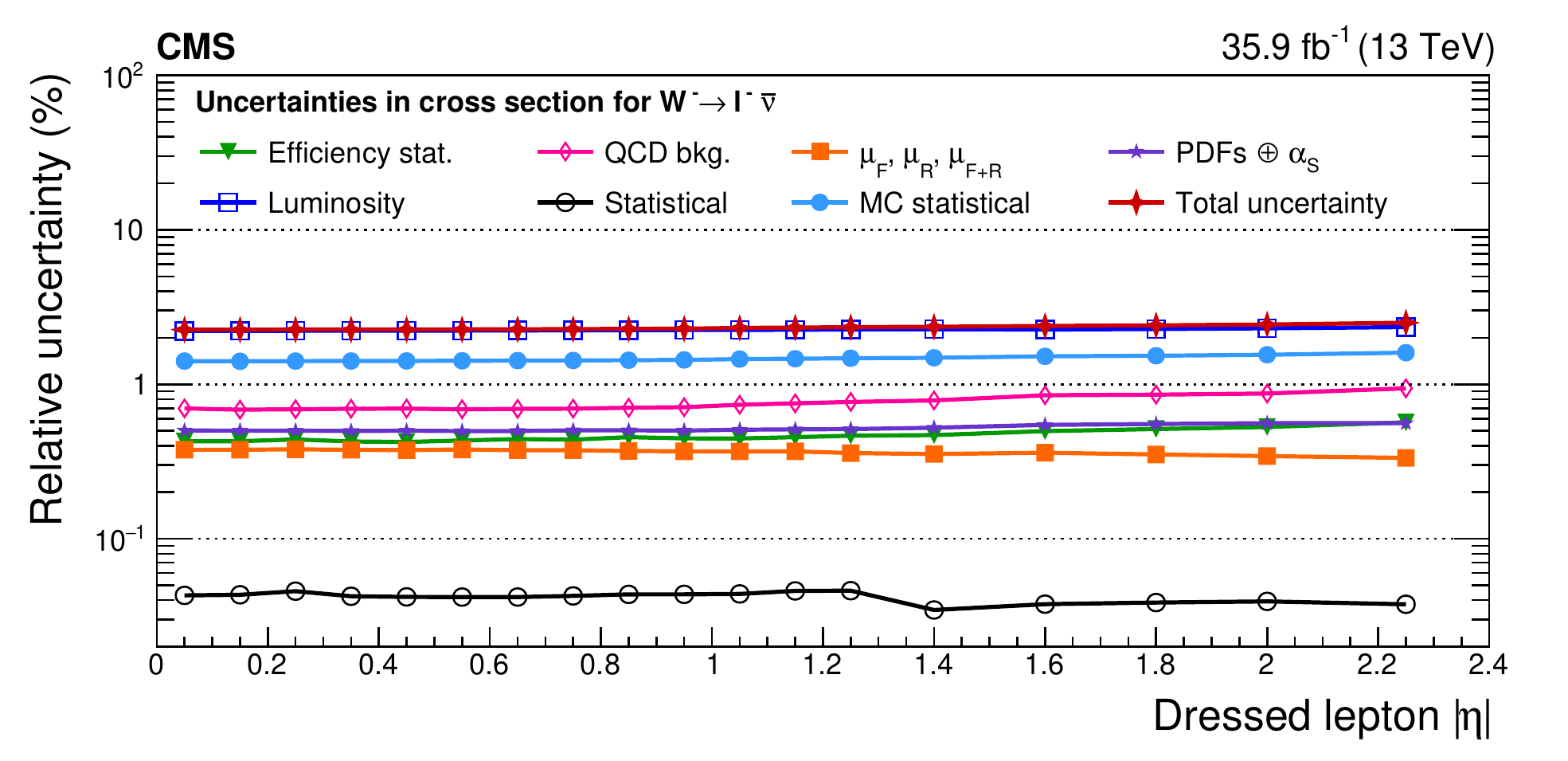} \\
\caption{Remaining impacts of the nuisance groups on the absolute
  cross sections as functions of \absleta, integrated in \ptl, for
  \wplus (upper) and \wminus (lower) bosons in the double-differential
  cross section fit. The groups of uncertainties subleading to
  the ones shown are suppressed for
  simplicity. \label{fig:impactsAbsEta}}
\end{figure}

\begin{figure}[h!tbp]
\centering
\includegraphics[width=0.97\linewidth]{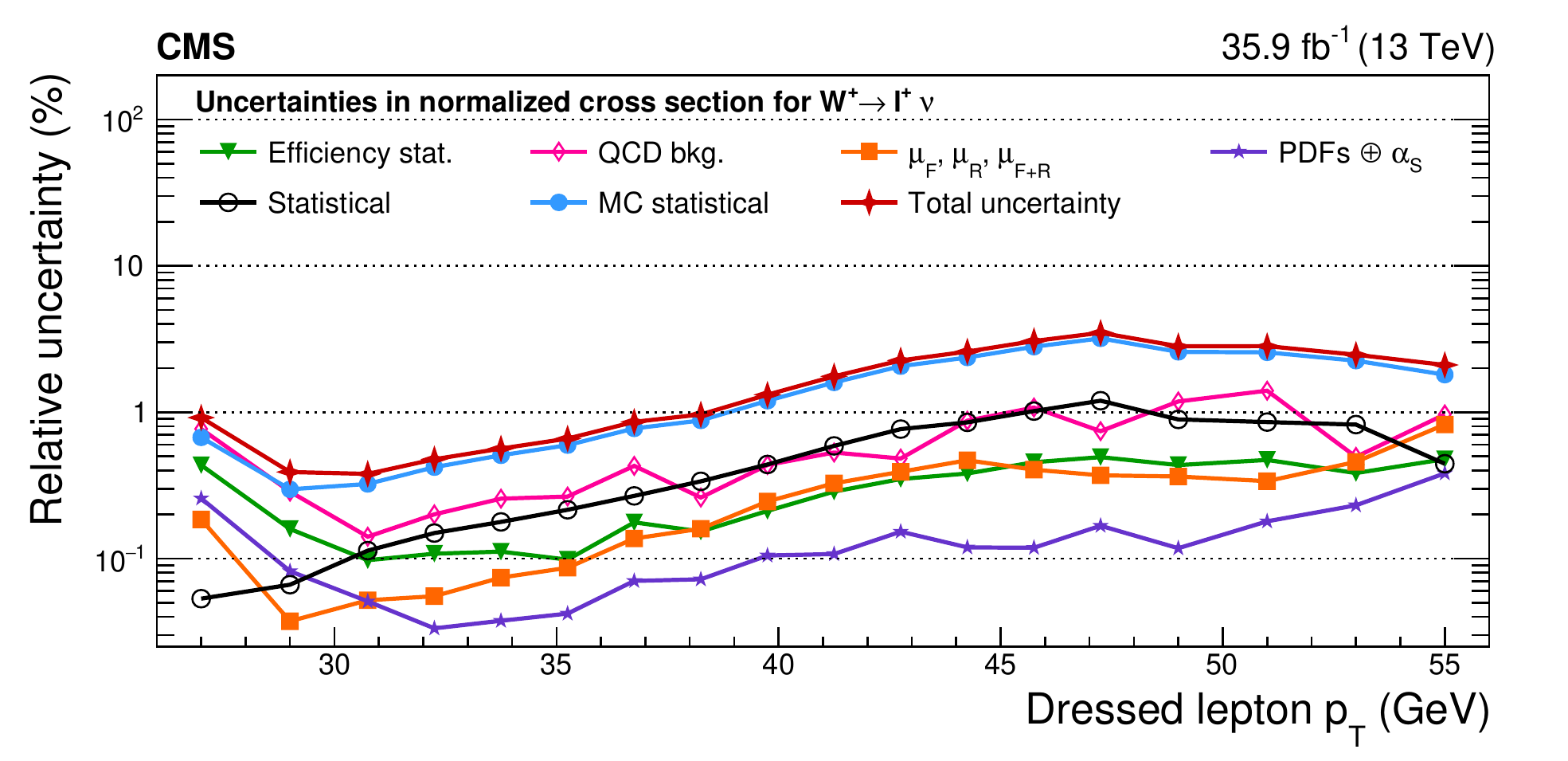} \\
\includegraphics[width=0.97\linewidth]{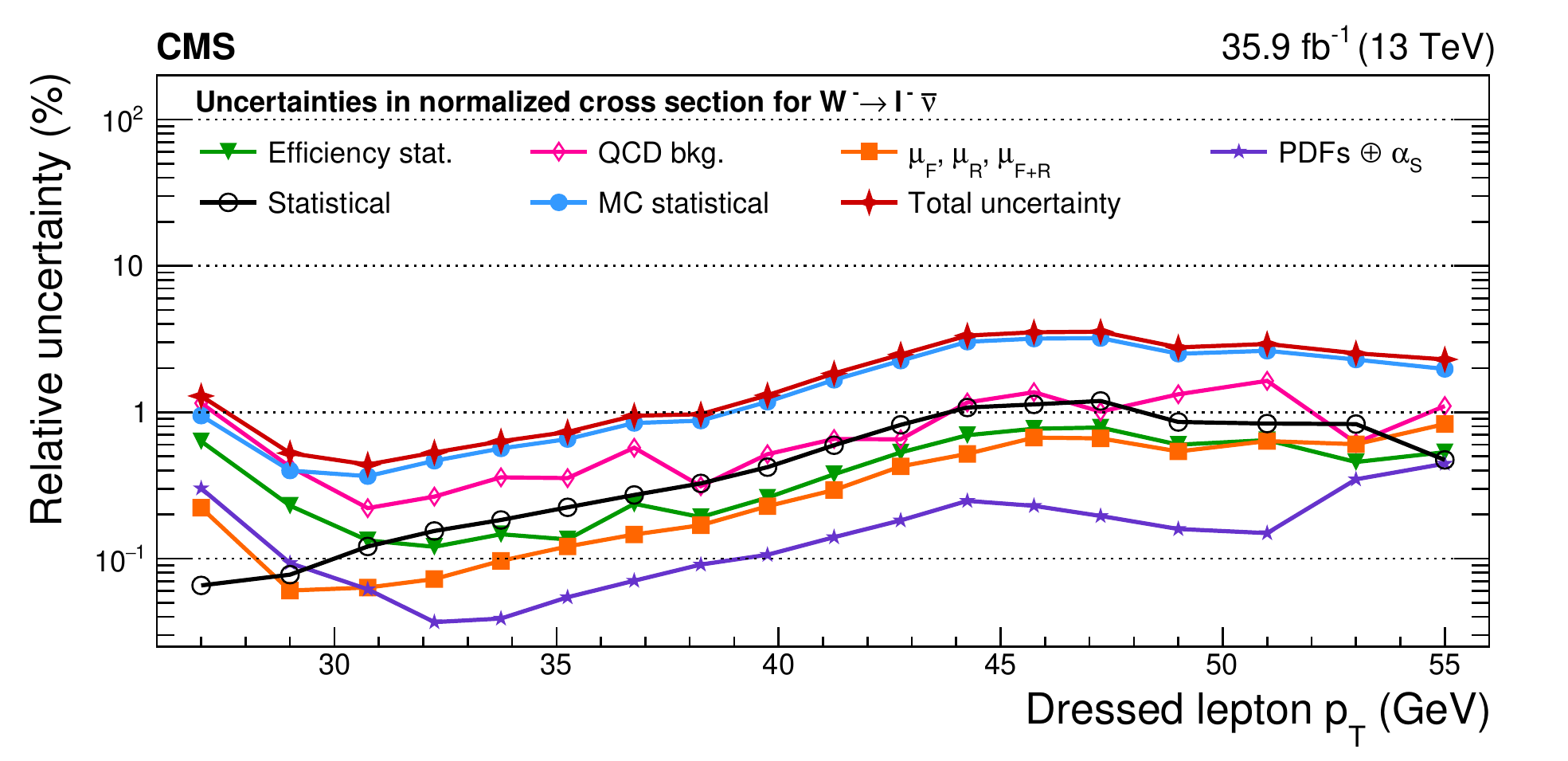} \\
\includegraphics[width=0.97\linewidth]{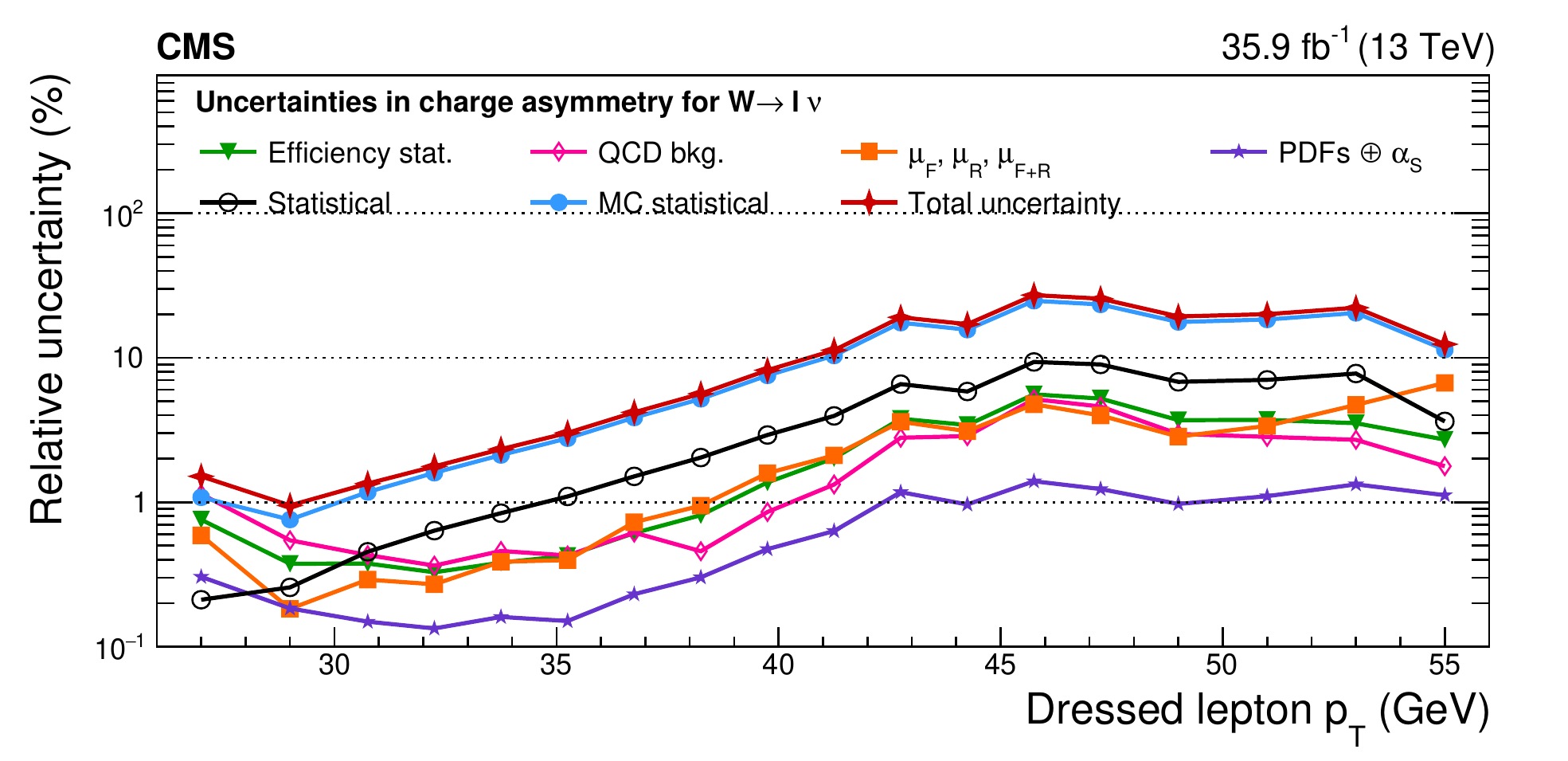} \\
\caption{Remaining impacts of the nuisance groups on the normalized
  cross sections as functions of \ptl, integrated over \absleta, for
  \wplus (upper), \wminus (middle) bosons, and the resulting charge
  asymmetry (lower) in the double-differential cross section fit. The
  groups of uncertainties subleading to the ones shown are
  suppressed for simplicity. \label{fig:impactsNormPt}}
\end{figure}

\begin{figure}[h!tbp]
\centering
\includegraphics[width=0.97\linewidth]{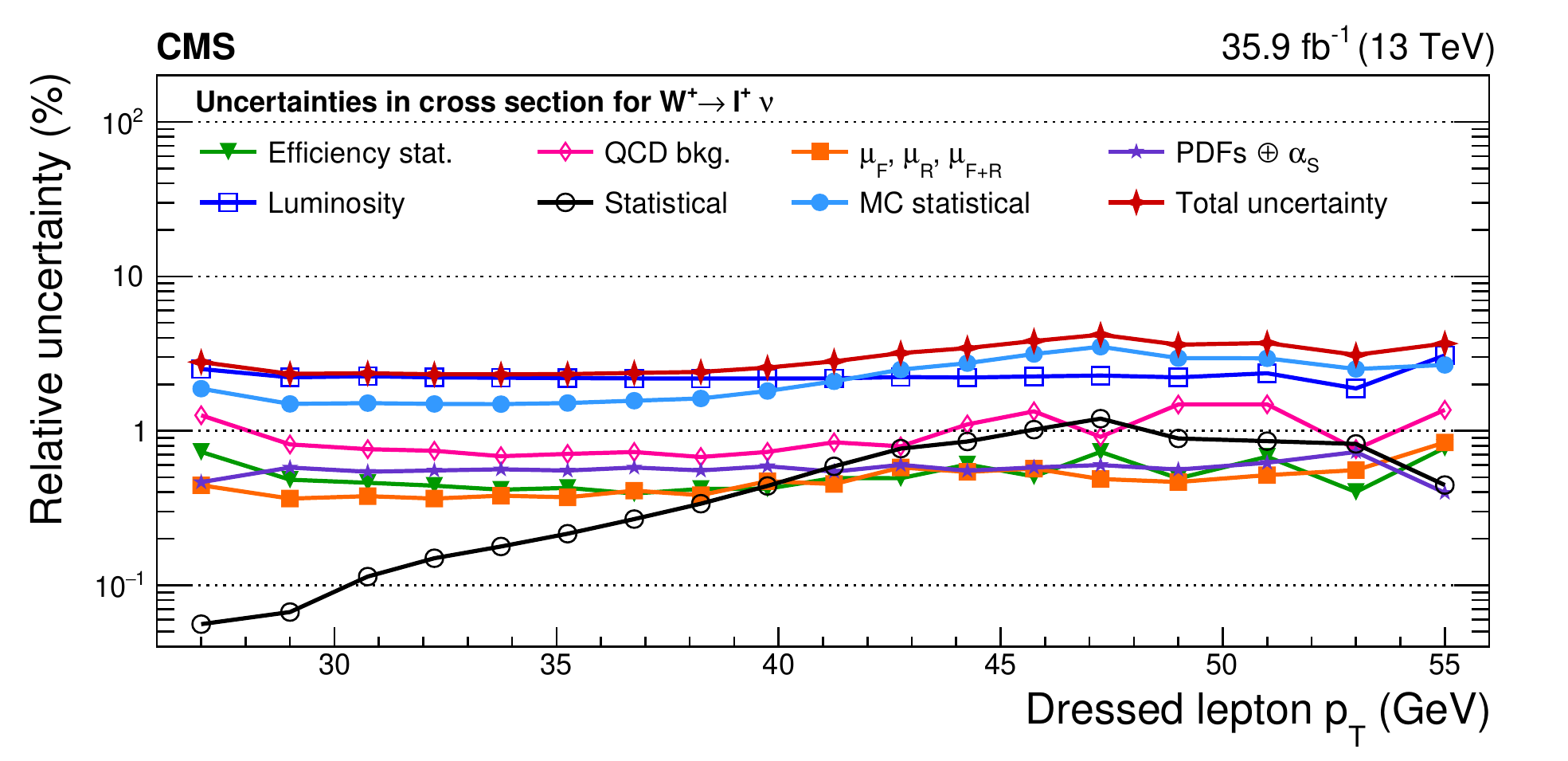} \\
\includegraphics[width=0.97\linewidth]{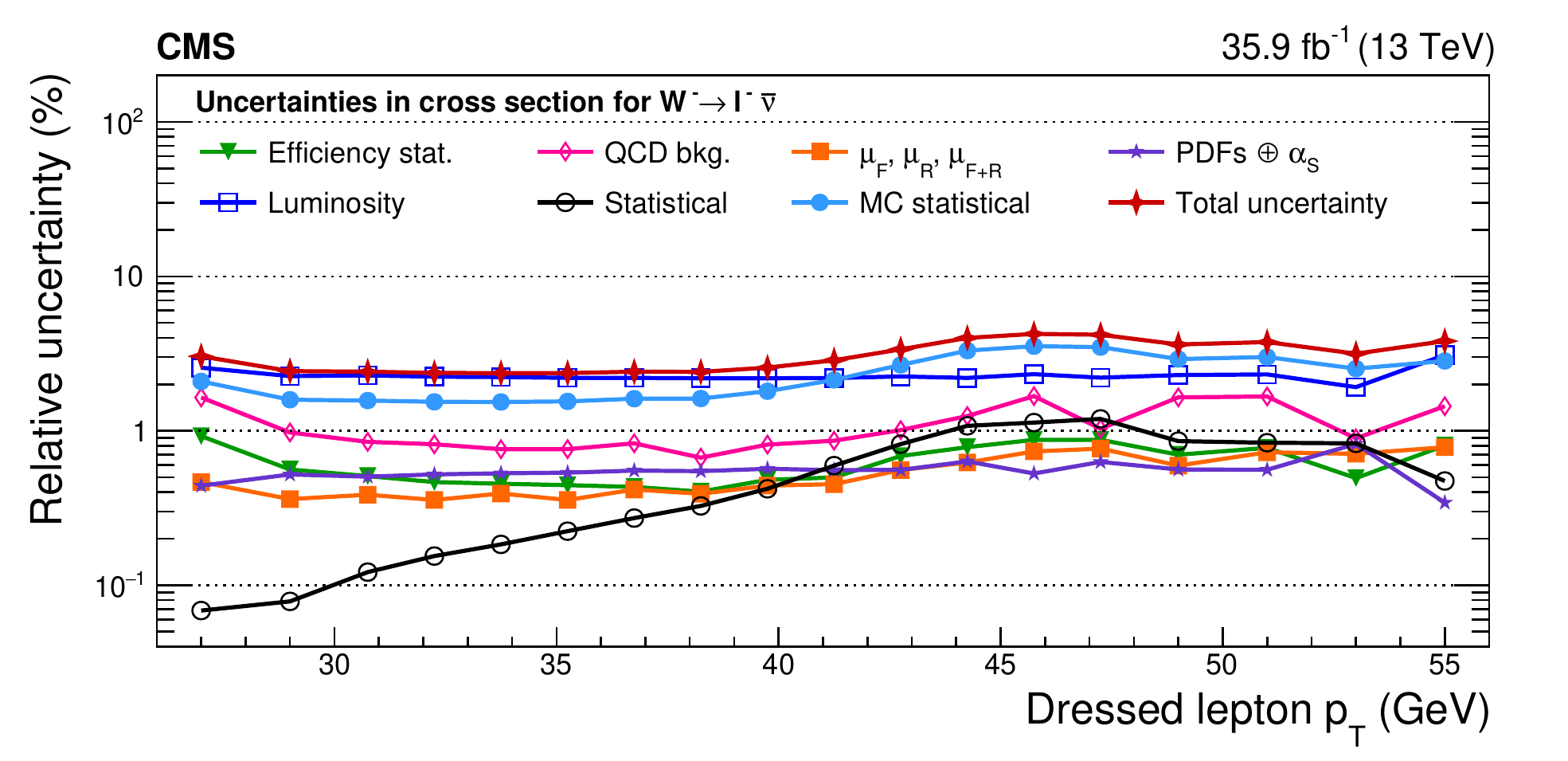} \\
\caption{Remaining impacts of the nuisance groups on the absolute
  cross sections as functions of \ptl, integrated over \absleta, for
  \wplus (upper) and \wminus (lower) bosons in the double-differential
  cross section fit. The groups of uncertainties subleading
  to the ones shown are suppressed for
  simplicity. \label{fig:impactsAbsPt}}
\end{figure}

Figures~\ref{fig:impactsNormEtaWminus}-~\ref{fig:impactsAbsPt} show the remaining impacts of the 2D differential
cross sections analysis, which were omitted in the main paper: the impacts on the normalized \PW cross 
sections as a function of \absleta for \PW bosons with negative charge in Fig.~\ref{fig:impactsNormEtaWminus}; the impacts
on the absolute cross sections for both charges as a function of \ptl in Fig.~\ref{fig:impactsAbsEta}; the impacts on 
the normalized \PW boson production cross sections as a function of \ptl for both charges, along with the impacts
on the charge asymmetry, in Fig.~\ref{fig:impactsNormPt}; and, finally, the impacts on the absolute \PW boson production
cross sections as a function of \ptl, for both charges of the \PW boson in Fig.~\ref{fig:impactsAbsPt}.

\cleardoublepage \section{The CMS Collaboration \label{app:collab}}\begin{sloppypar}\hyphenpenalty=5000\widowpenalty=500\clubpenalty=5000\input{SMP-18-012-authorlist.tex}\end{sloppypar}
\end{document}

%% file: SMP-18-012-authorlist.tex
\vskip\cmsinstskip
\textbf{Yerevan Physics Institute, Yerevan, Armenia}\\*[0pt]
A.M.~Sirunyan$^{\textrm{\dag}}$, A.~Tumasyan
\vskip\cmsinstskip
\textbf{Institut f\"{u}r Hochenergiephysik, Wien, Austria}\\*[0pt]
W.~Adam, F.~Ambrogi, T.~Bergauer, M.~Dragicevic, J.~Er\"{o}, A.~Escalante~Del~Valle, R.~Fr\"{u}hwirth\cmsAuthorMark{1}, M.~Jeitler\cmsAuthorMark{1}, N.~Krammer, L.~Lechner, D.~Liko, T.~Madlener, I.~Mikulec, N.~Rad, J.~Schieck\cmsAuthorMark{1}, R.~Sch\"{o}fbeck, M.~Spanring, S.~Templ, W.~Waltenberger, C.-E.~Wulz\cmsAuthorMark{1}, M.~Zarucki
% ...
\vskip\cmsinstskip
\dag: Deceased\\
1:  Also at Vienna University of Technology, Vienna, Austria\\
% ...
